\documentclass[aps,twocolumn,prd]{revtex4}
\usepackage{hyperref}
 \usepackage{breakurl}
\usepackage{url}
\usepackage{latexsym}
\usepackage{epsfig}
\usepackage{scrextend}

\usepackage{color}
\usepackage{amssymb}
\usepackage{amsmath}
\usepackage{graphicx}

\def\be{\begin{equation}}
\def\ee{\end{equation}}
\def\bea{\begin{eqnarray}}
\def\eea{\end{eqnarray}}
\def\ba{\begin{array}} 
\def\ea{\end{array}}
\def\bc{\begin{center}}
\def\ec{\end{center}}

\def\ghost#1{}

\def\simge{\mathrel{%
   \rlap{\raise 0.511ex \hbox{$>$}}{\lower 0.511ex \hbox{$\sim$}}}}
\def\simle{\mathrel{
   \rlap{\raise 0.511ex \hbox{$<$}}{\lower 0.511ex \hbox{$\sim$}}}}

\def\dis{\displaystyle}

\begin{document}

\title
{\Large The Supersymmetric Standard Model \vspace{2.5mm} \\}

\author{Pierre {\sc FAYET}
\vspace{4.5mm} \\ \small }

\affiliation{Laboratoire de Physique Th\'eorique de l'\'Ecole Normale Sup\'erieure
\vspace{.1mm}\\ {\footnotesize \hspace*{-1.5mm}{} UMR 8549 CNRS-ENS, associ\'ee \`a l'Universit\'e Paris~6 UPMC$\!\!\!$}
\vspace{.1mm}\\ 24 rue Lhomond, 75231 Paris cedex 05, France\vspace{.8mm}\\
\hbox{(and D\'epartement de physique, \'Ecole polytechnique, 91128 Palai\-seau cedex, France)}\vspace{-2mm}\\}

\begin{abstract}
The Standard Model may be included within a supersymmetric theory, postulating 
new {\it sparticles} that differ by half-a-unit of spin from their standard model partners, 
and by a new quantum number called $R$-parity. The lightest one, usually a neutralino, 
is expected to be stable and a possible candidate for dark matter.

\vspace{.6mm}

The electroweak breaking requires two doublets, leading to several charged and neutral 
Brout-Englert-Higgs bosons. This also leads to gauge/Higgs unification by providing 
extra spin-0 partners for  the spin-1 $W^\pm$ and $Z$.
It  offers the possibility to view, up to a mixing angle,  the new 125 GeV \linebreak 
boson as the spin-0 partner of the $Z$ under two supersymmetry transformations, 
i.e.~as a  $Z$ that would be deprived of its spin. Supersymmetry then relates {\it two existing particles of different spins}, in spite of their different gauge symmetry properties, through supersymmetry transformations acting on physical fields in a non-polynomial way.

\vspace{.8mm}

ÊWe also discuss how the compactification of extra dimensions, relying on $R$-parity  and other discrete symmetries, may determine both the
supersymmetry-breaking and grand-unification scales.

\vspace{2.5mm}

\hfill Preprint LPTENS 15/03

\vspace*{-5mm}
{ \ }

\end{abstract}

\maketitle

Is there a ÒsuperworldÓ of new particles\,? Could half of the particles at least have escaped our observations\,? 
Do new states of matter exist\,? After the prediction of antimatter by Dirac, supersymmetric extensions of the standard model 
lead to anticipate the possible existence of spin-0 squarks and sleptons, with the gluons, $W^\pm,\, Z$ and photon associated with 
gluinos, charginos and neutralinos \cite{R,ssm,ff,fayet79}. 
These new states  differ from ordinary particles by half-a-unit of spin and are distinguished by a $R$-parity quantum number 
related to baryon and lepton numbers,  making the lightest superpartner stable, and a possible candidate for the dark matter 
of the Universe. 
Spontaneous electroweak breaking is induced by two \hbox{spin-0} doublets instead of one in the standard model,
leading to several charged and neutral \hbox{spin-0} BEH bosons. These may even be related to the massive gauge bosons, 
with the possibility that the new 125 GeV boson recently discovered at CERN \cite{higgs,higgs2} be a spin-0 partner of the $Z$ under {\it two} supersymmetry transformations \cite{R,gh,prd14}.
But, where is all this coming from\,?
\vspace{-3mm}

\section{Fundamental \vspace{1.5mm}  interactions, \vspace{1.5mm} \hbox{\ \ \ symmetry breaking} \hbox{\ \ \ and the new spin-0 boson}}

\vspace{-1mm}

Special relativity and quantum mechanics, operating within quantum field theory, led to the Standard Model of particles and  interactions (SM).
It has met a long series of successes with the discoveries of weak neutral currents (1973), charmed particles (1974-76), gluons mediators of  strong interactions (1979),
 $W^\pm\!$ and $Z$'s mediators of weak interactions (1983), and the sixth quark known as the top quark (1995).
 Weak, electromagnetic and strong interactions are well understood from the exchanges of spin-1 mediators
 between \hbox{spin-$\frac{1}{2}$} quarks and leptons, generically referred to  as the constituents of matter  (cf. Table \ref{tab:sm0}).

\begin{table}[t]
\vspace{-2mm}
\caption{Particle content of the standard model.
\vspace{-1mm}
}
\label{tab:sm0}
\bc
\vspace{1mm}
\begin{tabular}{|ccc|}
\hline  && \vspace{-1mm}\\
\ \ spin-1 gauge bosons\,:  
&&
{ gluons, \ $W^+\!,\ W^-\!,\,Z$, \,photon}\ \ \ \ 
\\[3mm] 
\ \ spin-$\frac{1}{2}$ fermions\,: 
&&
$\left\{\ba{l}
\,  \hbox{6 quarks:} \ \,  \left( \! \ba{c}\vspace{-5mm}\\   u\vspace{-.5mm}\\ d\ea\!  \right)$    $\left(\!  \ba{c}\vspace{-5mm}\\  c\vspace{-.5mm}\\ s\ea\!  \right)$  $\left( \! \ba{c} \vspace{-5mm}\\ t \vspace{-.5mm}\\ b \ea \! \right)\!\!\!
\vspace{2.5mm}\\
 \hbox{6 leptons:} \  \left(\!\!  \ba{c}\vspace{-5mm}\\ \nu_e\vspace{-.5mm}\\ e^-\ea \!\!\!\right)$     $\left( \!  \! \ba{c} \vspace{-5mm}\\ \nu_\mu\!\vspace{-.5mm}\\ \mu^-\!\ea\!\! \right)\!$   
  $\left(\!\!  \ba{c} \vspace{-5mm}\\ \nu_\tau\!\vspace{-.5mm}\\ \tau^-\!\ea\! \! \right)\!\!\!\!
 \vspace{1mm}\\
\ea\ \  \right.$
\\[9mm] 
 spin-0  &&
{\color{black} scalar BEH boson}
\\[-2mm] && \\
\hline
\end{tabular}
\vspace{-6mm}{}
\ec
\end{table}

\vspace{2mm}

The eight gluons mediate the strong interactions, invariant under the color $SU(3)$ gauge group.
The $W^\pm,\,Z$ and photon
are associated with 
the $SU(2)\times U(1)$ electroweak gauge group  \cite{ew1,ew2,ws,ws2}.
$\!\!$The $W^\pm$ and $Z$ masses, $m_W\simeq 80$ GeV$/c^2$ and $m_Z\simeq 91$ GeV$/c^2$,  are generated through the spontaneous breaking of the electro\-weak symmetry, 
induced in the standard model by a doublet of spin-0 fields $\varphi$  \cite{ws,ws2}.
Three of its four real components, instead of being associated with  unwanted massless  Goldstone bosons \cite{g}, 
are eliminated by the Brout-Englert-Higgs mechanism \cite{be,h,ghk} to provide the extra degrees of freedom
\vspace{-.5mm}
 for the massive $W^\pm$ and $Z$.
The fourth component,  taken as $\phi= $ $\sqrt{2\,\varphi^\dagger\varphi}$, \,adjusts so that  the potential 
\be
\label{V}
V(\varphi)\,=\,\lambda_{\hbox{\tiny SM}}\,(\varphi^\dagger\varphi)^2-\,\mu_{\hbox{\tiny SM}}^{\,2}\,\varphi^\dagger\varphi
\ee
\noindent 
is minimum, for $\phi=v=\sqrt{\mu_{\hbox{\tiny SM}}^{\,2}/\lambda_{\hbox{\tiny SM}}}\ $ \cite{ws,ws2,g,be,h}. 

\pagebreak
\vspace{2mm}

The electroweak symmetry,
said to be ``spontaneously broken'', is in fact simply hidden, with $\phi$ being gauge-invariant.
\vspace{-.6mm}
The $W^\pm\!$ and $Z$ acquire masses 
$
m_W= gv/2\,$, $m_Z=\sqrt{g^2+g'^2}\ v/2 = m_W/\cos\theta
$, with
$\tan\theta=g'/g$. 
The elementary charge and the Fermi coupling of weak interactions
are given by $\,e=g\,\sin\theta\,$ and  
\vspace{-.3mm}
$\,G_F/\sqrt 2=g^2/\,8m_W^2=1/\,2v^2$, 
so that  
$v= (G_F\sqrt 2)^{-1/2}$ $\simeq 246$ GeV.
Charged lepton and quark fields 
interact with
$\varphi$ with coupling constants $\lambda_{l,q}\,$, 
\vspace{-.5mm}
so that the corresponding particles, sensitive to the physical BEH field
\vspace{-.5mm}
$\phi=\!\sqrt{2\,\varphi^\dagger\varphi}\,$ with $\,<\!\!\phi\!\!>\ =v$, acquire masses 
$\,m_{l,q} = \lambda_{l,q}\,v/\sqrt 2$, 
neutrinos
remaining massless at this stage.
\vspace{-.3mm}
The waves corres\-ponding
 to the space-time variations of $\,\phi= \!\sqrt{2\,\varphi^\dagger\varphi}$, when quantized,  are associated with \hbox{spin-0} Brout-Englert-Higgs bosons,  commonly referred to as Higgs bosons. Their mass, 

\vspace{-5mm}

\be
\label{mh}
m_h= \,\sqrt{2\mu_{\hbox{\tiny SM}}^{\,2}}\,=\,\sqrt{2\lambda_{\hbox{\tiny SM}} v^2}\,,
\ee
is fixed by the quartic coupling $\lambda_{\hbox{\tiny SM}}$ in the scalar potential $V(\varphi)$ in (\ref{V}), a mass of 125 GeV$/c^2$ corresponding to a coupling 

\vspace{-5mm}

\be
\lambda_{\hbox{\tiny SM}}\,=\,\frac{m_h^2}{2v^2}\,=\ \frac{g^2+g'^2}{8}\ \frac{m_h^2}{m_Z^2}\ = \,\frac{G_F\,m_h^2}{\sqrt 2}\,\simeq \,0.13\,.
\ee

\vspace{2mm}

The possible origin of this 
coupling  will be discussed later, within supersymmetric theories.
They  lead to consider several BEH bosons originating from the two spin-0 doublets

\vspace{-6mm}

\be
\label{h1h20}
h_1=\left(\ba{c} h_1^0\vspace{.5mm}\\ h_1^- \ea\right),\ \ \ h_2=\left(\ba{c} h_2^+\vspace{.5mm}\\ h_2^0\ea\right),
\ee
 relating their quartic couplings to the squares of the electroweak gauge couplings, in particular through 
 
 \vspace{-4mm}
\be
\label{susy}
\hbox{\em Supersymmetry}\ \ \ \Rightarrow\ \ \ \lambda_{\hbox{\tiny SM}}\ \to \ \frac{g^2+g'^2}{8}\ ,
\ee
with $\,(g^2+g'^2)/8 = G_F m_Z^2/\sqrt 2\simeq .069$.
 Here we first focus for simplicity on $h_2$ and on the ``large $\tan\beta$ limit'', for which $h_2$ acquires a non-vanishing v.e.v.
 much larger than for $h_1$.
We then get a neutral BEH boson that would have {\it  the same mass as the $Z$} \cite{R}, according to
\be
\label{mhZ}
\framebox [8.55cm]{\rule[-.78cm]{0cm}{1.75cm} $ \dis
\ba{l}
\ \ \ \hbox{\em Supersymmetry}\ \ \ \Rightarrow\ \ \ 
\vspace{3mm}\\
 m_h=\sqrt{2\lambda_{\hbox{\tiny SM}} v^2}\,=\hbox{$\dis \frac{\sqrt{g^2+g'^2}\ v}{2}$}=\,m_Z\simeq \, 91\ \hbox{GeV}/c^2,
\ea
$}
\ee
up to supersymmetry-breaking effects. 

\vspace{2mm}

This mass equality results from an unbroken supersymmetry in the sector of neutral particles, with the spin-1 $Z$ and the spin-0 $h$ in the same massive multiplet of supersymmetry. It remains valid even {\it independently of the value of the mixing angle $\beta$} defined from the ratio of the two doublet v.e.v.'s by
\be
\tan \beta =\frac{v_2}{v_1}\,,
\ee
with $\,<\!h_i^0\! >\ = v_i/\sqrt 2\,$, as long as supersymmetry remains unbroken in this sector \cite{R}. The corresponding spin-0 
boson then appears as {\it the spin-0 partner of the $Z$ under two supersymmetry transformations} \cite{gh,prd14}. It was even originally denoted by $z$ to make this association explicit.

\vspace{2mm}

Finding such a spin-0 boson with a mass of 125 GeV$/c^2$ \cite{higgs,higgs2},  not much higher than the $Z$ mass,  may thus be considered,
 at least,  as a very encouraging sign for supersymmetry.  This is especially true as the value $m_h\simeq m_Z$ required by unbroken supersymmetry  may be increased up to 125 GeV$/c^2$ by supersymmetry-breaking effects. This is the case, most notably, in models such as the N/nMSSM or USSM, that include an extra singlet next to the two doublets in (\ref{h1h20}), with a trilinear $\lambda\,H_2 H_1 S$ superpotential coupling \cite{R}. 
 
\vspace{2mm}

 The lightest spin-0 mass may then easily reach 125 GeV, without having to rely on very large effects from radiative corrections. This is a much better  situation than in the usual MSSM for which no electro\-weak breaking is obtained in the absence of the supersymmetry-breaking terms, $m_h$ is required to be less than $m_Z$ at the classical level, and it is difficult to obtain such a 125 GeV spin-0 boson from sufficiently large radiative corrections involving very heavy stop quarks.
 
\vspace{2mm}

The scalar boson of the standard model has  long remained its last missing particle after the discovery of the top quark in 1995.
The new boson found at CERN in 2012 shows the properties expected  from a scalar boson associated with the differentiation 
between electromagnetic and weak interactions,  and the generation of masses. 
It may well be identified with the one of the standard model, which may then be  considered as complete.

\vspace{1mm}

 Still, it would be presumptuous to imagine 
 that our knowledge of particles and interactions is now complete, without new particles or interactions remaining to be discovered.
The standard model does not answer many fundamental questions, concerning the origin of symmetries and symmetry breaking, the quark and lepton mass spectrum and mixing angles, etc..
Gravitation,  classically described by general relativity, cannot easily be cast into a consistent quantum theory. 
This is why string theories were developed, which seem to require supersymmetry for consistency.
\vspace{2mm}

The nature of dark matter 
and dark energy which govern the evolution of the Universe and its accelerated expansion remains unknown, 
as the origin of the predominance of matter over antimatter. 
Dark matter may be composed, for its main part, non-baryonic, of new particles, such as the neutralinos of 
supersymmetric theories.
There may also be new forces or interactions beyond the four known ones.
And maybe, beyond space and time,  new hidden dimensions, extremely small
or even stranger, 
 like the anticommuting quantum dimensions of supersymmetry.

\section{Introducing supersymmetry}

\vspace{-.5mm}

In contrast with pions, kaons and other spin-0 mesons,
composed of quarks and antiquarks,  the new 125 GeV boson presents at this stage all the characteristics of an elementary \hbox{spin-0} particle, the first one of its kind. 
The possible existence of such a scalar has long been questioned, 
many physicists having serious doubts about the very existence of fundamental spin-0 fields.
More specifically  in a theory involving very high mass or energy scales much larger than the electroweak scale, such as a grand-unification scale \cite{su5,su52}  (now usually believed to be 
of the order of $10^{16}$ GeV), or the Planck scale $\simeq 10^{19}$ GeV possibly associated with quantum gravity,  such \hbox{spin-0} fields tend to acquire very large mass terms. They would then  disappear from the low-energy theory, no longer being 	available to provide an appropriate breaking of the electroweak symmetry.

\vspace{1.5mm}

Many efforts were thus devoted  to replace fundamental \hbox{spin-0} fields by composite fields built from spin-$\frac{1}{2}$ ones,  without however much success at this point.
These \hbox{spin-$\frac{1}{2}$} subconstituent fields could have been, for example, techniquark fields interacting through a new interaction specially introduced for this purpose \cite{tc1,tc2,tc3,tc4}, in view of ultimately avoiding fundamental spin-0 fields and particles associated with the electroweak breaking, like the one discovered recently. Furthermore it would still remain difficult to completely avoid considering fundamental  \hbox{spin-0} fields, e.g.~to trigger the breaking of the initial extended technicolor gauge group.

\vspace{2.5mm}

In the meantime however, and even before these increased questionings about
fundamental spin-0 bosons, the situation concerning our view of spin-0 fields had already changed  considerably with the introduction of supersymmetry, in the early 1970's.
This one provides a natural framework for fundamental spin-0 fields. They may now be treated on the same footing as spin-$\frac{1}{2}$ ones, 
also benefiting from the same mass terms when supersymmetry is unbroken; and of mass terms which may remain moderate as compared to very large scales if supersymmetry is not too badly broken, then remaining available to trigger the electroweak breaking.

\vspace{2mm}
The supersymmetry algebra involves a self-conjugate  (Majorana) spin-$\frac{1}{2}$  generator $Q$ satisfying the 
anticommutation and commutation relations   \cite{gl,va,wz,revue,martin,ramond}
\be
\label{alg}
\left\{ \  
\begin{array}{ccc}
\{ \ Q\, , \, {\bar Q} \ \} \!&=&\! 
- \, 2\,\gamma_{\mu}   P^{\mu} , \vspace {1mm} \cr 
[ \ Q\,, \, P^{\mu} \,] \!&=& \ 0\ \ .
\end{array}  \right.      
\ee
They express that supersymmetry transformations may be combined to generate translations, and commute with them.
This algebra was originally introduced as a parity-violating one that might help understanding 
why weak interactions violate parity \cite{gl}, or  the mass\-lessness of  a neutrino by trying to view 
it as a Goldstone fermion \cite{va}, two possible motivations that soon appeared inadequate. It may also be obtained 
by generalizing to 4 dimensions the algebra of super\-gauge transformations 
acting in the 2-dimensional string world\-sheet \cite{wz}.

\vspace{2mm}

But what physical implications may really be  extracted from the consideration of this algebraic structure\,?
According to common knowledge, supersymmetry should relate
bosons, of integer spin, with fermions, of half-integer spin, as follows:

\vspace{-3mm}
 \be
 \label{bf}
 \ba{c}
\,\hbox{\color{black} \em bosons} \ \ 
\stackrel{\stackrel{\hbox{\normalsize supersymmetry}}{\phantom{a}}}
{\hbox{\Large $\longleftrightarrow$}}
\ \  \hbox{\color{black}\em  fermions}\,.
\ea
\ee
But  even this is not always valid, as there are supersymmetric theories involving only fundamental fermions, with supersymmetry transformations acting in a non-linear way \cite{va}.
Strictly speaking the algebraic structure of supersymmetry does not even require any boson at all, not to mention the superpartners that we shall introduce later.
But let us leave aside such unconventional situations. Let us add, also, that supersymmetry transformations are usually expected to relate bosons and fermions with the same gauge symmetry properties.

 \vspace{2mm}

Then, can this algebra be of any help in understanding  the real world  of particles and interactions\,?
If supersymmetry is to act at the fundamental level
the natural idea would be to use it to relate the known bosons and fermions in Table \ref{tab:sm0}. 
More precisely, can one relate the bosons (gluons, $W^\pm\!, \,Z$ and photon) messengers of interactions to the  fermions,
quarks and leptons, constituents of matter\,?
This would lead  to a sort of unification
 \be
   \label{bf2}
\ \hbox{\em Forces} \ \ \ 
\stackrel{\stackrel{\hbox{\normalsize supersymmetry\,?}\!\!\!}{\phantom{a}}}
{\hbox{\Large $\longleftrightarrow$}}
\ \ \  \hbox{\color{black} \em Matter}\,. 
\ee
 The idea looks attractive, even so attractive that supersymmetry is frequently presented as uniting forces with matter.
 This is however misleading at least at the present stage, and things do not work out that way.
  \vspace{2mm}
 
 Indeed the algebraic structure of supersymmetry  did not seem applicable to particle physics at all, in particular as known fundamental bosons and fermions do not seem to have much in common.
 There are also a number of more technical reasons, dealing with: 1) the difficulties of spontaneous supersymmetry breaking, 
 originating from the presence of the hamiltonian within the algebra; \,2) the fate of the resulting Goldstone fermion, after one has succeeded in breaking supersymmetry spontaneously \cite{R,fi,for,for2}, and as it may well continue to interact, 
 even after getting eaten away by the \hbox{spin-$\frac{3}{2}$} gravitino, according to the ``equivalence theorem'' of supersymmetry \cite{grav}; \,3) the presence of self-conjugate Majorana fermions, unknown in Nature; \,4) the requirements of baryon and lepton number conservation, which got associated with the definition of $R$-symmetry and the requirement of $R$-parity, etc..
 
 \vspace{2mm}

 {Relating bosons and fermions, yes, but how\,?}
 One  has to find out which of them might be related under supersymmetry, first considering possible associations between mesons and baryons. Or, at the fundamental level, exploring as a necessary exercise tentative associations like
 
\vspace*{-4.5mm} 
\be
 \label{bf3}
\left\{\ \ \ 
\ba{ccc}
 \hbox{photon}\ \ &\stackrel{\hbox{\normalsize ?}}{\longleftrightarrow}&\ \ \hbox{neutrino} 
\vspace{-.8mm}\\ 
 W^\pm\ \ &\stackrel{\hbox{\normalsize  ?}}{\longleftrightarrow}&\ \  e^\pm
\vspace{-.8mm}\\ 
\hbox{gluons}\ \ &\stackrel{\hbox{ \normalsize ?}}{\longleftrightarrow}&\ \ \hbox{quarks} \vspace{-.5mm}\\ 
& ...& 
 \vspace{-.8mm}\\ 
\ea \right.
\ee
But we have no chance to realize in this way systematic associations of known fundamental bosons and fermions.
This is also made obvious as we know 90 fermionic field degrees of freedom
(for 3 families of 15 chiral quark and lepton fields)
as compared to 28 only  for bosonic ones (16 + 11 + 1 including
the new scalar). Furthermore these fields have different gauge and 
$B$ and $L$ quantum numbers, preventing them from being directly related.

 \vspace{2mm}
 
 In supersymmetry we also have to deal with the systematic appearance of self-conjugate Majorana fermions, while Nature seems to know Dirac fermions
 only (with a possible exception for neutrinos with Majorana mass terms).
 How can we obtain Dirac fermions, and attribute them conserved quantum numbers like $B$ and $L$\,? 
 And if we start attributing $B$ and $L$ also to bosons (now known as squarks and sleptons), how can we be sure that their exchanges won't spoil the $B$ and $L$ conservation laws, at least to a sufficiently good approximation\,?
It is thus far from trivial to consider applying supersymmetry to the real world. 
But if this program can be realized and if Nature has ``chosen'' being supersymmetric,  consequences promise being spectacular.

 \vspace{2mm}
Addressing the difficult questions of spontaneous supersymmetry breaking, and electroweak breaking, will lead us, through the definition of a new symmetry called $R$ symmetry with its discrete remnant known as $R$-parity, to the Supersymmetric Standard Model.
The way to see supersymmetry now is to view it as {\it an extension of the standard model} that introduces a new  {\it sparticle} for each one in the standard model \cite{R,ssm,ff,fayet79}, in particular through
\be
\left\{
\begin{tabular}{ccc}
{quarks,\hspace{.7mm}leptons} \ & $\leftrightarrow$&\  {spin-0 \it squarks and sleptons},\\[.5mm]
{gluons} & $\leftrightarrow$&  {spin-$\frac{1}{2}$ \it gluinos}, \vspace{1mm}\\
{$W^\pm,Z,\ \gamma$} & $\leftrightarrow$&\,{spin-$\frac{1}{2}$ \it charginos and neutralinos},
\end{tabular} \right.
\\
\ee

\vspace{2mm}
\noindent
with  more to say about spin-0 BEH bosons, including charged and several neutral ones.

\vspace{2mm}

While this is now often presented as obvious, the necessity of postulating that {\it every known particle has its own image under supersymmetry}
\,(SM bosons having fermionic superpartners and SM fermions bosonic ones) was long mocked as a sign of the irrelevance of supersymmetry.
The introduction of a color octet of spin-$\frac{1}{2}$ Majorana fermions 
 called {\it gluinos} was also, at the time, forbidden by the principle of triality \cite{tria}.
 This one, however, gets systematically violated within supersymmetric theories.

\vspace{2mm}
 
 The necessity of {\it charged spin-0 BEH bosons} ($H^\pm$),
 \vspace{-.1mm}
 required by the 2-doublet structure of supersymmetric theories, was also taken as an argument against  supersymmetry and supersymmetric extensions of the standard model, on the grounds that even a single doublet, although possibly necessary as in the standard model, was already undesirable.
These charged spin-0 bosons, which have not been discovered yet \cite{hat,hcms}, also appear as the spin-0 partners of the $W^\pm$ under {\it two}  supersymmetry transformations, very much as the new 125 GeV boson may now also be interpreted as the spin-0 partner of the $Z$, up to supersymmetry-breaking effects.

  \vspace{1mm}

    \section{\boldmath  \vspace{1.5mm} \hbox{Supersymmetry} \hbox{breaking}  
\hbox{\ \ and $\,R\,$ symmetry}}

  \vspace{-.5mm}

  \subsection{Is spontaneous supersymmetry \vspace{1.5mm}  breaking possible at all\,?}
       
\label{sec:diff}

 \vspace{-1mm}
 
If bosons and fermions are directly related by supersymmetry they should have equal masses. 
Supersymmetry may then be only, at best, a broken symmetry. Considering terms breaking explicitly the supersymmetry (as frequently done now) would certainly make the task much easier, but may be considered only as a temporary substitute for a solution to the problem of supersymmetry breaking.
If supersymmetry is to be a genuine symmetry for the theory and its equations of motion, it should be broken {\it spontaneously}, as for the electroweak symmetry in the standard model. This is also necessary for supersymmetry to be realized as a local fermionic gauge symmetry \cite{vs}. It must then include general relativity, leading to supergravity theories \cite{sugra,sugra2}.

\vspace{1.5mm}

To trigger a spontaneous breaking of an ordinary  (global or gauge) symmetry, one simply  has to arrange for the symmetric vacuum state to be unstable, e.g.~by choosing a negative value for the mass$^2$ parameter $\,-\,\mu_{\hbox{\tiny SM}}^2 $ in the potential (\ref{V}), which is easily realized.
\vspace{1mm}

The situation concerning supersymmetry is, however, completely different. The hamiltonien $H$, which governs the energy of the possible vacuum states and thus determines which one is going to be stable, may now be expressed from the squares of the four components of the supersymmetry generator, as
 \be
\label{h}
H=\ \frac{1}{4}\ \,\hbox{\small$\dis \sum_\alpha$}\ \,Q_\alpha^{\,2}\,.
\ee

\vspace{-2mm}

\noindent
This implies that {\it a supersymmetric vacuum state} $\,|\,\Omega>$ (verifying $\,Q_\alpha|\,\Omega\!>\ = 0$) {\it must have a vanishing energy,}
with $H\,|\,\Omega\!>\ =0\,$. On the other hand 
any non-super\-symmetric state $\,|\,\Omega'\!>\,$ would have, within global supersymmetry, a larger, positive, energy density, and thus would be unstable. 
This was originally thought to prevent any spontaneous breaking of the supersymmetry to possibly occur \cite {iz},
apparently signing the impossibility of applying supersymmetry to the real world.

\vspace{-1mm}

\subsection{In search for a minimum \vspace{1.2mm} of the potential \hbox{\ \ \ \ \ breaking the supersymmetry}}

\vspace{-1mm}

In spite of this general argument however, which soon got circumvented, spontaneous supersymmetry breaking turned out to be possible, although  in very specific circumstances.  It is severely constrained and usually hard to obtain, at least within global supersymmetry, as any supersymmetric candidate for the vacuum state ($\,|\,\Omega>$) is necessarily stable. Furthermore in the presence of many spin-0 fields, there are usually many opportunities for them to adjust so as to provide such a stable supersymmetric vacuum, with a vanishing value of the potential $V=0$.

\vspace{2mm}

To obtain a spontaneous breaking of the global supersymmetry,  one cannot just attempt to make a supersymmetric vacuum 
unstable.  One must instead arrange for such a symmetric state to be {\it totally absent}, 
as it would otherwise be stable owing to expression (\ref{h}) of the hamiltonian.

\vspace{2mm}
In the usual langage of global supersymmetry \cite{revue,martin} involving gauge superfields $V_a(x,\theta,\bar\theta)$ and (left-handed) chiral superfields   $\Phi_i(x,\theta)$ 
\vspace{-.5mm}
with physical spin-0 and spin-$\frac{1}{2}$ components $\phi_i$
and $\tilde \phi_{iL}$
\cite{sf1,sf2}, the potential of scalar fields is
expressed as
\vspace{-4mm}

\be
\label{vpot}
V=\frac{1}{2}\ \hbox{\footnotesize$\dis\sum$}\,(D_a^2+F_i^2+G_i^2)=\ \hbox{\footnotesize$\dis\sum_a$} \  \frac{D_a^2}{2} \ +\, \hbox{\footnotesize$\dis\sum_i$}\,\left|\,\frac{\partial\cal W}{\partial \phi_i}\,\right|^2\!.\,
\ee
$D_a$ and $(F_i,\,G_i)$ stand for the auxiliary components of gauge and chiral superfields. The contribution from the $D$ terms is given by
\be
\label{vpotd}
V_D=\,\hbox{\footnotesize$\dis\sum_a$} \ \,\frac{D_a^2}{2}\,=\,\hbox{\small$\dis\frac{1}{2}\ \hbox{\footnotesize$\dis\sum_a$}\,$\,\Large$\left[\right.$ \normalsize $\!\xi_a+
\,g_a$ \footnotesize$\dis\sum_{ij}$}
\ \phi_i^*\,(T_a)_{ij}\,\phi_j \,
\hbox{\Large$\dis\left.\right]$}^2\,,
\ee
with the $\xi_a$ parameters relative to abelian $U(1)$ factors in the gauge group \cite{fi}. The superpotential ${\cal W}(\Phi_i)$ is an analytic function of the chiral superfields.

\vspace{2mm}

For a supersymmetric vacuum state $|\,\Omega\!>$ to be, not unstable but totally absent, the potential $V$ must be {\it strictly positive everywhere}. 
One at least of these auxiliary components must then have a non-vanishing v.e.v., which is indeed the signal for a spontaneously broken supersymmetry (except for trivial situations with a free superfield). Finding a spontaneously broken supersymmetry then amounts to finding situations for which the set of equations
\be
<D_a\!>\ = \ <F_i\!>\ =\ <G_i\!>\ =\, 0\ \  \hbox{must have \it no solution.}
\ee

\vspace{2mm}
How this may be realized, as well as the definition and role of the $R$ symmetry, leading to $R$-parity,
to appropriately constrain the superpotential, will be further discussed in the rest of this Section.
 The reader mostly interested in the construction of supersymmetric extensions of the standard model (MSSM, N/nMSSM, USSM, etc.) and in the relations between massive spin-1 gauge bosons and spin-0 BEH bosons 
  may choose to proceed directly to Sections \ref{sec:MSSM} and \ref{sec:gh}.

\subsection{\boldmath $D$ and $F$ supersymmetry \vspace{1.5mm}  breaking mechanisms, \,\,\,in connection with $R$ symmetry}
\label{subsec:df}

\vspace{-.5mm}

To avoid having a vanishing minimum  of $V$ when all physical fields $\phi_i$ vanish, there are two possibilities, which may be combined:

\vspace{2mm}

 \underline{\bf1}) The Lagrangian density may
include a linear term 
\be
{\cal L}_\xi \,=\, \xi D\,,
\ee
 associated with an abelian $U(1)$ factor in the gauge group   \cite{R,fi}. This term is indeed supersymmetric, up to a derivative which does not contribute to the variation of the action integral, and gauge invariant for an abelian gauge group.
It may lead to a spontaneous breaking of the supersymmetry, just by itself as in the presence of a single chiral superfield $S$ (with a charge $e$ such that \hbox{$\xi e>0$}) \cite{fa76},
or by making the set of equations $\{D_a=0\}$ without solution, as with a $SU(2)\times U(1)$ gauge group \cite{R}.
The Goldstone spinor is then a gaugino, corresponding for example to the photino in a $\,SU(2)\times U(1)$ theory, even if such a feature  cannot persist in a physically realistic theory \cite{fayet79}.

\vspace{2mm}

One may also arrange so that the set of equations $\{D_a=0,\ F_i=G_i=0\}\,$ be without solution, as done in the presence of chiral superfields $S$ and $\bar S$ with a mass term 
 $\,\mu\, \bar S S$~\cite{fi}; or with a suitable trilinear superpotential $\,\lambda\,H_2H_1 S$, 
the electroweak gauge group being extended to an extra $U(1)$ factor, as in the USSM \cite {ssm}.
 In all cases one has to make sure that no supersymmetric minimum of the potential exists
anywhere, otherwise supersymmetry would remain (or return to) conserved.

\vspace{2mm}

 \underline{\bf 2}) The Lagrangian density may appeal to a term proportional to the auxiliary  ($F$ or $G$) components
of a singlet chiral superfield $S (x,\theta)$. In this case the superpotential 

\vspace{-7mm}

\be
\label{sing}
{\cal W}\,=\,\sigma S\, +\, ...
\ee
includes a term linear in the singlet superfield $S$.  One can then try to make the system of equations $\{F_i=G_i=0\}$, i.e.
${\partial\cal W}/{\partial \phi_i}=0$, without solution. 

\vspace{2mm}
This first looks as an impossible task. Indeed with the superpotential
\be
\label{superpot0}
{\cal W}\,=\,  \frac{\lambda_{ijk}}{3}\ \Phi_i\Phi_j\Phi_k + \frac{\mu_{ij}}{2}\ \Phi_i\Phi_j + \sigma _i\,\Phi_i\,,
\ee
 taken for simplicity as a cubic function of the chiral superfields $\Phi_i$ for the theory to be renormalisable, the set of equations ${\partial\cal W}/{\partial \phi_i}=0$ reads 
\be
\label{equa}
\lambda_{ijk}\ \phi_j\phi_k + \mu_{ij}\ \phi_j + \sigma _i\,=0\,.
\ee
With $n$ equations for $n$ complex field variables it is expected to have  {\it almost always\,} solutions, for which supersymmetry is conserved.

\vspace{2mm}

Still it is possible choosing very carefully 
the set of interacting superfields and the superpotential $\cal W$ for spontaneous supersymmetry breaking to occur.
This appeals to a new symmetry called $R$ symmetry \cite{R}.
$R$ transformations, further discussed in the next subsection, act on (left-handed) chiral superfields according to
\be
\Phi (x,\theta)\  \stackrel{R}{\rightarrow}  \ e^{i\,R_\Phi\,\alpha}\ \Phi(x,\theta\,e^{-\,i\,\alpha})\,.
\ee
The superpotential must transform with $R=2$  for the theory to be invariant under $R$. It is then be said to be ``$R$-symmetric''.
This symmetry can be used to select and constrain appropriately the superpotential $\cal W$
so that the set of equations (\ref{equa}) has no solution and the corresponding breaking of the supersymmetry is obtained  {\it in a generic way}, not just for very specific values of the parameters \cite{for,for2}.

\vspace{2.5mm}

An interesting example
is obtained with a $R$ symmetric nMSSM-type superpotential \cite{R},  extended to a chiral triplet $T$.  It involves, as in the MSSM without the $\mu$ term,  the two doublets 
$H_1$ and $H_2$ with $R=0$. They are coupled to a singlet $S$ through a $\lambda \,H_2H_1 S\,$ trilinear term as in the nMSSM,  and similarly  to a triplet $T$, both  with $R=2$. The corresponding $R=2$ superpotential reads~\cite{for} 

\vspace{-4mm}
\be
\label{fbr0}
{\cal W}\,=\,\frac{1}{\sqrt 2}\ H_2\,  (g \, \tau .T-g'S)\,H_1\,+\sigma S\,,
\ee 
$H_1$ and $H_2$ having weak hypercharges $Y=-1$ and $+1$, respectively.
This one is also, in addition, the superpotential for a $N=2$
supersymmetric gauge theory (or ``hypersymmetric'' theory)  \cite{hyper} when the {$SU(2)\times U(1)$} symmetry is made again local,
with the superpotential couplings fixed in terms of the electroweak gauge couplings as in (\ref{fbr0}).
Its two terms may be written as proportional to 

\vspace{-6mm}

\be
H_2\, \Phi H_1 \ \ \ \,\hbox{and}\ \ \ \,\hbox{Tr}\, \Phi\,, 
\ee
with $\Phi=\frac{1}{2}\,(g \,\tau .T\!  -g'S)$.
Or  $\,\Phi=g\,\Lambda .T\!-\frac{g'}{2}\,S\,$
where the chiral superfields $T$ are in the adjoint representation of the gauge group,  and the matrices  $\Lambda$ are relative to the hypermultiplet representation of the gauge group described by $H_1$ and $H_2$.

\vspace{2.5mm}

For a non-abelian $\,N\!=\!2\,$  theory the superpotential (\ref{fbr0}) 
reads
 \vspace{-6mm}

\be
\label{lambdag-1}
{\cal W}\,=\,g\,\sqrt 2\ H_2\, \Lambda .T \,H_1\,,
\ee
 leading to the $N=4$ supersymmetric Yang-Mills theory when $H_1$ and $H_2$ are also taken in the adjoint representation of the gauge group \cite{Z}. The adjoint gauge superfield then interacts with 3 adjoint chiral ones, now denoted by $S_1,S_2 $ and $S_3$,
 coupled through the trilinear superpotential
\be
\label{sss}
{\cal W}\, =\ g\,\sqrt 2\ f_{ijk}\, S_{1}^i\, S_{2}^j \,S_{3}^k\,.
\ee
But we shall return to this later.

\vspace{2mm}

 Let us come back to the superpotential (\ref{fbr0}) for a non-gauged $SU(2)\times U(1)$ theory involving at the moment chiral superfields only, in view of generating a spontaneous breaking of the supersymmetry through $F$ terms \cite{for}.
The conjugates of the 4 (complex) auxiliary components of the $R=0$ superfields $H_1, H_2$ have $R=2$, and vanish with the 4 components of  {$t$} and $s$ (also with $R=2$).
The conjugates of the 4 auxiliary components of the  $R=2$ superfields {$T$} and $S$ have $R=0$, and depend quadratically on the 4 components of  $h_1$ and $h_2$.  One might naively expect that they should  also, ``generically'', be able to vanish simultaneously, so that supersymmetry  would be conserved.
Still this does not happen, as it would require

\vspace{-4mm}

\be
 \label{fsft}
\left\{\  \ba{cclcc}
\hbox{\small $\dis \frac{\partial{\cal W}}{\partial t}$}\!&=&\! \hbox{\small $\dis \frac{g}{\sqrt 2}$}\  h_2\,\tau  h_1\!&=&\! 0\,,
\vspace{2mm}\\
\hbox{\small $\dis \frac{\partial{\cal W}}{\partial s}$}\!&=&\!  \hbox{\small $\dis \frac{g'}{\sqrt 2}$}\ h_2 h_1+\sigma\!&=&\! 0\,,
 \ea\right.
 \ee
  which  are incompatible for $\,\sigma\!\neq 0\,$. 
 
 \vspace{2mm}
 Supersymmetry is spontaneously broken, with a massless Goldstone spinor $\zeta_\gamma$, taken as left-handed,  carrying $R=1$. It is described, together its $R=2$ spin-0 partner, which remains classically massless, by a combination of the $R=2$ chiral superfields. 
 
\vspace{2mm}
 
 This $R\!=\!2$ chiral superfield  is in this example the photon-like combination $\,\sin\theta \, T_3 + \cos\theta \, S$. It will be understood further by turning again the theory into a $N=2$ supersymmetric one \cite{hyper}, with a $\,SU(2)\times U(1)$ gauge group spontaneously broken into $U(1)_{\rm QED}$. The present Goldstone spinor $\zeta_\gamma$ associated with this $F$-breaking of the supersymmetry is described by the $R=2$ chiral superfield $\,\sin\theta \, T_3 + \cos\theta \, S$. It then gets interpreted as the second photino field within $N=2$, \,both photino fields $\lambda_\gamma$ and $\zeta_\gamma$, now related by a global $U(2)_R$ symmetry of $N=2$, being the two Goldstone spinors of $N=2$ supersymmetry.
 
\vspace{-2.5mm}

\subsection{\boldmath On the role of \vspace{1.5mm} $R$ symmetry to allow for supersymmetry breaking through $F$ terms}
 \label{subsec:role}
 
\vspace{-.5mm}

 Without $R$ symmetry, 
 $\,S^2$ or $S^3$ terms would be allowed in the superpotential, and we would lose the benefit of  having introduced a linear $\sigma S$ term, which may then be eliminated by a translation of $S$. 
 Once $\sigma$ is eliminated the potential has a vanishing minimum when all physical fields vanish, and supersymmetry is conserved.
 This shows {\it the crucial role played by $R$ symmetry} \,to render possible {\it a generic breaking of the supersymmetry} \,through $F$ terms \cite{for}.

\vspace{2mm}

This mechanism leads to a classically-massless $R=2$ spin-0 field, superpartner of the $R=1$ Goldstone spinor (goldstino). Both are described by a $R=2$ chiral superfield. A translation of its $R=2$ spin-0 component, if it had to be performed,  would lead to a spontaneous breaking of $R$ symmetry (or a quasi-spontaneous breaking if $R$ is anomalous). The imaginary part of this $R=2$ field would then describe a massless 
$R$ Goldstone boson or, for an anomalous symmetry, a classically-massless $R$-axion.

\vspace{2mm}
But is there a tighter connection between spontaneous supersymmetry breaking by this method, and the possible occurrence of spontaneous $R$-symmetry breaking through the v.e.v.~of such a $R=2$ scalar, superpartner of the $R=1$ goldstino\,? The above  example indicates that
there is no need for $R$-symmetry to be spontaneously broken,  to get spontaneous supersymmetry breaking. 
Conversely, for $\sigma \!=\!0$ $\,R$ symmetry may  indeed be spontaneously broken, owing the $R=2$ flat directions of the potential, with a massless $R$ Goldstone boson, and a conserved  supersymmetry. Thus spontaneous $R$-symmetry breaking is not a sufficient condition either, for spontaneous supersymmetry breaking to occur.
 \vspace{2mm}
 
 It thus appears that while spontaneous $R$-symmetry breaking may occur, {\it it is neither necessary nor sufficient} to lead to such a spontaneous breaking of the supersymmetry through $F$ terms.
 What is indeed essential is the presence of $R$ symmetry to restrict appropriately the superpotential as in (\ref{fbr0}) \cite{for}.

  \vspace{2mm}
 
Furthermore, and in contrast with a current belief,
 spontaneous supersymmetry  breaking occurs here, for $\sigma \neq 0$, in spite of having equal numbers of $R=2$ and $R=0$ superfields. There is thus no excess of $R=1$ over $R=-1$ (left-handed) spinors, that would facilitate having a massless left-over $R=1$ spinor that could become a Goldstone spinor. 
 In fact with the same number of $R=2$ 
 and $R=0$ superfields one might usually expect all spinors to acquire masses. Then there would be no candidate for a massless Goldstone spinor, and supersymmetry would have to remain conserved. Indeed the auxiliary components of the $R=2$ superfields, which have $R=0$, then depend on the same number of $R=0$ physical fields, and might be expected to all vanish simultaneously.
  \vspace{2mm}
  
Still this additional obstruction could be bypassed so as  to render the system (\ref{fsft}) of 4 equations for 4 variables generically without solution,  and obtain spontaneous supersymmetry breaking  \cite{for}.  This has been made possible, in particular, thanks to the spontaneously broken  global $SU(2)\times U(1)$ $\to \,U(1)$ \,symmetry  generated by $\,<\!h_1\!>$ and $<\!h_2\!>$, leading to exactly massless 
 $R=0$ spin-0 Goldstone fields associated with $R=-1$ spinors.  One of the latter balances a massless $R=1$ spinor that is going to be the goldstino.

\vspace{-.5mm}

\subsection{\boldmath Unifying $D$- and $F$- breakings \vspace{1.5mm}  within $N=2$, \hbox{\ \ \ \ and going to $N=4$ \,supersymmetry}}

\vspace{-.5mm}

Beyond that, when $SU(2)\times U(1)$  is gauged again as in the nMSSM, with the gauge superfields $V$ and $V'$
associated to the chiral triplet and singlet  $T$ and $S$, the theory based on  the superpotential (\ref{fbr0}) acquires an enhanced symmetry, namely extended $N=2$ supersymmetry
(or hypersymmetry), with $H_1$ and $H_2$ jointly describing a $N=2$ hypermultiplet
\cite{hyper}. Of course no superpotential term proportional to $S^2,\,S^3$ as in the general NMSSM, $T^2$ or $ST^2$ may be allowed here. Such terms, which would ruin the possibility of having a $N=2$ supersymmetry, were already excluded by means of $R$-symmetry, which showed the way to extended supersymmetry, and subsequently extra dimensions.

\vspace{2mm}

 The $D$- and $F$-breaking mechanisms then become equivalent, getting unified within $N\!=\!2$ supersymmetry.
Indeed  the set of auxiliary components $\{-G,-F,\,D\}$ for a $N\!=\!2$ gauge multiplet transform as the three components of a $SU(2)_R$ isotriplet within a $U(2)_R=[SU(2)\times U(1)]_R$ global symmetry group. The $\xi D$ term for a $U(1)$ gauge superfield can then be turned into a $\xi F$ term for its associated chiral superfield through a $SU(2)_R$ transformation turning the first supersymmetry generator into the second.
This $N\!=\!2 $ supersymmetry breaking generates two massless Goldstone spinors, both with $R=1$.
 A $SU(2)\times U(1)$, or more generally  $G_{\hbox{\footnotesize non-abelian}}\!\times U(1)$ gauge group is then required if we intend to get a spontaneous breaking of the extended supersymmetry rather than just of  the gauge 
 symmetry~\cite{hyper}.  

\vspace{2mm}

 For a non-abelian $N=2$ gauge theory the superpotential reads
$
\,{\cal W}=g\,\sqrt 2\ H_2\, \Lambda .T \,H_1\,$
as in (\ref{lambdag-1}), with $T$ in the adjoint representation and 
the  $\Lambda$ representing the gauge group for the
hypermultiplet described by $H_1$ and $H_2\,$.
A $N=2$ super\-symmetric theory with a massless matter hypermultiplet in the adjoint representation provides the $N\!=\!4$ supersymmetric Yang-Mills theory \cite{Z}.
The adjoint gauge superfield interacts with 3 adjoint chiral ones $S_1,S_2 $ and $S_3$
 coupled through the trilinear superpotential (\ref{sss}),
$
\,{\cal W}=g\,\sqrt 2\ f_{ijk}\, S_{1}^i\, S_{2}^j \,S_{3}^k\,.
$
This provides the

\vspace{-5mm}

\be
\ba{l} N=4 \ \ \hbox{\it supersymmetric Yang-Mills theory,}   \  \ \ \hbox{with}\   \vspace{1mm}\\ \ \ 
(\,1 \ \hbox{spin-1}  + 4 \  \hbox{spin-\small$\dis \frac{1}{2}$} + 6 \ \hbox{spin-0}\,)\   \hbox{\it adjoint gauge fields.}
\ea
\ee

\vspace{2mm}

The $R$ symmetry acting chirally on the $N=1$  supersymmetry generator (see later eq.\,(\ref{R}))  gets promoted from $U(1)_R$ in $N=1$ up to $SU(2)_R$ or $U(2)_R$ in $\,N=2\,$, and
$\,SU(4)_R \sim O(6)_R\,$ in $\,N=4\,$ supersymmetry. This corresponds to the following chain
\be
\ba{ccl}
R{\hbox{-parity}}\,\subset\, U(1)_R \!&\subset&\, \underbrace{SU(2)_R \, \subset\, U(2)_R}_{\hbox{\small $N=2$}}\, 
\vspace{2.5mm}\\
 \!&\subset&\, \underbrace{SU(4)_R \, \sim \,O(6)_R}_{\hbox{\small $N=4$}}\,.  
\ea
\ee

\vspace{-1mm}
 
 The spontaneous breaking of the gauge symmetry in a $N=2$ theory will be very useful, providing larger associations between massive 
 spin-1 gauge bosons, \hbox{spin-$\frac{1}{2}$} charginos and neutralinos and spin-0 BEH bosons \cite{2guts}, and leading us to a description of particle physics in a higher-dimensional space-time 
 \cite{gutbis}.

 \pagebreak

\subsection{\boldmath Origin \vspace{1.5mm} of $R$ symmetry,  
\,and of the extra $U(1)$ rotating $\,h_1$ and $h_2$}

$R$ symmetry originates from an earlier $Q$ symmetry acting within a precursor of a supersymmetric theory including two spin-0 doublets (now called $h_1$ and $h_2$), and a Dirac spin-$\frac{1}{2}$ doublet, 
\vspace{-.6mm}
subsequently providing the corresponding higgsinos
$\tilde h_{1L}$ and $\tilde h_{2L}$ \cite{f74}.
The $Q$ symmetry restricting both the possible form of the potential $V$ and of the Yukawa couplings responsible for fermion masses
was already a $R$-type symmetry. It acts according to
    \be
  \label{q}
   H_1\ \stackrel{Q}{\to} \ e^{i\alpha}\ H_1(x,\theta\,e^{-\,i\,\alpha}),\ \  H_2\ \stackrel{Q}{\to} \ e^{i\alpha}\ H_2(x,\theta\,e^{-\,i\,\alpha})\,,
   \ee
allowing for a $\mu \,H_2H_1$ superpotential mass term for $H_1$ and $H_2$.
It  was then turned into the $R$ symmetry familiar to us today, defined as $R=Q \,U^{-1}$
(or equivalently $Q=R\,U$)
and acting according to \cite{R}
   \be
  \label{oldr}
   H_1\ \stackrel{R}{\to} \ H_1(x,\theta\,e^{-\,i\,\alpha}),\ \  H_2\ \stackrel{R}{\to} \ H_2(x,\theta\,e^{-\,i\,\alpha})\,.
   \ee
This $R$ symmetry leaves $h_1$ and $h_2$ invariant so as to survive the electroweak breaking.

   \vspace{2mm}
   
Here $U$  denotes a $U(1)$ symmetry transformation commuting with supersymmetry, acting on the two electroweak doublets $h_1$ and $h_2$ according to
   \be
  \label{u0}
   h_1\ \stackrel{U}{\to} \ e^{i\alpha}\ h_1,\ \  \ h_2\ \stackrel{U}{\to} \ e^{i\alpha}\ h_2\,,
   \ee
   or in terms of superfields,
     \be
  \label{u0bis}
   H_1\ \stackrel{U}{\to} \ e^{i\alpha}\ H_1,\ \ \, H_2\ \stackrel{U}{\to} \ e^{i\alpha}\ H_2\,.
   \ee
   This definition was immediately extended in \cite{R} to the extra nMSSM singlet $S$ transforming according to
    \be
    \label{us}
   S\ \stackrel{U}{\to} \ e^{-2i\alpha}\ S\,.
   \ee
   
  The  transformation (\ref{u0}) was first introduced  as a way to constrain the potential in a two-doublet model by {\it allowing  for independent phase transformations of $h_1$ and $h_2$},
  jointly with the weak hypercharge $U(1)_Y$ ($h_1$ and $h_2$ having $Y=-1$ and $+1$)   \cite{f74}.
   This does not lead to the appearance of an axion or axionlike particle as long as we are dealing with an {\it inert-doublet model}, keeping an  unbroken symmetry combining a $U(1)$ transformation (\ref{u0}) with a $U(1)_Y$ transformation, under which 
   \be
   h_1\, \to \, e^{2i\alpha}\,h_1\,,\,\ \ h_2\,\to \,h_2\,.
   \ee
  This residual $U(1)$ includes {\it a $Z_2$ discrete symmetry under which the inert doublet 
   $h_1$ changes sign}, $h_1\to -\,h_1$, \,and allows for a non-vanishing v.e.v. $\!<\!h_2\!>\ \neq \,0\,$,\, which breaks spontaneously the electroweak symmetry. Such inert-doublet models can thus also provide, from the stability of the lightest component of $h_1$, a possible dark matter candidate.
   
   \vspace{2mm}

   In supersymmetric extensions of the standard model, however, both $h_1$ and $h_2$ must acquire non-vanishing v.e.v.'s.
  A classically massless particle ($A$)  would then appear in the spectrum as a consequence of the additional $U(1)$ symmetry (\ref{u0},\ref{u0bis}), if this one remains indeed present. This particle is immediately apparent in the spectrum in the absence of the extra singlet superfield $S$ (i.e. for $\lambda =0$).
   Such a feature, considered as undesired, was {\it avoided 
 from   the beginning\,} by breaking explicitly the extra-$U (1)$ symmetry (\ref{u0},\ref{u0bis}) 
 through the introduction of the singlet $S$ transforming as in (\ref{us}). This singlet is coupled to $H_1$ and $H_2$ by a trilinear superpotential term
  $\lambda\,H_2H_1S\,$, invariant under the extra-$U(1)$.
  
  \vspace{2mm}
  
  The introduction of the linear term $\sigma S$ in the nMSSM superpotential  $\,\lambda \,H_2H_1S+\sigma S\,$  breaks explicitly the extra-$U(1)$ symmetry (\ref{u0}-\ref{us}),
   providing a mass $\,\lambda v/\sqrt 2\,$ for the would-be ``axion'' $A$~\cite{R}. Its mass vanishes with $\lambda$, the extra-$U(1)$ symmetry with its associated Goldstone (or pseudo-Goldstone) boson $A$ getting recovered for $\lambda=0$.
  The same $U(1)$ transformation  (\ref{u0}) acting on the two doublets $h_1$ and $h_2$ became useful  later in a different context, to rotate away the $CP$-violating parameter $\theta$ of QCD \cite{pq}. The resulting presence of an axion $\!A$, after having escaped attention in \cite{pq}, was pointed out in \cite{ax1, ax2}.

  \vspace{2mm}
  But no such axion as been observed yet. This may be understood if the extra-$U(1)$ symmetry is broken at a high scale through 
  a large v.e.v. $<\!s\!>\,$ for a singlet transforming non-trivially under the  extra $U(1)$, as in (\ref{us}). We shall return to this in subsection {\ref{subsec:ussm}}, 
  when dealing with the interactions of a very light neutral \hbox{spin-1} gauge boson $Z'$ (or $U$) as may be present in the USSM, in which the extra-$U(1)$ symmetry 
  (\ref{u0bis},\ref{us}) is gauged \cite{ssm}. This light spin-1 boson would behave 
  very much as the corresponding eaten-away axionlike 
  pseudo\-scalar $a$, then mostly an electroweak singlet and interacting very weakly, thus largely ``invisible'' \cite{pl80b,U}.

\vspace{-1mm}

\subsection{\boldmath Action of $R$ symmetry}
   \vspace{-1mm}

Let us now return to  $R$ symmetry. It enlarges the initial supersymmetry algebra (\ref{alg}) by introducing the new symmetry generator $R$ corresponding to an abelian group $U(1)_R$. It acts chirally on the supersymmetry generator $Q$ according to 

   \vspace{-5mm}
\be
\label{R}
Q\ \stackrel{R}{\to}  \ e^{-\gamma_5 \alpha}\,Q\,,
\ee 
or equivalently $\,Q_L \to e^{-i\alpha}\,Q_L$,
transforming gauge and (left-handed) chiral superfields according to
\be
\label{R1}
\left\{
\ba{ccc}
V(x,\theta,\bar\theta) \ &\stackrel{R}{\rightarrow}&\ V(x,\theta\,e^{-\,i\,\alpha},\bar\theta\,e^{\,i\,\alpha} )\,,
\vspace{1.5mm}\\
\Phi (x,\theta)\  &\stackrel{R}{\rightarrow} & \ e^{i\,R_\Phi\,\alpha}\ \Phi(x,\theta\,e^{-\,i\,\alpha})\,.
\ea \right.
\ee

\vspace{.5mm}

The spin-0 components $\phi=\Phi(x,0)$ of chiral superfields
 transform with  $R$ quantum numbers $R_\Phi$.  Their associated spin-$\frac{1}{2}$ components $\tilde \phi_L$, proportional to  $\,[\,Q_L,\,\phi\,]\,$ (or 
 equivalently to the linear term in the expansion of $\Phi$ with respect to the Grassmann coordinate $\theta$), \,have 
$R=R_\Phi-1$. The $R$ symmetry transformations (\ref{R1}) thus act on field components as
\be
\label{R2}
\left\{\ 
\ba{cccccc}
V^\mu&\stackrel{R}{\rightarrow}& \ V^\mu,\ &\ 
\lambda &\stackrel{R}{\rightarrow}& e^{\gamma_5\alpha}\ \lambda\,,
\vspace{1mm}\\
\phi &\stackrel{R}{\rightarrow}& \ e^{iR_\Phi\alpha}\ \phi\,,\ &\
\tilde \phi_L&\stackrel{R}{\rightarrow}& e^{i(R_\Phi-1)\alpha}\ \tilde\phi_L\,,
\ea \right.
\ee
$\lambda$ denoting the Majorana gaugino fields associated with the gauge fields $V^\mu$. The (complex) auxiliary components
\linebreak
$(F+iG)/\sqrt 2$ of the chiral superfields $\Phi$ transform with $R=R_{\Phi} -2$, according to
\be
\hbox{\small$\dis \frac{F+iG}{\sqrt 2}$}\ \ \stackrel{R}{\rightarrow}\  \ e^{i(R_\Phi-2)\alpha}\ \hbox{\small$\dis \frac{F+iG}{\sqrt 2}$}\,.
\ee
The auxiliary components of  $R=2$ superfields are invariant under $R$. This  was used in (\ref{sing},\ref{fbr0}) to include a linear contribution $\sigma S$ within the superpotential $\cal W$ of a $R$-symmetric theory, as in the nMSSM \cite{R}.

 \vspace{1mm}

\subsection{\boldmath Constructing {\em Dirac} charginos \vspace{1.5mm} and neutralinos  with a conserved $R$ symmetry}

$R$ symmetry (i.e.~$U(1)_R$) allows in particular for $R$-invariant Yukawa couplings of gauginos to spin-$\frac{1}{2}$
	and spin-0 fields described by chiral superfields, that may be expressed as
	\be
	\label{LY}
{\cal L}_Y\,=\, \hbox{\small$\dis\sum_a$}\ \,	(i)\,g_a\,\sqrt 2\,\ \bar\lambda_{aR}\ \,\phi_i^\dagger\,(T_a)_{ij}\,\tilde\phi_{jL}+\,\hbox{h.c.}\,,
	\ee	
	
\vspace{-3mm}

\noindent
with 
\be
\lambda_{aR}\, \stackrel{R}{\to} \,e^{-i\alpha}\, \lambda_{aR}\,, \ \ \phi_i^\dagger\,(T_a)_{ij}\,\tilde\phi_{jL} \, \stackrel{R}{\to}\, e^{-i\alpha} \
\phi_i^\dagger\,(T_a)_{ij}\,\tilde\phi_{jL}\,.
\ee

\vspace{1mm}\noindent
The phase factor $\pm 1$ or $\pm i$ that may appear in front of  the first term in (\ref{LY}) is convention-dependent and may be modified by a chiral redefinition of the  gaugino fields $\,\lambda_a$, or a relative phase redefinition of $\,\phi_i$ and $\,\tilde\phi_{iL}$\,.
\vspace{2mm}

This leads to the possibility of generating, through a spontaneous breaking of the gauge symmetry, $R$-in\-variant non-diagonal mass terms connecting gauginos with higgsinos, transforming as 
  \be
   \label{trh}
   \left\{\ \ba{ccc}
     \hbox{\it gauginos} \    \  \lambda &\stackrel{R}{\to}& e^{\gamma_5\alpha}\ \lambda\,,
   \vspace{2mm}\\
  \hbox{\it higgsinos} \ \  \psi &  \stackrel{R}{\to} & e^{-\gamma_5 \alpha}\,\psi\,.
   \ea \right.
   \ee 
   One thus gets Dirac spinors known as charginos and neutralinos -- even if denominations like winos, zino, etc.~could be more appropriate as we shall see. They may be expressed as \cite{R}
\be
\label{dirac}
\left\{
\ba{ccc}
R=+1\ \ \hbox{\it Dirac ino} &=&\hbox{\it gaugino}_L+\hbox{\it higgsino}_R\ ,
\vspace{-.5mm}\\
&\hbox{or}&
\vspace{-.5mm}\\
R=-1\ \,\hbox{\it Dirac ino} &=&\hbox{\it higgsino}_L+\hbox{\it gaugino}_R\ .
\ea\right.\!
\ee

\noindent
They have the same masses $m_W,\,m_Z$, etc. as the corresponding spin-1 gauge bosons, as long as supersymmetry is unbroken.
This already hints at ``gauge/BE-Higgs unification'', a crucial property that may be the prime motivation for supersymmetry
 \cite{gh,prd14}.
 
 \vspace{2mm}
 
The introduction of direct gaugino ($m_1, m_2$) and higgsino ($\mu$) mass terms then modifies these $R$-conserving chargino and neutralino mass matrices by including supersymme\-try-breaking $\Delta R=\pm 2$ contributions. 
The  $\mu$  parameter may be considered as ``supersymmetric'' as a $\mu \,H_2H_1$ mass term may be included directly in the superpotential, or regenerated from the $\lambda\,H_2H_1 S$ coupling through the translation of the $R=2$ spin-0 component of the singlet $S$, leading to

\vspace{-4mm}
\be
\label{mus}
\mu\,=\, \lambda \, <\!s\!>\,.
\ee
Still the $\mu$ term generates a super\-symmetry-breaking contribution to the mass matrices when the spin-0 doublets $h_1$ and $h_2$ acquire non-vanishing v.e.v.'s, by contributing to non-vanishing v.e.v.'s for the auxiliary components of $H_1$ and $H_2$.

\vspace{-.5mm}

  \subsection{\boldmath From $R$ symmetry to $R$ parity}

$R$ symmetry was introduced for reasons related with the triggering of the electroweak breaking
induced by $h_1$ and $h_2$, which must both acquire non-van\-ishing v.e.v.'s.. Otherwise we would stay with an unwanted massless chargino, even before thinking about introducing quarks and leptons and generating their masses. $R$ symmetry  was also introduced with the desire of defining a conserved quantum number $R$ attributed to massless or massive Dirac spinors as in (\ref{trh},\ref{dirac}),
 with differences $\Delta R=\pm 1$ between fermions and bosons within the multiplets of supersymmetry.
 
 \vspace{2mm}
This was done in a toy-model attempt at relating the photon with a ``neutrino'' carrying one unit of $R$, and the $W^-$ with a light chargino that might have been an ``electron'' candidate (or even in 1976, at the time of the $\tau$ discovery, a $\tau$ candidate, with the fermionic partner of the photon  as a $\nu_\tau$ candidate).
 But the previous ``neutrino'',
 called a gaugino in modern language, must in fact  be considered as a new photonic neutrino  within supersymmetric extensions of the standard model \cite{ssm}. It was called the photino, with, similarly, 
 the spin-$\frac{1}{2}$ partners of the gluons called the gluinos \cite{grav,glu}, so that
 \be
 \left\{
 \ba{ccc}
 \hbox{\it photon} &\ \stackrel{\it SUSY}{\longleftrightarrow} \ &\hbox{\it photino}\,,
 \vspace{1mm}\\
  \hbox{\it gluons} &\ \stackrel{\it SUSY}{\longleftrightarrow} \  &\hbox{\it gluinos}\,.
 \eaÊ\right.
 \ee
 
 \vspace{1mm}
  
 The parity of the new quantum number $R$ carried by the supersymmetry generator, 
 \be
 \label{rp}
 R_p\,=\,(-1)^R\,,
 \ee
 plays an important role. It  distinguishes between ordinary particles, with $R_p=+1$, and superpartners, also called {\it sparticles}, with $R_p=-1$, while allowing for the generation of masses
 \vspace{-.5mm}
  for the Majorana spin-$\frac{3}{2}$ gravitino and spin-$\frac{1}{2}$ gluinos, which transform chirally under $R$ symmetry \cite{grav,glu}. Their mass terms break explicitly the continuous $R$ symmetry, reducing it to $R$-parity. This one may then be 
 identified as~\cite{ssm,ff}
 \be
  \label{rp1}
 R_p\,=\,(-1)^R\,=(-1)^{2S}\,(-1)^{3B+L}\,.
 \ee
As $R_p=(-1)^{2S}(-1)^{3(B-L)}$, its conservation follows from the conservation of $B-L$, even only modulo 2, ensuring the stability of the lightest supersymmetric particle, or LSP. This remains valid even in the presence of neutrino Majorana mass terms.

\vspace{2mm}
All superpartners are then expected to decay so as to ultimately provide, at the end of the decay chain, a stable LSP, usually taken to be 
a neutralino or a light gravitino \cite{grav}, although other possibilities may also be considered. The neutralino, in particular, turns out to be a good candidate for the non-baryonic dark matter of the Universe.
 
  \vspace{2mm}
 
 Conversely, should $R$-parity necessarily be conserved\,? A non-conservation of $R$-parity, as in $R_p$-vio\-lating theories  \cite{rpv}, requires $B$ and/or $L$ violations. It usually leads to severe difficulties with unobserved effects 
 such as a much-too-fast proton decay mediated by squark exchanges, or too large neutrino masses, unless the corresponding products of $R_p$-violating couplings are taken sufficiently small. Also, if $R$-parity is no longer conserved, we generally lose the possibility of having a stable LSP as a candidate for the non-baryonic dark matter of the Universe.

\section{\boldmath N/{\lowercase {n}}MSSM \vspace{1.5mm} and MSSM superpotentials and potentials}
\label{sec:MSSM}

\subsection{Superpotentials}
\label{subsec:u1r}
 
 \vspace{-1mm}

Let us precise  the role of $R$ symmetry in restricting adequately the superpotentials considered. 
 The last component of the superpotential ${\cal W}$ provides a contribution to the Lagrangian density invariant under supersymmetry, up to a derivative which does not contribute to the action integral.
For the theory to be invariant under $R$  its superpotential $\cal W$
 must transform according to
 \be
 \label{wr}
{\cal W} (x,\theta) \ \  \stackrel{R}{\rightarrow}\ \ e^{2\,i\,\alpha}\ {\cal W}(x,\theta\,e^{-\,i\,\alpha})\,,
\ee
 so that its last component, which appears as the coefficient of the $\theta\theta$ term in its expansion and contributes to $\cal L$, be $R$-invariant.

 \vspace{2mm}
 
 A product of chiral superfields $\Pi\, \Phi_i$ transforms with $R=\sum R_{\Phi_i}$, and is allowed in the superpotential if and only if
 
 \vspace{-6.5mm}
\be
\hbox{\large $\sum$}\ R_{\Phi_i}=\,2\,.
 \ee
 The parameters $\lambda_{ijk}, \,m_{ij}$ and $\sigma_i$ in the superpotential
\be
\label{superpot}
{\cal W}\,=\,  \frac{\lambda_{ijk}}{3}\ \Phi_i\Phi_j\Phi_k + \frac{\mu_{ij}}{2}\ \Phi_i\Phi_j + \sigma _i\,\Phi_i\,
\ee
are required by $R$ symmetry to vanish, unless the corresponding products of superfields verify
 $\,R_{\Phi_i}\!+R_{\Phi_j}\!+R_{\Phi_k}\!=2$,  $\,R_{\Phi_i}\!+R_{\Phi_j}\!=2$, \,or $R_{\Phi_i}\!=2$. 
  
 \vspace{2mm}
 
 These restrictions from $R$ symmetry are used to select the nMSSM superpotential for the two electroweak doublets $H_1$ and $H_2$ interacting with an extra singlet $S$ through a trilinear superpotential coupling $\lambda \,H_2H_1 S$ 
 \cite{R},\be
 \label{wnmssm}
{\cal W}_{\rm nMSSM}\,=\,S\,(\lambda \,H_2H_1+\,\sigma) \,.
 \ee
 The terms involving quarks and leptons will be considered later \cite{ssm}.
 This superpotential is obtained by imposing $R$ symmetry on the general NMSSM superpotential, also including a 
 $\mu \,H_2 H_1$ mass term 
 as well as mass and self-interaction terms for $S$,
 \be
 \label{NMSSM}
{\cal W}_{\rm NMSSM}=\,S\,(\lambda \,H_2H_1+\,\sigma) + \mu\,H_2H_1 \,+\frac{\kappa}{3}\ S^3+ \frac{\mu_S}{2}\ S^2.
 \ee
$H_1$ and $ H_2$ transform as in (\ref{oldr}) so that $R$ sym\-metry can survive the electroweak breaking, extended to $S$ according to
  \be
  \label{oldrbis}
   H_{1,2}\ \stackrel{R}{\to} \ H_{1,2}(x,\theta\,e^{-\,i\,\alpha}),\ \  S\ \stackrel{R}{\to} \ e^{2i\alpha }\ S(x,\theta\,e^{-\,i\,\alpha})\,.
   \ee
Both $\lambda \,H_2H_1 S$ and $\sigma S$ are allowed by $R$ in the nMSSM superpotential (\ref{wnmssm}).
 The other NMSSM terms  in (\ref{NMSSM}), proportional to $H_2H_1,\ S^2$ and $S^3$, are excluded.

 \vspace{2mm}

Another way to restrict the general NMSSM superpotential (\ref{NMSSM}) into the nMSSM one (\ref{wnmssm}) is to ask for $\cal W$ to be invariant under the extra-$U(1)$ symmetry (\ref{u0bis},\ref{us})   \cite{R},

\vspace{-3mm}
   \be
  \label{u0ter}
   H_i\ \stackrel{U}{\to} \ e^{i\alpha}\ H_i,\ \  
   S\ \stackrel{U}{\to} \ e^{-2i\alpha}\ S\,,
   \ee
simply broken by the  dimension-2 linear term $\sigma S$, thus automatically avoiding a classically massless 
\hbox{spin-0} ``axion'', before this notion was even put into light. This extra-$U(1)$ symmetry also excludes NMSSM 
self-inter\-action and mass terms  proportional to $S^3$ and $S^2$ in the superpotential, as well as $\mu \,H_2H_1$.  
The latter  may still be subsequently regenerated through a translation of $S$ as in (\ref{mus}).

 \vspace{2mm}
Incidentally, the $\mu$ parameter, coefficient of the $\mu\,H_2H_1$ superpotential mass term in the MSSM, is ``supersymmetric'' 
(in the sense that $\mu \,H_2H_1$ may be present in the superpotential) but comes in violation of both the $R$-symmetry (\ref{oldrbis}) and the extra-$U(1)$ symmetry (\ref{u0ter}). It may thus remain naturally small or of moderate size, as compared to very large mass scales like the grand-unification or the Planck scales.

\vspace{2mm}
A special version of the above general NMSSM superpotential (\ref{NMSSM})
involves trilinear terms only in the superpotential \cite{NMSSM,NMSSM2}, with 
\be
 \label{NMSSM30}
{\cal W}_{\rm NMSSM}=\,\lambda \,H_2H_1 S\,+\,\frac{\kappa}{3}\ S^3\,,
 \ee
$\lambda$ and $\kappa$ being dimensionless. Most of its interesting properties rely on the same trilinear $\lambda \,H_2H_1 S$ coupling as in the nMSSM. 
 \,\,In the limit $\kappa\to 0$, both the $U(1)_R$  (\ref{oldrbis}) \linebreak and the extra-$U(1)$  (\ref{u0ter}) would be restored. The latter being broken by $\,<\!h_1\!>\ $ and $\,<\!h_2\!>\,$ (and $<\!s\!>\,$ if  also present)  a classically massless axionlike boson ($a$) would then reappear
in this limit, that was 
precedently avoided in the nMSSM by the  linear $\sigma S$  term (and in the above version of the NMSSM by $\frac{ \kappa}{3}\,S^3$).
Such a particle, which has not been observed, may also acquire a mass, possibly small, 
through the soft supersymmetry-breaking terms breaking explicitly  the extra-$U(1)$ symmetry.

\vspace{-1mm}

\subsection{The USSM, with a new neutral gauge boson}
\label{subsec:ussm}

\vspace{-1mm}

 Another option  is to gauge  the above extra-$U(1)$ symmetry (\ref{u0ter}), assuming the corresponding anomalies appropriately cancelled, usually through the introduction of extra fermion fields. These may involve, for example, mirror fermions,  or exotic fermions as would be present in an $E(6)$ theory. The would-be (axionlike) Goldstone boson ($a$) is then ``eaten away'' when the additional neutral gauge boson $Z'$ acquires a mass. This leads to the USSM,  with the trilinear superpotential 
 \be
 \label{wumssm}
{\cal W}_{\rm USSM}\,=\,\lambda \,H_2H_1 S\,,
 \ee
 the theory being at this stage invariant under both the $R$ symmetry (\ref {oldrbis}) and the extra-$U(1)$ symmetry  (\ref{u0ter}), now promoted to a local gauge symmetry  \cite{ssm}.
  \vspace{2mm}
  
 The gauging of an additional $U(1)$, possibly appearing as a subgroup of a non-abelian grand-unification group like $E(6)$, 
 with (anti)quark and (anti)lepton chiral superfields transforming axially according to
 \be
 \label{fax}
 (L,Q\,;\,\bar E,\bar D,\bar U)\ \stackrel{U}{\to} \ e^{-\frac{i\alpha}{2}}\, (L, Q\,;\,\bar E,\bar D,\bar U)\,,
 \ee
requires a new spin-1 gauge boson $Z'$. More generally the extra-$U(1)$ symmetry generator to be gauged may involve a linear combination of the axial $U(1)$ quantum number defined from (\ref{u0ter},\ref{fax}) as 
\be
\label{fua}
\left\{
\ba{ccc}
F_A(L,Q\,;\,\bar E,\bar D,\bar U)\,=\,-\,\hbox{\small$\dis\frac{1}{2}$}\,,
\vspace{1mm}\\
F_A(H_1,H_2)\,=\,1\,,\, \ \ F_A(S)\,= \,-\,2\,.
\ea\right.
\ee
with the weak hypercharge $Y$ and the $B$ and $L$ (or $B-L$) quantum numbers.
 A large  v.e.v.~for an extra singlet like $s$, already present in the theory and  transforming as in (\ref{u0ter}), 
 $s\to e^{-2i\alpha}\,s$,  may  make the new gauge boson much heavier than the $W^\pm$ and $Z$, giving it a large mass
 $\simge $ TeV scale \cite{pl80b}. But no new heavy boson corresponding to an enlargement of the gauge group has been discovered yet.
 
 
   \subsection{\boldmath A new\vspace{1,5mm} light gauge boson $U$\,? or a light pseudoscalar $a$\,?}
    \vspace{-1mm}
    
   There is also another interesting possibility.
   An additional $U(1)$ factor in the gauge group, if not embedded within a grand-unification group like $O(10)$ or $E(6)$, ...~,  would have its own gauge coupling constant $g"$, next to $g$ and $g'$.
   This one may be much smaller than  $g$ and $g'$, in which case the mass of the new neutral gauge boson may well be small.
This $Z'$,  also called a $U$ boson, would then have, for its longitudinal polarisation state, effective interactions fixed by  $g" k^\mu/m_U$. It would behave very much as the ``eaten-away'' 
 Goldstone boson $a$, acquiring effective axionlike pseudoscalar couplings to quarks and leptons recovered from its axial couplings $f_A$ (proportional to $g"$),  as~\,\cite{U}
 \be
 f_p\,=\,f_A\ \frac{2m_{l,q}}{m_U}\,.
 \ee

   \vspace{0.5mm}

 This is very similar to the situation for  a massive but light spin-$\frac{3}{2}$ gravitino, 
 \vspace{-.5mm}
 with a very small gravitational coupling 
 $\kappa= \sqrt{8\pi\,G_N}\,\simeq 4\,10^{-19}$ GeV$^{-1}$, and a small mass 
 
 \vspace{-7mm} 
 \be
 \label{mgrav}
 m_{3/2}= \frac{\kappa d}{\sqrt 6}\,=\,\frac{\kappa F}{\sqrt 3}\,.
 \ee
$F$, or $\sqrt F$, is usually referred to as the super\-symmetry-breaking scale parameter.
 Such a light gravitino 
 would have its $\pm 1/2$ polarisation states  interacting  
 propor\-tionally to $\kappa \,k^\mu/m_{3/2}$, or $k^\mu/F$. 
   \vspace{-.4mm}
 It would still behave very much as the ``eaten-away''
  \hbox{spin-$\frac{1}{2}$} goldstino, according to the ``equivalence theorem'' of supersymmetry. The strength of its interactions then depends on the scale at which supersymmetry is spontaneously broken, getting very small if the supersymmetry-breaking scale 
  \vspace{-.4mm}
  ($\sqrt d$ or $\sqrt F$) is large enough \cite{grav}.

  \vspace{2mm}
  
 Let us return to a light spin-1 $U$ boson. As it would behave very much like the corresponding equivalent Goldstone boson
 \cite{U}, it would certainly be excluded if it could be produced,
 most notably in the radiative decays of the $\psi$ and the $\Upsilon$,  much like a standard axion ($A$). Fortunately the \hbox{singlet}  $s$ already present in these theories, transforming  under $U$ as in  (\ref{u0ter}) according to $s\to e^{-2i\alpha}\,s$,  may acquire a large v.e.v., significantly above the weak scale. The extra-$U(1)$ symmetry is then  broken ``at a large scale'' $F_U$,
 where the mass $\,m_U\propto g" F_U$ may still be small when the extra $U(1)$ is gauged with a very small coupling.
 The corresponding particle (either the very light \hbox{spin-1} $U$ boson or its  ``equivalent''  \hbox{spin-0} pseudoscalar $a$) \,is then coupled effectively very weakly, proportionally to $ g"/m_U$, or $1/F_U$  \cite{pl80b,U}. This pseudoscalar $a$ is mostly an electroweak singlet, largely inert.
 
  \vspace{2mm}
  
 Dealing with a spin-0 particle this also provided, as a by-product, a very early realization of  the ``invisible axion'' mechanism that became popular later, in which  the ``invisible axion'' is mostly an electroweak singlet  \cite{pl80b}. 
Furthermore the doublet and singlet $U(1)$ quantum numbers are here appropriate to the supersymmetry framework, 
with an invariant   $\lambda \,H_2H_1S$ trilinear coupling, resulting in the $U(1)$ quantum numbers 
$+1$ for $h_1$ and $h_2$, $-1/2$ and $+1/2$ for left-handed and right-handed quarks and leptons, and $-2$ for the extra singlet $s$ as in (\ref{fua}).

  \vspace{2mm}
  
 In a similar way a light spin-1 gauge boson  $U$, interacting very much as the eaten-away Goldstone boson 
 $a$ i.e.~as an ``invisible axion'' (except  for the $\gamma\gamma $ coupling of the latter), 
 also becomes  largely ``invisible''  if the extra-$U(1)$ symmetry is broken at a sufficiently high scale. 
 But the hunt for such a light spin-1 $U$ boson is another story \cite{prodU}.

 \subsection{N/nMSSM \vspace{1mm} and MSSM potentials}
 
 \vspace{-2mm}
 
The nMSSM superpotential (\ref{wnmssm})  leads to the potential \cite{R}

\vspace*{-5mm}
 \be  
  \label{lhh}
  \ba{r}
V_{\rm \tiny nMSSM}\,= \ \hbox{$\dis \frac{g^2+g'^2}{8}$} \ (h_1^\dagger \,h_1-h_2^\dagger\, h_2)^2\,+\, 
\hbox{$\dis \frac{g^2}{2}$}\ |h_1^\dagger \,h_2|^2
\vspace{1mm}\\
+\,
\hbox{$\dis \frac{\xi\, g'}{2}$}\,(h_1^\dagger \,h_1-h_2^\dagger\, h_2)\,+\,\hbox{$\dis \frac{\xi^2}{2}$}\hspace{10mm}
  \vspace{2.5mm}\\
+ \,|\,\lambda \,h_2 h_1+\sigma|^2 + \lambda^2\, |s|^2Ê\ (|h_1|^2+Ê|h_2|^2)\,.
 \ea
 \ee
 The $D$-term contributions  take into account an abelian  $-\xi D'$ term in $\cal L$ (this sign choice, different from the usual one in (\ref{vpotd}), being made to have $\xi>0$ for $v_2>v_1\,$ i.e.  $\tan\beta>1$).
\vspace{2mm}

 This also applies to the general NMSSM through the replacements 
 \be
\! \ba{lc}
 \sigma S\,\to\ f(S)\,=\hbox{\small $\dis \frac{\kappa}{3}$}\, S^3+ \hbox{\small $\dis \frac{\mu_S}{2}$}\,S^2+\sigma S\,,& \ \lambda S\,\to\, \mu+\!\lambda S\,,
 \vspace{2mm}\\
\ \  \sigma\,\to \,\hbox{$\dis \frac{df(s)}{ds}$} =\,\kappa\, s^2+\mu_S\,s +\sigma\,,& \ \lambda s\,\to\, \mu+\!\lambda s\,,
\ea
\ee
 in the superpotential and potential, respectively, leading to
 \be  
  \label{lhhNMSSM}
  \ba{c}
V_{\rm \tiny NMSSM}\,= \ \hbox{$\dis \frac{g^2+g'^2}{8}$} \ (h_1^\dagger \,h_1-h_2^\dagger\, h_2)^2\,+\, 
\hbox{$\dis \frac{g^2}{2}$}\ |h_1^\dagger \,h_2|^2
\vspace{1mm}\\
+\,
\hbox{$\dis \frac{\xi\, g'}{2}$}\,(h_1^\dagger \,h_1-h_2^\dagger\, h_2)\,+\,\hbox{$\dis \frac{\xi^2}{2}$}\hspace{10mm}
  \vspace{2.5mm}\\
\! + \,|\,\lambda \,h_2 h_1+\kappa\, s^2+\mu_S\,s +\sigma|^2+\,  |\mu+\lambda s|^2Ê\, (|h_1|^2\!+Ê|h_2|^2)\,.
 \vspace{1mm}\\
 \ea
 \ee

 \vspace{2mm}

The translation (\ref{mus}) of the singlet $S$ restores the (N)MSSM mass term $\mu \,H_2H_1$ from the nMSSM superpotential (\ref{wnmssm}).
Furthermore in the $\lambda\to 0, \sigma \to \infty $ limit, with $\lambda\sigma$ fixed,  for which $S$ decouples, we recover the 
MSSM potential, in the conceptually-interesting situation of a MSSM potential with \hbox{dimension-2} {\it soft-breaking terms generated from a supersymmetric Lagrangian density}. It reads (up to a very large or infinite constant term, irrelevant at the moment) 
 \be  
  \label{lhhMSSM}
  \ba{c}
V_{\rm \tiny MSSM}\,= \ \hbox{$\dis \frac{g^2+g'^2}{8}$} \ (h_1^\dagger \,h_1-h_2^\dagger\, h_2)^2\,+\, 
\hbox{$\dis \frac{g^2}{2}$}\ |h_1^\dagger \,h_2|^2
  \vspace{2.5mm}\\
  (\mu^2+ \hbox{\small$\dis \frac{\xi g'}{2}$} )\,h_1^\dagger h_1\,+\,(\mu^2-\hbox{\small$\dis \frac{\xi g'}{2}$} )\,h_2^\dagger h_2 \,+\,2\,\lambda\sigma\, \Re \, h_2h_1\,.
  \ea
  \ee
  
  \vspace{1mm}
  
  \noindent
The last term, $\propto  \Re\, h_2h_1$, forces $h_1$ as well as $h_2$ to acquire a non-vanishing v.e.v..
But this does not lead to an unwanted classically-massless axion or axionlike pseudoscalar $A$, as this term 
$\propto  \Re\, h_2h_1$ 
breaks explicitly the extra-$U(1)$ symmetry (\ref{u0},\ref{u0bis}),  $h_1\to e^{i\alpha}\,h_1,\ h_2\to e^{i\alpha}\,h_2$\,.

  \vspace{2mm}
If the (extremely weak) interactions of the singlet $S$ were reconsidered again, with an extremely small coupling $\lambda$, the vacuum state corresponding to (\ref{lhhMSSM}), which then has an extremely large  energy density $\,\simeq \,\sigma^2\propto1/\lambda^2$, would be destabilized, but still staying effectively quasi-metastable.

\vspace{2mm}

These expressions of the N/nMSSM and MSSM potentials illustrate how spin-0 interactions may now be viewed as  
{\it part of the electroweak gauge interactions}, 
with their quartic couplings fixed  by
   \be
    \hbox{$\dis \frac{g^2+g'^2}{8}$}\ \ \ \hbox{and}\ \ \hbox{$\dis \frac{g^2}{2}$}\,.
   \ee
 They lead to a spontaneous  breaking of $\,SU(2)\!\times U(1)$ into $U(1)_{\rm QED}$,
with non-vanishing v.e.v.'s for both $h_1$ and $h_2$, 

\vspace{-4mm}

 \be
\label{h1h2}
<h_1>\ =\left(\ba{c} \frac{v_1}{\sqrt 2}\vspace{.5mm}\\ 0\ea\right),\ \ \ <h_2>\,=\left(\ba{c} 0\vspace{.5mm}\\ \frac{v_2}{\sqrt 2}\ea\right),
\ee
where $\,v_1=v\cos\beta,\,v_2=v\sin\beta\,$ with $v\simeq 246$ GeV. 
This also leads us in the direction of gauge-Higgs unification already alluded to in (\ref{mhZ}) \cite{fa76,gh,prd14}, discussed in the next Section.

\vspace{2mm}

The nMSSM potential (\ref{lhh}), in particular, forces $\,v_1$ and $v_2$ to verify at this stage
 (before the introduction of extra terms  breaking supersymmetry explicitly)
$\,\sigma - \frac{1}{2} \,\lambda \,v_1 v_2=0$,
which ensures the vanishing of the $F$ terms in the potential. 
Minimizing the $D$ terms leads (without a $\mu$ term yet) to $<\!D_Z\!>\ =0\,$ with $<\!D_\gamma\!>\ \neq 0\,$ so that the photino is the Goldstone spinor, 
then fixing, in the absence of other soft-breaking terms, 
\be
m_Z^2\ (-\cos 2\beta)\,=\,\xi g'\, ,
\ee
 i.e.~$\sqrt \xi  \simeq  {m_Z}/{\sqrt {g'}} \simeq 155  \ \hbox{GeV}$,
 in the large $\tan\beta$ limit~\,\cite{prd14}.

 \vspace{2mm}

The structure of the nMSSM superpotential (\ref{wnmssm})  (and resulting potential as in (\ref{lhh})) is useful in many circumstances, 
and most notably to trigger gauge symmetry breaking by
rendering the gauge-symmetric vacuum state unstable. It also leads to  inflationary potentials useful in the description of the very early Universe, 
with an initial energy density  such as $\sigma^2+ {\xi^2}/{2}$, providing the necessary fuel for inflation.
Additional soft-breaking terms, of dimension $\leq 3$  \cite{gg}, possibly induced from supergravity
\cite{cremmer,grbr1,grbr2,cfg,grbr3,grbr4}, may also be added to the (N/n)MSSM potentials 
(\ref{lhh},\ref{lhhNMSSM},\ref{lhhMSSM}).

 \bc
  \vspace{0mm}
  
  *\hspace{8mm}*
  
   \vspace{2mm}
   
 *
 \ec
 
    \vspace{-3mm}
    
  \pagebreak

\section{\boldmath Gauge/BE-Higgs \vspace{1.8mm} unification \hbox{in the \,(N/{\lowercase {n}})MSSM}}
\label{sec:gh}

\vspace{-.5mm}

\subsection{\boldmath An index $\Delta$ for counting massless chiral spinors}

\vspace{-1mm}
 
These theories make use of gauge superfields, describing (left-handed) gaugino fields $\lambda_L$ carrying $R=1$, and 
$R=2$ or 0  chiral superfields  describing (left-handed) chiral spinors with $R=1$ and $-1$ respectively, as seen from (\ref{R2},\ref{dirac}).
With $n_g$ gauge superfields and $n_2$ and $n_0$ chiral superfields with $R=2$ or 0, we get $n_g+n_2$ and $n_0$ 
(left-handed) spinors with $R=1$ and $-1$, respectively. The former are in excess by the difference
\be
\Delta \,= \underbrace{n_g+n_2}_{\hbox{\small $R=1$\  \rm spinors}}\ 
-\underbrace{\ \ \ n_0\phantom{n_g}\ }_{\hbox{\small $R=-1$\  \rm spinors}}\!\!\!\!.
\ee
$R=\!1$ and $-1$ left-handed spinors may combine as in (\ref{dirac}), in a way compatible with $R$ symmetry, into massive 
Dirac spinors carrying $R=\pm 1$. For $\Delta \geq 0$,  $\Delta$ chiral spinors with $R=1$, at least, must remain massless
if $R$ symmetry is conserved, 
as staying unpaired with \hbox{$R=-1$} counterparts.

\vspace{2mm}
With the nMSSM superpotential  (\ref{wnmssm}) $\,n_g=4$ for $SU(2) \times U(1)$, $n_0=4$ for $H_1$ and $H_2$ 
and  $\,n_2=1$ for $S$, so that 
$\Delta=1$. 
One left-handed spinor with $R=1$, neutral,  must remain massless, which is here the photino.
This one becomes the Goldstone spinor when supersymmetry is spontaneously broken. 
An early version of the USSM, with one additional extra-$U(1)$ gaugino, had $\Delta= 5+1-4= 2$, leading to two massless $R=1$ spinors, with a
goldstino (eaten-away by the spin-$\frac{3}{2}$ gravitino) different from the photino, superpartners ultimately decaying into gravitinos or photinos carrying away missing energy-momentum \cite{ssm,grav}.

\vspace{-1mm}

\subsection{\boldmath The goldstino must transform with $R=1$}

\vspace{-1mm}

The (left-handed) massless Goldstone spinor $\lambda_g$ associated with spontaneous
supersymmetry breaking must have $R = 1$ in a $R$ symmetric theory, as follows from the $R$
transformation properties (\ref{R}) of the supersymmetry generator and vector-spinor current, such that
\be
J^\mu_{\,\alpha}\ \stackrel{R}{\to}\ e^{-\gamma_5\alpha}\ J^\mu_{\,\alpha}\ .
\ee
For spontaneously  broken supersymmetry the vector-spinor current may be expressed as 
\be
J^\mu_{\,\alpha}\,=\,d\,\gamma^\mu \gamma_5\,\lambda_g+ \,...\ ,
\ee
 where $d/\sqrt 2=F $ is  the supersymmetry-breaking scale parameter 
which determines 
\vspace{-.4mm}
the gravitino mass $m_{3/2}=\kappa d/\sqrt 6= \kappa F/\sqrt 3\,$ in (\ref{mgrav})\,\,\cite{grav}. 
\vspace{-.8mm}
The Goldstone spinor must transform according to
$
\lambda_g\, \stackrel{R}{\to}\, e^{\gamma_5\alpha}\ \lambda_g$,
or equivalently

\vspace{-6mm}

\be
\label{gold}
\,\lambda_{gL}\ \stackrel{R}{\to}\ e^{i\alpha}\ \lambda_{gL}\,,
\ee
i.e. it should transform with $R=1$\,.

\pagebreak

 It should be either a gaugino
as in pure $D$-breaking \cite{R,fi,fa76}, or a \hbox{spin-$\frac{1}{2}$} fermion field described by a $R=2$ chiral superfield as in $F$-breaking \cite{for,for2}, or a mixing of both as in \cite{ssm,fi}. In the nMSSM at the present stage, with an unbroken $R$ symmetry and without any addition of soft supersymme\-try-breaking terms yet, 
the massless goldstino field, with $R=1$,  coincides with the photino field, supersymmetry remaining unbroken within neutral multiplets.
This degeneracy  gets broken later through terms breaking explicitly (although softly) the supersymmetry \cite{martin,gg}, possibly obtained from gravity-induced supersymmetry breaking \cite{cremmer,grbr1,grbr2,cfg,grbr3,grbr4}. Still the
$R$-symmetric nMSSM considered at the present stage is essential in the understanding of the gauge/BE-Higgs unification and of the resulting mass spectrum for the various versions of the MSSM or N/nMSSM, as we shall see.

\subsection{\boldmath $U(1)_R$ symmetric nMSSM mass spectrum}

 \vspace{-1mm}

With the gauge and chiral superfields transforming under the continuous $R$ symmetry ($U(1)_R$) according to
\be
\label{r}
\left\{
\ba{ccc}
V_a & \stackrel{R}{\to} &V_a(x,\theta e^{-i\alpha},\bar\theta e^{i\alpha})\,,
\vspace{1mm}\\
H_{1,2}& \stackrel{R}{\to} &    H_{1,2}(x, \theta e^{-i\alpha})\,,
\vspace{1mm}\\
S& \stackrel{R}{\to} &    e^{2i\alpha }\ S(x, \theta e^{-i\alpha})\,,
\ea\right.
\ee
$R$ symmetry leads to 1 chiral spinor remaining massless, with 4 massive Dirac ones. The $R$-symmetric and quasi-supersym\-metric fermion spec\-trum for the nMSSM is at this stage, with ${gv_1}/{\sqrt 2}=m_W\sqrt 2\,\cos\beta,\  {gv_2}/{\sqrt 2}=m_W\sqrt 2\,\sin\beta$ \cite{R}:
\be
\label{spec}
\ba{c}
\!\framebox [8.2cm]{\rule[-1.7cm]{0cm}{3.55cm} $ \dis
\left\{\ \ba{lcl}
\hbox{1 massless \it photino:} &&m= 0\,,
\vspace{2mm}\\
\hbox{2 Dirac \it winos:} &&m= 
\left\{\ba{c}
m_W\sqrt 2\,\cos\beta\,, \vspace{1mm}\\ 
m_W\sqrt 2\,\sin\beta\,,\ea\right.
\vspace{2mm}\\
\hbox{1 Dirac \it zino:} &&m_Z=\sqrt{g^2+g'^2}\ v/2\,,
\vspace{2mm}\\
\hbox{1 Dirac \it neutralino:} &&m=\lambda v/\sqrt 2\ ,
\ea \right.
$}
\vspace{1.5mm}\\
\ea
\ee

\vspace{1.5mm} \noindent
all spinors carrying $R=\pm 1$, in agreement with their expressions from gaugino and higgsino fields as in (\ref{dirac}).
The corresponding $5\times 5$ neutralino mass matrix expressed in a gaugino-higgsino basis will be given later in subsection \ref{subsec:neut}.

\vspace{2mm}

The charged and neutral \hbox{spin-0} masses, obtained from the potential (\ref{lhh}), are 
\be
\label{boson}
\left\{
\ba{c}
m_{H^\pm}=m_W\,,\ \ \ m_z\,=\,m_Z\,, 
\vspace{2mm}\\
m\,(\hbox{\,4 \it neutral spin-0 bosons}\,) \,=\, \hbox{\small$\dis\frac{\lambda v}{\sqrt 2}$}\,.
\ea\right.
\ee
All neutral spin-0 bosons have the same mass $m_Z$ or ${\lambda v}/{\sqrt 2}$ as their fermionic partners in (\ref{spec}), thanks to the unbroken supersymmetry  in the neutral sector, the photino being here the Goldstone spinor.
Two of the  four neutral bosons of mass $\lambda v/\sqrt2$ are described by the singlet superfield $S$, with $R=2$.
The two others are described by the $R=0$ superfield
\be
\label{HA}
H_A\,=\,H_1^0\,\sin\beta + H_2^0\,\cos\beta\,.
\ee 
As $\,<\!h_1^0\!>\ = v\cos\beta/\sqrt 2, \ <\!h_2^0\!>\ = v\sin\beta/\sqrt 2$, 
 $H_A$ acquires the mass ${\lambda v}/{\sqrt2}$ from the $\lambda \,H_2H_1S$ superpotential term, by combining with the singlet $S$
according to
\be
\label{mix}
\lambda \ H_2H_1 S\,=\, -\,\hbox{\small$\dis \frac{\lambda v}{\sqrt 2}$}\ (H_1^0\,\sin\beta + H_2^0\,\cos\beta)\,S\,+\,...\,,
\ee
in a way compatible with $R$ symmetry (with $H_2 H_1 =-\,H_1H_2=  H_2^+H_1^- -H_2^0H_1^0$).

\vspace{3mm}

All four scalars  would return to massless
for $\lambda \to 0$, for which $S$ decouples,
$H_A$ returning to massless. Indeed in the $\lambda \to 0$ limit one recovers at the classical level a
spontaneously broken extra-$U(1)$  acting as in (\ref{u0},\ref{u0bis}), generating a classically-massless axion or axionlike particle. This one, which has here the mass $m_A=\lambda v/\sqrt 2\,$,  is described by  the imaginary part of the  spin-0 component of $H_A$,
\vspace{-6mm}

\be
\label{A}
A\,=\,\sqrt 2\  \,\hbox{Im}\ (h_1^0\,\sin\beta + h_2^0\,\cos\beta)\,.
\ee


\subsection{\boldmath Gauge/BE-Higgs \vspace{1.5mm} unification}
\vspace{-2mm}

\bc
{\it Relating gauge and BEH bosons, \vspace{.5mm}in spite of different electroweak properties}
\ec

\vspace{1mm}

The superfield orthogonal to $H_A$ in (\ref{HA}) is
\be
\label{HZ}
H_z=\,- \,H_1^0\,\cos\beta + H_2^0\,\sin\beta\,.
\ee
The imaginary part of its spin-0 component,
\be
z_g\,=\,\sqrt 2\  \,\hbox{Im}\ (-\,h_1^0\,\cos\beta +h_2^0\,\sin\beta)\,,
\ee
orthogonal to the $A$ field in (\ref{A}),
describes  the would-be Goldstone boson eaten away by the $Z$. Indeed this Goldstone field $z_g$ originates from the imaginary part of 
SM-like combination 
$\varphi_{\rm sm}^{\circ}\!=h_1^{\circ}\,\cos\beta + h_2^{\circ *}\,\sin\beta\,$
\vspace{-.5mm}
responsible for the electroweak breaking,
with \hbox{$<\!\varphi_{\rm sm}^{\circ}\!>\ =v/\sqrt 2\,$}.

\vspace{2mm}

The real part of the spin-0 component of $H_z$ is
\be
\label{z0}
z\,=\,\sqrt 2\  \,\Re\ (-\,h_1^0\,\cos\beta +h_2^0\,\sin\beta)\,.
\ee
The signs are chosen 
\vspace{-.4mm}
for convenience so that $H_z \to H_2^0$ and $\,z\to\sqrt 2\ \,\Re \ h_2^0$ 
in the large $\tan\beta$ limit.
This field, suitably translated so that $\,<\!z\!>\ =0\,$, describes in this formalism the spin-0 boson partner of the $Z$ within a massive gauge multiplet of supersymmetry. Its mass is obtained from the   
$D_Z^2/2$ contribution to the potential, as expressed in the nMSSM superpotential (\ref{lhh}), with 
\cite{prd14}
\be
\label{dz}
\ba{ccl}
D_Z\!&=&\!\!\dis \frac{\sqrt{(g^2+g'^2)}\ v}{2}\ [\sqrt 2\ \Re\, (-h_1^0\cos\beta+h_2^0\sin\beta)]+ ...\,
\vspace{1mm}\\ 
&=&\ \ m_Z\ z\,+\, ...\ ,
\vspace{-1mm}\\ 
\ea
\ee
and
\vspace{-6mm}

\be
\label{dz2}
\hbox{\small$\dis \frac{1}{2}$}\ \,D_Z^2\,= \,\hbox{\small$\dis \frac{1}{2}$}\ m_Z^2\ z^2\,+\,...\,.
\ee
One thus has
\be
\label{eqzZ}
\framebox [8.55cm]{\rule[-.65cm]{0cm}{1.45cm} $ \dis
\ba{c}
m_z\,=\,m_Z\simeq \ 91\  \hbox{GeV}/c^2\  
\vspace{1.5mm}\\
\hbox{\it \small  (up to supersymmetry-breaking mass and mixing effects)\,,}\
\ea
$}
\ee 
independently of the value of $\,\tan\beta$, in agreement with the unbroken supersymmetry in the neutral sector \cite{R}.

\vspace{2mm}

More precisely when the BEH mechanism operates within a supersymmetric theory, it provides {\it massive gauge multiplets} \cite{fa76}. Each of them 
describes  a massive spin-1 gauge boson, two spin-$\frac{1}{2}$ inos constructed from gaugino and higgsino components
as in (\ref{dirac}), and a \hbox{spin-0} BEH boson associated with the spontaneous breaking of the gauge symmetry.
We get systematic associations between massive gauge bosons and \hbox{spin-0} BEH bosons, {\it a quite non-trivial feature owing to 
their different gauge symmetry properties},  and very different couplings to quarks and leptons \cite{gh,prd14,epjc}.

\vspace{2mm}

We have in particular the association  
\be
\label{gh20}
\framebox [8.55cm]{\rule[-.23cm]{0cm}{.75cm} $ \dis
Z  \stackrel{\ba{c}\hbox{\footnotesize \it SUSY}\ea}{\longleftrightarrow}  \hbox{2 Majorana zinos} 
 \stackrel{\ba{c}\hbox{\footnotesize \it SUSY}\ea}{\longleftrightarrow}  \hbox{spin-0 BEH boson}.
$}
\ee
Independently of $\tan\beta$, and of $\lambda$ in the presence of the N/nMSSM singlet $S$, the neutral spin-0 boson described by the $z$ field  in (\ref{z0})
becomes the spin-0 partner of the $Z$ within a massive multiplet of supersymmetry.
It has  the same mass $m_Z$ as long as supersymmetry is unbroken in this sector, in agreement with (\ref{mhZ},\ref{spec},\ref{boson},\ref{dz2}) \cite{R,prd14,gh}. 
 \vspace{2mm}

This also applies to the $W^\pm $, according to
\be
\label{gh3deb}
\framebox [8.55cm]{\rule[-.23cm]{0cm}{.75cm} $ \dis
W^\pm \stackrel{\ba{c}\hbox{\footnotesize \it SUSY}\ea}{\longleftrightarrow}\hbox{2 Dirac winos}
 \stackrel{\ba{c}\hbox{\footnotesize \it SUSY}\ea}{\longleftrightarrow}
\hbox{spin-0 boson} \ H^\pm.
$}
\ee
The $W^\pm $ is associated with  two Dirac winos (usually known as charginos), obtained as in (\ref{dirac}) with masses given in (\ref{spec}), and a charged spin-0 boson 
$H^\pm$ (or $w^\pm$), with
\be
\label{wh}
w^\pm\, \equiv\, H^\pm=\,\sin \beta\  h_1^\pm +\cos\beta \ h_2^\pm\,,
\ee
approaching $h_1^\pm$ at large $\tan\beta$.
 This one was originally called $w^\pm$ in \cite{R} to emphasize its relation with the $W^\pm$, leading in a model-independent way to 
\be
\label{eqhw}
\framebox [6.55cm]{\rule[-.55cm]{0cm}{1.25cm} $ \dis
\ba{ccc}
m_{H^\pm}\,=\ m_{W^\pm} 
\simeq \ 80\  \hbox{GeV}/c^2\  
\vspace{1mm}\\
\hbox{\it \small  (up to supersymmetry-breaking effects)\,.}\
\ea $}
\ee 
 We shall see later how these mass equalities get modified in the presence of supersymmetry-breaking effects, in the MSSM and N/nMSSM.

\vspace{2mm}

As seen on (\ref{gh20},\ref{gh3deb}) the supersymmetry generator $Q$ has become able to {\it relate bosons and fermions
with  different electroweak gauge symmetry properties}, a quite non-trivial feature, in contrast with the usual belief. This makes supersymmetry a very powerful symmetry, much beyond the simple replication of degrees of freedom by associating bosons and fermions with the same gauge symmetry properties.

\vspace{2mm}

While massive gauge bosons and \hbox{spin-0} BEH bosons have different symmetry properties for the electroweak gauge group, that is spontaneously broken, they do have the same  properties for the $SU(3)_{\rm QCD}\times U(1)_{\rm QED}$ symmetries, that remain unbroken.

\vspace{2mm}

When supersymmetry is broken the lightest neutral spin-0 boson should presumably be identified with  the 125 GeV boson found at CERN
\cite{higgs,higgs2} (unless a lighter one has escaped attention).
\vspace{-.3mm}
This one may well correspond to the above $z$ field
(approaching
$\sqrt 2\  \Re\ h_2^0$ in the large $\tan\beta$ limit),
up to a mixing angle, possibly small, induced by supersymmetry breaking.

\vspace{2mm}
However, the non-observation, at this stage, of a charged \hbox{spin-0} BEH boson $H^\pm$ seems to indicate (unless such a boson is found, with a moderate mass)
that the effects of supersymmetry breaking are more important in the $W^\pm$ than in the $Z$ multiplet.  This may be understood from the possible form of the supersymmetry-breaking terms.


\subsection{Describing  \vspace{1.5mm} 
spin-0 BEH bosons by massive gauge superfields}

\vspace{-.5mm}

This association between the spin-1 $W^\pm$  and $Z$ and the spin-0  $ H^\pm$ (also called $w^\pm$) and $z$ can be made explicit in a different superfield formulation.
Spin-0 BEH bosons will now be
described by the spin-0 components of {\it massive gauge superfields} \cite{fa76,prd14,gh}, after all components of the
 superfields
$H_1^-, H_2^+$ and  $H_z$ in (\ref{HZ}), then considered as chiral Goldstone superfields,  get completely gauged away through  the generalized gauge choices 
\be
H_1^-\equiv H_2^+\equiv 0, \ \ \ H_z\equiv \ <H_z>\ =\,-\,\hbox{\small$\dis\frac{v}{\sqrt 2}$}\ \cos 2\beta\,.
\ee

\vspace{-1mm}

In this new picture 
these spin-0 bosons 
get described, in a manifestly supersymmetric formulation, by the lowest ($C$) spin-0 components of {\it massive} $Z$ and  $W^\pm $ superfields, expanded as
$\,Z(x,\theta,\bar\theta)=C_Z + ...  -\,\theta\sigma_\mu \bar\theta\   Z^\mu + \,...\,,\ 
W^\pm(x,\theta,\bar\theta)=C_W^\pm +\, ... \, -\,\theta\sigma_\mu \bar\theta\ W^{\mu\,\pm } +\, ...\ $.
Their \linebreak spin-0 $\,C$ components now describe,
through {\it non-polyno\-mial field transformations}\,
linearized as $
\,z=-\,m_Z \,C_Z+\,...\ $, $w^\pm =$ $ \,m_W\,C_W^{\pm} +\,... \,$, the same spin-0 fields $z$ and $w^\pm$ 
as in the usual formalism (with signs depending on previous choices for the definitions of $z$ and $w^\pm$). 
\vspace{2mm}

We thus have
\be
\label{expsf}
\left\{\ba{ccl}
Z(x,\theta,\bar\theta)\!&=&(\,\hbox{\small$\dis\frac{-z}{m_Z}$}+ ...\,)+\, ... \,-\,\theta\sigma_\mu \bar\theta\    Z^\mu\  + \,...\ , 
\vspace{2mm}\\
W^\pm(x,\theta,\bar\theta)\!&=& (\,\hbox{\small$\dis\frac{w^\pm}{m_W}$}+ ...) +\,... \,
-\,\theta\sigma_\mu \bar\theta\      W^{\mu\,\pm} + ...\ .
\ea  \right.
\ee
{\it The spin-0 components of massive gauge superfields now describe spin-0 BEH bosons\,!} 
Their subcanonical ($\chi$)  spin-$\frac{1}{2}\!$ components, instead of being gauged-away as usual, now also correspond to physical degrees of freedom
describing
the \hbox{spin-$\frac{1}{2}$} fields usually known as higgsinos.

\vspace{2mm}

Supersymmetry transformations act in a linear way on the components 
($C,\chi,M,N,V^\mu\!,\lambda, D$) 
of a massive gauge superfield $V(x,\theta,\bar\theta)$, including auxiliary as well as physical components. But they act in a more complicated way when they are formulated in terms of the usual canonically-normalized 
spin-0 BEH and spin-$\frac{1}{2}$ higgsino fields, in particular as their  expressions  involve  the dimensionless $C$ components in a non-polynomial way.

\section{(N/\lowercase {n})MSSM \,mass spectra}

\vspace{-2mm}
\bc
{ \em with gauge/BE-Higgs unification}
\ec

\vspace{-6mm}

\subsection{Spin-0 masses in the MSSM}

\vspace{-1mm}

The non-observation, at this stage, of a charged \hbox{spin-0} BEH boson seems to indicate
that the effects of supersymmetry breaking should be more important in the $W^\pm$ than in the $Z$ multiplet. 
This may be an effect of a significant supersymmetry-breaking term,  possibly generated spontaneously from the decoupling limit of an extra singlet as indicated in  (\ref{lhhMSSM}),
 or from soft  gravity-induced terms.

 \vspace{1.5mm}

Let us define
\be
\left\{\ba{ccc}
\varphi_{\rm sm}\!&=&\!   h_1 \,\cos\beta +\,h_2^c \,\sin\beta\,,
\vspace{1.5mm}\\
\varphi_{\rm in}\!&=&\!   h_1\, \sin\beta -\,h_2^c\, \cos\beta\,,
\ea\right.
\ee
so that $\varphi_{\rm sm}$ appears as a SM-like doublet responsable for the electroweak breaking and  $\varphi_{\rm in}$
as an ``inert doublet'', with $<\!\varphi_{\rm sm}\!\!>\ =v/\sqrt 2,\ <\!\varphi_{\rm in}\!\!>\ =0\,$. 
Viewing for convenience $\beta$ as a fixed parameter unaffected by super\-symmetry-breaking terms, these  terms may be viewed as   providing a mass term for the ``inert'' doublet $ \varphi_{\rm in}$,
without modifying the vacuum state defined by $<\!h_1\!>$ and $\,<\!h_2\!>\, $.

 \vspace{2mm}
 One has, using (\ref{A}) and (\ref{wh}),
\be
\label{ma2deb}
\!\!\ba{ccl}
|\varphi_{\rm in}|^2 \!\!&=&\! |\, h_1 \sin\beta -\,h_2^c \cos\beta\,|^2 
\vspace{2mm}\\
\!&=&\!  
 |H^+|^2\! + \frac{1}{2}\,A^2 \!+ 
 \frac{1}{2} \,|\sqrt 2  \ \Re \,( h_1^0 \,\sin\beta -h_2^0 \,\cos \beta)\,|^2.
\ea
\ee
 Furthermore, if  these di\-mension-2 supersymmetry-break\-ing terms expressed as $\,m_A^2\,|\varphi_{\rm in}|^2\,$ were generated as in (\ref{lhhMSSM}) from a decoupling limit of the singlet $S$ we would have
\vspace{-7mm}

\be
m_A^2\,=\ 2\,\mu^2\,.
\ee
Or in a more general way, allowing for extra soft-breaking contributions for $h_1$ and $h_2$,
\be
m_A^2\,=\,2\,\mu^2+\,\Delta m^2(h_1)+\Delta m^2(h_2)\,\,.
\ee

\vspace{1mm}

The mass term for $\varphi_{\rm in}$ provides equal contributions to $m_A^2$ and $m_{H^\pm}^2$, and leads to a further mixing between the neutral scalars described by the real parts of $\varphi_1^0$ and $\varphi_2^0$.
It provides in particular,
in the large $\tan\beta$ limit for which $v_1$ is small,  a rather large mass$^2$ term for $h_1$ contributing to $m^2(H^\pm \!\simeq h_1^\pm)$, $m_A^2$ and to a small mixing between the neutral scalars. 

\vspace{2mm}
Specializing in the MSSM,
adding the supersymmetric ($m_W^2,$ $m_Z^2$) and supersymmetry-breaking contributions to the mass$^2$  matrices 
implies immediately,  in this specific model,

\vspace{-7mm}

\be
m_{H^\pm}^2\,=\,m_W^2\,+ m_A^2\,.
\ee
In the large $\tan\beta$ limit,  $\,h\simeq z \simeq \sqrt 2 \ \Re\ \varphi_2^0$ and $H\simeq \sqrt 2\ \Re\ \varphi_1^0\,$
have masses close to $m_Z$ and $m_A$, respectively.

\begin{table}[t]
\vspace{-2mm}
\caption{Minimal content of the Supersymmetric Standard Model (MSSM).
Neutral gauginos and higgsinos mix into a photino, two zinos and a higgsino, further mixed into four neutralinos.
Ordinary particles, including additional BEH bosons, in blue, have $R$-parity $+1$.
Their superpartners, in red, have $R$-parity $-1$. The N/nMSSM includes an extra singlet with a trilinear $\lambda \,H_2H_1S$ superpotential coupling, describing a singlino and two additional neutral spin-0 bosons. The USSM also includes an extra neutral gauge boson
$Z'$ (or $U$) and its associated gaugino.
\label{tab:mssm}}
\vspace{2mm}
\begin{center}
\begin{tabular}{|c|c|c|} 
 \hline  &&\\ [-2.5mm] 
 Spin 1       &Spin 1/2     &Spin 0 \\ [1.2mm]\hline  
\hline 
&&\\ [-1.8mm]
\color{blue}gluons       	 &\color{red}gluinos ~$\tilde{g}$        &\\[.8mm]
\color{blue}photon           &\color{red}photino ~$\tilde{\gamma}$   &\\ [.3mm]
--------------&$- - - - - - - - -$&-------------------------------- \\ [-3.2mm]
 

$\begin{array}{c}
\\ \color{blue} W^\pm\\ [1.1mm]\color{blue}Z \ \  \\ [.8mm]
\\ \\
\end{array}$

&$\begin{array}{c}
\color{red}\hbox {winos } \ \widetilde W_{1,2}^{\,\pm} \\ [.8mm]
\color{red}\,\hbox {zinos } \ \ \widetilde Z_{1,2} \\ [1.3mm]
\color{red}\hbox {higgsino } \ \tilde h
\end{array}$

&$\hspace{-12mm}\left. \begin{array}{c}
\color{blue}H^\pm\\ [1.3mm]
\color{blue} h \\
[1.3mm]
\color{blue}H, \,A
\end{array}\ \right\} 
\begin{array}{c}\!\! \hbox {BEH bosons}\\ 
\end{array}\hspace{-12mm}$  \\ &&\\ 
[-5.5mm]
\hline &&

\\ [-2.5mm]
&\color{blue}leptons ~$l$       &\color{red}sleptons  ~$\tilde l$ \\[.8mm]
&\color{blue}quarks ~$q$       &\color{red}squarks   ~$\tilde q$\\ [-2.2mm]&&
\\ \hline
\end{tabular}
\ec
\end{table}

\vspace{2mm}

The mass$^2$ matrix for the neutral scalar fields $h_1^0$ and $h_2^0$ may be written as the sum of two supersym\-metry-conserving and supersymmetry-breaking contributions. It follows from (\ref{z0}) involving $\,z= \sqrt 2\  \Re\ (-\,h_1^0\,\cos\beta +h_2^0\,\sin\beta)$,  and for the non-supersymmetric part (\ref{ma2deb}) involving $\,\sqrt 2  \ \Re$ $ ( h_1^0 \,\sin\beta -h_2^0 \,\cos \beta)\, $:
\be
\label{2mass}
{\cal M}_\circ^2\,=\,
\underbrace{\left(\ba{cc}
m_Z^2& 0 \vspace{1mm}\\
0 & 0
\ea\,\right)_{\!-\beta}}_{\hbox{\small SUSY-conserving}} \!+\ \,
\underbrace{\left(\,\ba{cc}
0 & 0 \vspace{1mm}\\
0 & m_A^2
\ea\right)_{\!\beta}}_{\hbox{\small SUSY-breaking}}.
\ee

\vspace{.5mm}
\noindent
The two basis denoted by  $_{-\beta}$ and $_\beta$ \,are rotated from the ($h_1^0,h_2^0$) basis by angles $-\beta$ and $\beta$, and are at angle $2\beta$.  
The mass of the lightest eigenstate increases  from $0$ to $m_Z$ when $\beta$ increases from $\pi/4$ to $\pi/2$ (or decreases from $\pi/4$ to 0), assuming $\,m_A>m_Z$.
The second derivative of $V$, in the SM-like direction orthogonal to the $m_A^2$ eigenstate of the second matrix, at angle $2\beta$ with the direction of $z$, receives only a contribution
$m_Z^2 \cos^2 2\beta$ from the first term. This  implies a mass eigenstate verifying
\be
\label{mlight0}
m_h\,\leq m_Z\,|\cos 2\beta|\, \ \ (\,+\, \hbox{\small radiative corr.} ) \, .
\ee

\vspace{1mm}

More precisely (\ref{2mass}) reads
\be
{\cal M}_\circ^2\,=\,\left(\ba{cc}
c_\beta^2\,m_Z^2+ s_\beta^2\,m_A^2 &-\,s_\beta c_\beta \, (m_Z^2+m_A^2)   \vspace{2mm}\\
- \,s_\beta c_\beta\,  (m_Z^2+m_A^2)   & s_\beta^2\,m_Z^2+ c_\beta^2\,m_A^2 
\ea\right)\,,
\ee
and has the eigenvalues
\be
\ba{ccl}
m^2_{H,h}\!\!&=&\!\! \hbox{\small $\dis\frac{m_Z^2+m_A^2}{2}$}\pm 
\sqrt {\hbox{ \footnotesize$\dis\left(\frac{m_Z^2+m_A^2}{2}\right)^2 $}  -  m_Z^2 m_A^2\,\cos^22\beta }.
\vspace{-1mm}\\
\ea
\ee
The smallest one verifies (\ref{mlight0}), approaching $m_Z$ in the large $\tan\beta$ limit for which the two mass eigenvalues get close to $m_Z$ and $m_A$ as seen from (\ref{2mass}). The lighter scalar $h$ becomes close to being the spin-0 partner of the $Z$, $h\simeq z\simeq \sqrt 2\ \Re \ h_2^0$, 
with a mass close to $m_Z$,
\vspace{-.4mm}
 the heavier one $H\simeq \sqrt 2\ \Re \ h_1^0$  having a mass close to $m_A$.

 \vspace{2mm}

These formulas, leading back to $m_{H^\pm}\!=m_W$ and $m_H=m_Z$ when the supersymmetry-breaking parameter $m_A^2$ vanishes, in agreement with (\ref{eqzZ},\ref{eqhw}), illustrate  the implications of gauge/BE-Higgs unification, even in a situation of broken supersymmetry.
Large radiative corrections, involving most notably  very heavy and/or strongly mixed stop quarks,
are then required in the MSSM to keep a chance to get  $m_h$ sufficiently above $m_Z$, in view of identifying it with the 125 GeV spin-0 boson.

\subsection{Spin-0 masses in the N/nMSSM}

\vspace{-2mm}
\bc
{\em with heavier spin-0 bosons thanks to the extra singlet}
\ec

The situation is much better in the N/nMSSM (or also in the USSM) thanks to the trilinear coupling  $\lambda$ in the superpotential leading to an
additional quartic term in the potential $\,\lambda^2 |h_2h_1|^2$, and to a steepest potential 
allowing for larger masses, already at the classical level.
\vspace{2mm}

Indeed starting from the $R$-symmetric nMSSM spectrum (\ref{boson})  with $m_A=\lambda v/\sqrt 2$, $m_{H^\pm} =m_W$ and $m_h=m_Z$, independently of $\beta$
\cite{R},  the sum  of supersymmetric and supersymmetry-breaking contributions leads  to the mass formulas 
\be
\label{nmssmwaA}
\!\left\{\,
\ba{ccl}
m_A^2\!&=&\!\hbox{\small$\dis \frac{\lambda^2 v^2 }{2}$} +\, \delta m_A^2\ ,\ \ 
\vspace{1mm}\\
m_{H^\pm}^2 \!&=&\, m_W^2+\,\delta m_A^2 \ =\, m_W^2+ \,m_A^2- \hbox{\small$\dis \frac{\lambda ^2 v^2}{2}$}\,.
\ea\right.
\ee
Neutral scalars are also expected to be heavier than in the MSSM. Their $2\times 2$ mass$^2$ submatrix, restricted to the 
$h_1^0,h_2^0$ subspace by ignoring  the singlet scalar, now reads
\be
\label{3mass}
{\cal M}_\circ^2\,=\,
\underbrace{\left(\!\ba{cc}
m_Z^2& \!0 \vspace{1mm}\\
0 & \!\hbox{\small$\dis \frac{\lambda^2 v^2 }{2}$}
\ea\!\right)_{\!-\beta}}_{\ba{c}\hbox{\small SUSY-conserving} \vspace{-.5mm}\\ \hbox{\footnotesize (nMSSM)}\ea}
 \,+\ \,
\underbrace{\left(\,\ba{cc}
0 & \ 0 \vspace{3mm}\\
0 & \ \delta m_A^2
\ea\right)_{\!\beta}}_{\ba{c}\vspace{-3mm}\\ \hbox{\small SUSY-breaking}\ \ea}.
\ee

\vspace{-2mm}
Its lightest mass eigenstate is immediately seen to be in the range
$(m_Z$, $\lambda v/\sqrt 2)$. 
\vspace{-.8mm}
For 
\be
\label{lambdalim}
\lambda\,\geq \,\hbox{\small$\dis \sqrt{\frac{g^2+g'^2}{2}}$}\,=\,\hbox{\small$\dis \frac{m_Z\,\sqrt 2}{v
}$}\,=\, 2^{3/4}\,G_F^{1/2}m_Z\, \simeq\, .52\,
\ee 
the two mass eigenstates are already both heavier than $m_Z$, independently of $\tan\beta$ and before taking into account super\-symmetry-breaking effects from $\delta m_A^2$.~\,This makes it much easier to reach 125 GeV, without having to rely on large radiative corrections 
from very heavy 
stop quarks, as one must do in the MSSM.

\vspace{2mm}

For $\beta=\pi/4\,$
 i.e.~$v_1=v_2$ the $2\times 2$ matrix (\ref{3mass}) has $ {\lambda^2 v^2 }/{2}$ and $m_Z^2+\delta m_A^2$ for eigenvalues,
its lightest mass eigenvalue being large if $ {\lambda^2 v^2 }/{2}$ and $\delta m_A^2$ are both large.
We then get 
\be
\left\{\ \ba{ccc}  m_z^2\!&=\!&m_Z^2 + \delta m_A^2\,,
\vspace{1mm}\\
m_{H^\pm}^2\!&=\!& m_W^2 + \delta m_A^2\,
\ea \right.
\ee
as obtained for example with gravity-induced supersym\-metry-breaking,
\vspace{-.6mm}
 for $\delta m_A^2=4 \,m_{3/2}^2$ \cite{cfg}. 
This illustrates how the gauge/BE-Higgs unification may manifest  on the mass spectrum.
For example taking here $m_z$ at 125 GeV would imply a $H^\pm$ around 117 GeV.
\,In the absence of a relatively light $H^\pm$, the above relations
would lead us to view, for \hbox{$\tan\beta =1$},  the $z$ as a heavier spin-0 BEH boson close in mass to the $H^\pm$, rather than the one found at 125 GeV.

\vspace{2mm}

More precisely the second derivative of $V$, in the SM-like direction orthogonal to the $m_A^2$ eigenstate of the second matrix, at angle $2\beta$ with the direction of $z$, receives a contribution only from the first matrix in (\ref{3mass}), 
\vspace{-.6mm}
and is thus now equal to
$\,m_Z^2 \cos^2 2\beta+ \frac{\lambda^2 v^2}{2} \sin^2 2\beta$. This   implies a neutral mass eigenstate verifying
\be
\label{mlight}
m_h^2\,\leq m_Z^2\,\cos^2 2\beta + \!\hbox{\small$\dis \frac{\lambda^2 v^2 }{2}$}  \sin^2 2\beta\,\ \ (\,+\, \hbox{\small radiative corr.} ) \, ,
\ee
This upper bound may also be obtained directly by noting that the neutral SM-like combination 
$\varphi_{\rm sm}^{\circ}\!=h_1^{\circ}\,\cos\beta + h_2^{\circ *}\,\sin\beta\,$
\,has its quartic coupling  $\lambda_{\rm sm} |\varphi_{\rm sm}^{\circ}|^4$ in the N/nMSSM potential (\ref{lhh}) or (\ref{lhhNMSSM}) fixed by
\be
\label{lambdasm}
\lambda_{\rm sm} =\, \hbox{\small$\dis\frac{g^2+g'^2}{8}$} \,\cos^2 2\beta+\hbox{\small$\dis\frac{\lambda^2}{4}$}\, \sin^22\beta\, .
\ee
Expanding $V$ as a function of $\sqrt 2\  \Re\ \varphi_{\rm sm}^{\circ}$ 
provides for this field 
\vspace{-.8mm}
the mass$^2$ parameter 
$
2\mu_{\rm sm}^2\!= 
2\lambda_{\rm sm} v^2\!=m_Z^2\cos^2 2\beta+\frac{\lambda^2v^2}{2} \sin^2 2\beta
$.
\vspace{-.8mm}
Neutral spin-0 
bosons cannot all be heavier, the lightest having mass
$\leq \mu_{\rm sm}\,\sqrt 2$ at most, leading to the mass bound  (\ref{mlight}).

\vspace{2mm}

Let us explicitate  for completeness the mass$^2$ matrix (\ref{3mass})  for the neutral scalar
$h_1^0,h_2^0$ subspace:
\be
\label{4mass}
\hbox{\small$
\left(\!\ba{cc}
c_\beta^2\,m_Z^2+ s_\beta^2\,(\frac{\lambda ^2 v^2}{2}+\delta m_A^2) &-\,s_\beta c_\beta \, (m_Z^2-\frac{\lambda ^2 v^2}{2}\!+\delta m_A^2)   \vspace{2mm}\\
- \,s_\beta c_\beta\,  (m_Z^2-\frac{\lambda ^2 v^2}{2}\!+\delta m_A^2)   & s_\beta^2\,m_Z^2+ c_\beta^2\,(\frac{\lambda ^2 v^2}{2}+\delta m_A^2)
\ea\!\right)$}\!,
\ee

\vspace{2mm}\noindent
with $m_A^2=\frac{\lambda ^2 v^2}{2}\!+\delta m_A^2$ as in (\ref{nmssmwaA}).
$m_Z^2$ and $\frac{\lambda ^2 v^2}{2}$ correspond to the supersymmetric contributions in the $R$-symmetric nMSSM spectrum (\ref{boson}) \cite{R}, $\delta m_A^2$ being the su\-per\-symmetry-breaking contribution  from the ``inert doublet'' 
$\varphi_{\tiny \rm in}$ mass term,  as in (\ref{ma2deb}).
This  $2\times 2$ submatrix should then be included within a $3\times 3$ matrix taking into account mixing effects with the singlet  $\sqrt 2\ \Re\, s$, 
\vspace{-.3mm}
involving in particular $\,\mu\,\lambda v_1/\sqrt 2$ and $\,\mu\,\lambda v_2/\sqrt 2$ as seen from (\ref{lhhNMSSM}).

\vspace{2mm}

A more instructive expression  of the matrix (\ref{3mass},\ref{4mass}) is obtained in the SUSY basis $_{-\beta}$, by rotating by $-2\beta$ the matrix for the supersymmetry-breaking contribution;  or by writing 
\vspace{.3mm}
 $\,-\,\Re \, \varphi_{\rm in}=  \Re\, $ $(-h_1^0 \sin\beta +\,h_2^0 \cos\beta) =\sin2\beta $
$\Re \, \underbrace{(- h_1^0 \cos\beta+ h_2^0\sin\beta)}_{z} +\,\cos2 \beta\  \Re\, ( h_1^0 \sin\beta+ h_2^0\cos\beta)\,$:

 \vspace*{-2mm}
 
\bea
\label{5mass}
{\cal M}_\circ^2\,=\,
\left(\,\ba{cc}
m_Z^2+\delta m_A^2\,\sin^2 2\beta & \ \delta m_A^2\,\sin 2\beta\, \cos 2\beta \vspace{3mm}\\
\delta m_A^2\,\sin 2\beta\, \cos 2\beta  & \ \hbox{\small$\dis \frac{\lambda^2 v^2 }{2}$}\!+ \delta m_A^2\,\cos^2 2\beta
\ea\right)_{\!-\beta}\!\!.
\nonumber 
\eea
\vspace{-9mm}

\be
\ee

 \vspace{-4mm}

\subsection{\boldmath Charginos and neutralinos mass matrices}

 \vspace{-2mm}

\bc
{\it as understood\vspace{1mm} from $R$ symmetry \hbox{and gauge/BE-Higgs unification}}
\ec

 The higgsino fields $\tilde h_{1L}$ and $\tilde h_{2L}$, described by $H_1$ and $H_2$ with $R=0$,
 \vspace{-.5mm}
 transform according to  (\ref{R2}) as
   $
   \tilde h_{iL}\,{\to}$ $ \,e^{-i\alpha}\ \tilde h_{iL}\,,\ \tilde h_{2L}\,{\to} \  e^{-i\alpha}\ \tilde h_{2L}
   \,$.
   The Dirac higgsino doublet $\psi$ constructed from  $ \tilde h_{1L}$ and $  (\tilde h_{2L})^c\,$ and the Majorana gauginos $\lambda$ transform chirally in opposite ways as in  (\ref{trh},\ref{dirac}). Gaugino mass terms $m_{1/2}$  (denoted by $m_3,\,m_2$ and $m_1$ for the gluinos and $SU(2)\times U(1)$ gauginos) violate the continuous $R$ symmetry,  as for a $\mu$ term, reducing it to $R$-parity \cite{grav,glu}.
   
   \vspace{2mm}
   Gaugino and higgsino fields can combine through 
   $R$-invariant non-diagonal mass terms generated  from the Yukawa couplings (\ref{LY}). 
   The resulting charginos and neutralinos appear at this stage,
   with $\mu=m_i=0$,  as {\it Dirac particles} carrying $R=\pm 1$ as expressed in (\ref{dirac}), leading to the $U(1)_R$ symmetric nMSSM mass spectrum (\ref{spec},\ref{boson}).
   
   \vspace{2mm}
   The Dirac zino with $R=+1$ of mass $m_Z$, in agreement with the as-yet-unbroken supersymmetry still present in the neutral sector, is obtained by combining 
   \be
   \label{zinozino}
   \left\{
 \ba{lcc}  
 \hbox{the \it gaugino} \ \ \ \ \ \, \lambda_Z= \,\lambda_{3} \,c_\theta -\lambda' \,s_\theta\,,
 \vspace{2mm}\\
 \hbox{the \it higgsino} \ \  - \tilde h_z= \,\tilde h_1^0\,c_\beta-\tilde h_2^0\,s_\beta\,.
 \ea \right.
   \ee
   The gaugino $\lambda_Z$ is directly associated with the $Z$, and the higgsino $\tilde h_z$ is described by the chiral superfield $H_z$ in (\ref{HZ}). The Dirac zino may be expressed  as in
 \cite {R} as
\be
\label{expzino}
\lambda_{Z L}+ (- \tilde h_z)_R \,
= \,(\lambda_{3} c_\theta -\lambda's_\theta )_L+ (\tilde h_1^0\,c_\beta-\tilde h_2^0\,s_\beta )_{R},
\ee
or reexpressed in terms of the two Majorana spinors in (\ref{zinozino}). 
The corresponding $2\times 2$ mass matrix in a gaugino-higgsino basis,
\be
\label{mzino0}
{\cal M}_{\rm zinos} = \left(\ba {cc}
\vspace{-5mm}\\
0 & \!\!m_Z
\vspace{0mm}\\
m_Z& \!\!0
\ea\! \right),
\ee 
may be further unpacked into a   $4\times 4$ matrix as below in (\ref{mzino}).

\vspace{2mm}
      Including the $\Delta R=\pm \,2 $
gaugino and higgsino mass terms $m_i$ and $\mu$  breaking explicitly $R$ symmetry, we get, 
with $\mu\,H_2H_1=\mu\,(H_2^+ H_1^- -H_2^0 H_1^0$),
the chargino and neutralino mass matrices in the MSSM,
   \vspace{-3mm}
   
\be
\label{mwinos1}
 \hbox{$\dis
{\cal M}_{\rm winos} \,= \left(\ba {cc}
m_2 \!\!& m_W\sqrt 2\, s_\beta
\vspace{1mm}\\
m_W\sqrt 2 \, c_\beta \!\!& \!\!\!\!\!\mu
\ea\! \right),
$}
\ee 
\vspace{-1mm}

 \noindent
 and
 
  \vspace*{-4mm}
  \be
  \label{mzino}
{\cal M}_{\rm inos} =
\hbox{$\dis
\left( \ba {cccc}
\ m_1 &\ 0 &      \!\!-s_\theta c_\beta \,m_Z & \ s_\theta s_\beta \,m_Z   
\vspace{1mm}\\
\ 0&\ m_2 &       \  c_\theta c_\beta \,m_Z & \!\!-c_\theta s_\beta \,m_Z
\vspace{1mm}\\
\!\!-s_\theta c_\beta  \,m_Z & \ c_\theta c_\beta  \,m_Z & \ \,0&-  \mu
\vspace{1mm}\\
\ s_\theta s_\beta \,m_Z&\!\!-c_\theta s_\beta  \,m_Z & - \mu & \ \,0
\ea \right)
$}
.
\ee 

\vspace{2mm}
The part proportional to $m_Z$ in the neutralino mass matrix is the supersymmetric contribution, in a way compatible with gauge/BE-Higgs unification, while the $\Delta R=\pm \,2$ part  involving $m_1,\,m_2$ and $\mu$
is the supersym\-metry-breaking part. We  recall that the parameter $\mu$, although initially ``supersymmetric'', still leads to super\-symmetry-breaking effects in the presence of $v_1$ and $v_2$.

\vspace{2mm}

Their eigenvalues verify relations such as 
\be
\label{swino}
m^2(\hbox{\it wino}_1)+ m^2(\hbox{\it wino}_2)\,=\, 2\,m_W^2+\mu^2+m_2^2\,,
\ee
and, similarly,
\be
\label{szino}
\hbox{\small $\dis \sum_{1..4}$} \ m^2(\hbox{\it neutralino})\,=\, 2\,m_Z^2+2\,\mu^2+m_1^2+m_2^2\,.
\ee

Without the gaugino masses $m_1,\,m_2$, and with the dimension-2 soft-breaking mass terms for $h_1$ and $h_2$ generated 
spontaneously as in (\ref{lhhMSSM}) so that $m_A^2\!=2\,\mu^2$, the average mass$^2$ for bosons and fermions would be the same in the
multiplets considered, with
\be
\left\{\ba{c}
\ba{ccl}
3\,m_W^2+m_{H^\pm}^2 \!&=&\! 4\,m_W^2+ m_A^2=4\,m_W^2+ 2\mu^2
\vspace{1.5mm}\\
&=&\!2\,[m^2(\hbox{\it wino}_1)+ m^2(\hbox{\it wino}_2)]
\ea
\vspace{2.5mm}\\
\!\!\ba{ccl}
3\,m_Z^2+m_{h}^2\!+m_H^2\!+m_A^2= 4\,m_Z^2+ 2m_A^2=4\,m_Z^2+ 4\mu^2
\vspace{1.5mm}\\
=\ 2\ \hbox{\footnotesize $\dis \sum_{1..4}$} \ m^2(\hbox{\it neutralino})\,.
\vspace{-2mm}
\ea
\ea \right.
\ee

\vspace{1mm}

\subsection{Neutralinos in the N/nMSSM}
\label{subsec:neut}

 \vspace{-2mm}

\bc
{\it as understood\vspace{1mm} from $R$ symmetry \hbox{and gauge/BE-Higgs unification}}
\ec

   The N/nMSSM introduces an additional neutral sin\-glino described by $S$.
      The chargino mass matrix (\ref{mwinos1}) is simply affected by the replacement $\mu\to\mu_{\rm eff} =\mu\,+$ $\lambda <\!s\!>\,$. 
   The neutralino mass matrix (\ref{mzino}) gets embedded into a $5\times 5$ one.
   It now includes $R$-conserving non-diagonal mass terms  corresponding to the nMSSM mass spectrum in (\ref{spec}), 
   with 
       \vspace{-.5mm}
    $-\,\frac{ \lambda v }{\sqrt 2}\,\sin\beta$  and $-\,\frac{ \lambda v }{\sqrt 2}\,\cos\beta$ 
    contributions obtained     from the 
    $R$-invariant $\lambda\, H_2H_1 S$ coupling 
    (\ref{mix}) mixing the doublet higgsinos $\tilde h_1$and $\tilde h_2$ with the singlino $\zeta$\,.
    $\tilde h_{1R}, \,\tilde h_{2R}$  and the singlino $\zeta_L$ all have  $R=1$, in agreement with the $R$ transformation properties
    \be
 \hbox{\small\it gauginos:}\  \lambda \stackrel{R}{\to} \,e^{\gamma_5\alpha}\,\lambda\,;\ \ \ \tilde h_i\,\stackrel{R}{\to} \,e^{-\gamma_5\alpha}\,\tilde h_i\,\ \ \ \zeta \,  \stackrel{R}{\to} \,  e^{\gamma_5\alpha}\,\zeta\,.
    \ee
 
   \vspace{2mm}
   
   The $R$-conserving part of the mass matrix corresponds to a conserved supersymmetry. It is a rank-4 $\,5\times 5$ matrix obtained by unpacking the two matrices
\be
\label{mzino2}
\left(\ba {cc}
\vspace{-5mm}\\
0 & \!m_Z
\vspace{1.5mm}\\
m_Z& \!0
\ea\! \right) \ \ \ \hbox{and}\ \ \ 
\left(\ba {cc}
\vspace{-5mm}\\
0 & \!\!-\frac{\lambda v}{\sqrt  2}
\vspace{0mm}\\
-\frac{\lambda v}{\sqrt  2}& \!\!0
\ea\! \right).
\ee 
 This provides as in (\ref{spec}) 5 neutralinos organized as a Dirac zino of mass $m_Z$, a Dirac neutralino mixing the singlino 
with the left-over higgsino, carrying $R=\pm1$,  and a massless chiral photino with $R=1$.
The massive spinors involve, as in (\ref{expzino}) \cite{R}
\be
\left\{
\ba{ccc}
\lambda_Z=\lambda_{3} c_\theta -\lambda's_\theta\! & \hbox{\small combined with}& \!\!
-\tilde h_z= \tilde h_1^0\,c_\beta-\tilde h_2^0\,s_\beta ,
\vspace{2mm}\\
\tilde s& \hbox{\small combined with}& \ \ \
  -\,(\tilde h_1^0\,s_\beta+\tilde h_2^0\,c_\beta)\,.
\ea\right.
\ee
The neutralino mass spectrum (\ref{spec}) for the $U(1)_R$-invari\-ant nMSSM \cite{R} is reexpressed into the $5\times 5$ neutralino mass matrix 
 \bea
  \label{mzino5r}
\hbox{ $\dis
\left( \ba {ccccc}
\ 0 &\ 0 &      \!\!\!-s_\theta c_\beta m_Z & \ s_\theta s_\beta m_Z   &   0
\vspace{1mm}\\
\ 0&\ 0 &       \,  c_\theta c_\beta m_Z & \!\!-c_\theta s_\beta m_Z  &    0
\vspace{1mm}\\
\!\!-s_\theta c_\beta  m_Z & \ c_\theta c_\beta  m_Z & \ 0&0   & -\frac{\lambda v}{\sqrt 2}s_\beta \vspace{1mm}\\
\ s_\theta s_\beta m_Z&\!\!-c_\theta s_\beta  m_Z &0& \ \,0       & -\frac{\lambda v}{\sqrt 2}c_\beta
\vspace{2mm}\\
0&0&-\frac{\lambda v}{\sqrt 2}s_\beta& -\frac{\lambda v}{\sqrt 2}c_\beta &0
\ea \right)
$} \ .\!\!\!\!\!\!\!
\nonumber 
\vspace{4mm}\\ 
\eea

\vspace{2mm}

One completes this $R$-symmetric  $5\times 5$ mass matrix by the $\Delta R=\pm \,2$ contributions,
re-introducing the $\mu$ and $\mu_S$ doublet and singlet mass parameters previously discarded from the nMSSM superpotential \cite{R} to get $R$ symmetry. If a translation on $S$ has to be performed,  $\mu$ and $\mu_S$ get modified  into 
$\mu_{\rm eff}=\mu+\lambda \! <\!s\!>$ and  $\mu_{S\rm\, eff }=\mu_S+2\kappa\!<\!s\!>$. The gaugino mass parameters
$m_1$ and $m_2$ may be generated by radiative corrections, or from gravity-induced supersymmetry breaking as e.g. in \cite{cfg}.

\vspace{2mm}

The resulting $5\times 5$ neutralino mass matrix reads
\vspace{2mm}
 \bea
  \label{mzino5}
\hbox{ $\dis
\left( \ba {ccccc}
\ m_1 &\ 0 &      \!\!\!-s_\theta c_\beta m_Z & \ s_\theta s_\beta m_Z   &   0
\vspace{1mm}\\
\ 0&\ m_2 &       \,  c_\theta c_\beta m_Z & \!\!-c_\theta s_\beta m_Z  &    0
\vspace{1mm}\\
\!\!-s_\theta c_\beta  m_Z & \ c_\theta c_\beta  m_Z & \ 0&-  \mu_{\rm eff}     & -\frac{\lambda v}{\sqrt 2}s_\beta \vspace{1mm}\\
\ s_\theta s_\beta m_Z&\!\!-c_\theta s_\beta  m_Z & \!- \mu_{\rm eff} & \ \,0       & -\frac{\lambda v}{\sqrt 2}c_\beta
\vspace{2mm}\\
0&0&-\frac{\lambda v}{\sqrt 2}s_\beta& -\frac{\lambda v}{\sqrt 2}c_\beta &\mu_{S\rm\, eff }
\ea \right)
$} \ .\!\!\!\!\!\!\!
\nonumber 
\vspace{4mm}\\ 
\eea 
 It involves 12 non-diagonal terms $\propto m_Z$ and 
$\lambda v/\sqrt 2$, originating from the supersymmetric and $U(1)_R$-invariant nMSSM mass spectrum
(\ref{spec},\ref{mzino5r}), next to the 5 additional $\,\Delta R=\pm\, 2\,$ gaugino and doublet + singlet higgsino mass terms.
The chargino and neutralino masses  now verify
\be
m^2(\hbox{\it wino}_1)+ m^2(\hbox{\it wino}_2)\,=\, 2\,m_W^2+\mu_{\rm\, eff }^2+m_2^2\,,
\ee
and
\be
\hbox{\small$\dis\sum_{1...5}$} \, m^2(\hbox{\it ino})=\,2\left(m_Z^2+\hbox{\small$\dis \frac{\lambda^2v^2}{2}$}+\mu_{\rm\, eff }^2\right)+\,m_1^2+m_2^2
+\mu_{S\rm\, eff }^2\,,
\ee
reducing to (\ref{swino},\,\ref{szino}) for $\lambda=0$.

\subsection{Squarks, sleptons and supersymmetry breaking}

Left-handed quark and lepton fields are described by the left-handed chiral
quark and lepton doublet superfields, $Q$ and $L$.~\,Right-handed ones, viewed as the conjugates of antiquark and antilepton fields, are described by the left-handed singlet superfields $\bar U,\,\bar D$ and $\bar E$.~\,$H_1$ and $H_2$ generate charged-lep\-ton and down-quark masses, and up-quark masses,  
 from the trilinear superpotential couplings  ${\cal W}_{lq}$  of lepton and quark superfields to  $H_1$ and $H_2$ \cite{ssm}, 
   \be
\label{wlq}
{\cal W}_{lq}\,=\, \lambda_e \, H_1 \,\bar E L \, +\, 
\lambda_d \, H_1\,\bar D Q \,-\, 
\lambda_u \, H_2 \,\bar U Q\,.
\ee
$H_1$ and $H_2$  are separately responsible, from
$ <\!h_1^0\!>\ =$ $ {v_1}/{\sqrt 2}$, $ <\!h_2^0\!>\ = \,{v_2}/{\sqrt 2}$,
 for charged-lepton and down-quark masses, and up-quark masses, respectively, with
 
 \vspace{-3mm}
\be
m_e=\, \frac{\lambda_e v_1}{\sqrt 2}\,,\ \ \
m_d=\,\frac{\lambda_dv_1}{\sqrt 2}\,, \ \ \
m_u=\,\frac{\lambda_uv_2}{\sqrt 2}\,.
\ee
This tends to favor a smaller $v_1$ as compared to $v_2$
i.e.~a large $\tan\beta=v_2/v_1$, in view of the large mass of the $t$ quark as compared to the $b$.

\vspace{2mm}

The superpotential interactions resulting from (\ref{wlq}) are  invariant under the continuous $U(1)_R$ symmetry (\ref{R1},\ref{R2}),
with $R=+1$ for left-handed (anti)quark and (anti)lepton superfields,
so that leptons and quarks carry $R=0$ and sleptons ans squarks $+1$ (for $\tilde l_L,\tilde q_L$) or $-1$  (for $\tilde l_R,\tilde q_R)$.
Gauge and chiral superfields transform under $R$ according to (\ref{r}) so that, altogether
\be
\label{r1}
\ba{c}
\left\{
\ba{ccc}
V_a & \stackrel{R}{\to} &V_a(x,\theta e^{-i\alpha},\bar\theta e^{i\alpha})\,,
\vspace{1mm}\\
(L,Q;\bar E,\bar D,\bar U)& \stackrel{R}{\to} &\ e^{i\alpha}\,(L, Q;\bar E,\bar D,\bar U)(x, \theta e^{-i\alpha})\,,
\vspace{1mm}\\
H_{1,2}& \stackrel{R}{\to} &    H_{1,2}(x, \theta e^{-i\alpha})\,,
\vspace{1mm}\\
S& \stackrel{R}{\to} & e^{2i\alpha}\ S(x, \theta e^{-i\alpha})\,,
\ea\right.
\vspace{-4mm}\\
\ea
\ee

\vspace{4mm}

\noindent
the superpotential ${\cal W}$, including ${\cal W}_{lq}$, having $R=2\,$.

\vspace{2mm}
A mixing between left-handed doublet and right-handed singlet squarks, e.g.~$\tilde t_L$ and $\tilde t_R$, 
with $R=+1$ and $-1$ respectively, can be generated by the $\Delta R=\pm \,2$ terms in the Lagrangian density.
It leads for the $\tilde t$  squarks to a non-diagonal term  $A_t-\mu m_t \,\cot\beta$, combining the contributions  from the dimension-3 soft-breaking terms involving $h_2$ \,with those from $F$ terms.

\vspace{2mm}

One essential question is the way by which squarks and sleptons can acquire very large masses.
 A spontaneous breaking of the global supersymmetry generates a massless chiral Goldstone spinor, carrying $R=1$ as seen in (\ref{gold}). It may describe the photino, or more generally a neutral gaugino possibly involving an extra-$U(1)$ gaugino, combined with a neutral spin-$\frac{1}{2}$ field $\zeta $  described by a $R=2$  chiral superfield \cite{ssm}.
 
  \vspace{2mm}
  
An extension of the gauge group to include an extra $U(1)$ factor is necessary if one intends  to generate large masses for all squarks and sleptons, at the tree approximation, in a globally supersymmetric theory.  Indeed with $SU(3)\times SU(2)\times U(1)$ or $SU(5)$ as the gauge group the squarks of the first generation would verify, at the classical level \cite{fayet79},
 \be
\hbox{\footnotesize$\dis\sum_{1,2}$} \ \,m^2(\tilde u_i)+ m^2(\tilde d_i)\,=\,2\,(m_u^2 + m_d^2)\,.
 \ee
One at least should then have  a very small or even negative mass$^2$,  leading to a charge and color-breaking vacuum. 
This 
 leads either to consider an extension of the gauge group to include an extra-$U(1)$ factor, or to generate large masses from radiative corrections
as in the ``gauge-mediated'' supersymmetry-breaking  (GMSB) models, or to go to local supersymmetry. Or, finally, to move to extra dimensions to generate the required breaking of the supersymmetry using {\it discrete boundary conditions involving $R$-parity}, as discussed in Section \ref{sec:dim}.

 \vspace{2mm}
 
Making all squarks and sleptons heavy  was first done through the v.e.v.~of the $D$ component from an extra $U(1)$
with non-vanishing axial couplings to quarks and leptons, as in the USSM briefly introduced in subsection \ref{subsec:ussm}.
This may be done by taking advantage of the $U$ transformations  (\ref{u0bis},\ref{us},\ref{u0ter},\ref{fax}), under which

\vspace{-3mm}
\be
\label{ulq}
\ba{c}
\left\{
\ba{ccc}
(L,Q;\bar E,\bar D,\bar U)& \stackrel{U}{\to} &\ e^{-\frac{i\alpha}{2}}\, (L, Q;\bar E,\bar D,\bar U)\,,
\vspace{1mm}\\
H_{1,2}& \stackrel{U}{\to} &  e^{i\alpha} \  H_{1,2}\,,
\vspace{1mm}\\
S& \stackrel{U}{\to} & e^{-2i\alpha}\ S\,.
\ea\right.
\ea
\ee

\vspace{1mm}
\noindent
This is in particular  a symmetry of the trilinear superpotential ${\cal W}_{lq}$ responsable for quarks and lepton masses \cite{ssm}.
It acts axially on quarks and leptons, the corresponding (axionlike) Goldstone boson ($a$) getting eliminated as the extra neutral boson
$Z'$ (also called $U$) acquires a mass. This still leaves us with the question of how to generate a mass for gluinos, possibly from radiative corrections \cite{glu}, or from supergravity as anticipated in \cite{grav}, since supergravity requires abandoning the continuous $R$ symmetry in favor of the discrete $R$-parity.

 \vspace{2mm}

 Supersymmetry breaking is now usually realized by generating soft supersymmetry-breaking terms \cite{martin,gg} from radiative corrections, or supergravity. 
In  the first case, gauge-mediated models are generally caracterized by the possibility of  a light or very light gravitino LSP,  behaving very much like a goldstino \cite{grav}. 
 \vspace{2mm}
 
When the local supersymmetry is spontaneously broken, it generates a massive spin-$\frac{3}{2}$ Majorana gravitino. Its mass term $m_{3/2}$ breaks explicitly the continuous $R$ symmetry ($U(1)_R$), reducing it to $R$-parity. This  reallows, in the supergravity framework, direct gaugino mass parameters, 
whose mass scale may be naturally fixed  from the gravitino mass $m_{3/2}$.
Soft supersymmetry-breaking terms may then be generated  from supergravity \cite{cremmer}, leading to  gravity-induced supersymmetry-breaking models~\cite{grbr1,cfg,grbr2,grbr3,grbr4}, where the gravitino is generally taken to be heavy.

 \vspace{2mm}
 
 In all these cases (MSSM, N/nMSSM, USSM, ...)
supersymmetry-breaking and $R$-symmetry breaking contributions are added to the Lagrangian density. This includes the reintroduction of the 
``supersymmetric'' parameter $\mu$ (with $\Delta R=\pm\,2$ and also breaking the extra-$U(1)$ symmetry), possibly regenerated from a translation of $S$, and the inclusion of the $\Delta R=\!\pm \,2$ gaugino mass parameters, and of other terms of dimension $\leq 3$ breaking  supersymmetry explicitly but softly.

 \vspace{2mm}
This would be a natural place to stop this presentation of the Supersymmetric Standard Model, hoping  for supersymmetric particles to be discovered soon at LHC. 

 \vspace{2mm}
Still there may be more, and this is likely  not to be the end of the story. More symmetries may be jointly operating to provide a better understanding of the electroweak and grand-unification breakings, opening the way to new compact dimensions of space-time, next to the quantum anticommuting dimensions of supersymmetry.

\vspace{1mm}

\section{Beyond  the N/\lowercase {n}MSSM}

\vspace{1mm}

\subsection{\boldmath Towards a  $\,N=2\,$ supersymmetric spectrum}
\label{subsec:towards}

 Let us return to the $R$-symmetric  superpotential (\ref{wnmssm}), ${\cal W}_{\rm nMSSM}\,=\,S\,(\lambda \,H_2H_1+\sigma)$ \cite{R} and 
 resulting spectrum (\ref{spec},\ref{boson},\ref{mzino2},\ref{mzino5r}).
For $\lambda$ equal to the limiting value (\ref{lambdalim}),
\be
\label{lambdag1}
\lambda \,=\,\hbox{\small$\dis \sqrt {\,\frac{g^2+g'^2}{2}}$}\,\simeq\,.52\,
\ee
(up to a possible convention-dependent sign for a real coupling), 
so that

\vspace{-5.5mm}
\be
\label{mzlg}
m_Z\,=\, \frac{\sqrt {\,{g^2+g'^2}}\ v}{2}\, = \,\frac{\lambda v}{\sqrt 2}\,,
\ee
 the theory has in its $Z$ sector, with the superpotential
 \be
 \label{w0}
 {\cal W}_\circ\,=\,-\
 \hbox{\small$\dis \sqrt{\,\frac{g^2+g'^2}{2}}$}
 \ H_2^0H_1^0\,S +\sigma S\,,
 \ee
 an unbroken $N=2$ supersymmetry, independently of the value of the mixing angle $\beta$ \cite{hyper,invi}. 
 Its effects  may be observed in the two $2\times 2$ neutralino mass matrices  (\ref{mzino2}), which participate equally in the $5\times 5$ neutralino mass matrix  (\ref{mzino5r}). The $Z$ gets associated with 2 Dirac zinos (or 4 Majorana ones) and 5 neutral spin-0 bosons within a massive gauge multiplet of $N=2$ supersymmetry, according to
\be
\label{Z2}
\framebox [8.5cm]{\rule[-.88cm]{0cm}{1.85cm} $ \dis
\ba{l}
\hbox{{\it nMSSM} \small  \ \,with}\  \ \lambda =\hbox{\footnotesize$\dis \sqrt {\frac{g^2+g'^2}{2}}$}\ \ \ \Rightarrow
\vspace{.5mm}\\
Z\,\stackrel{\ba{c}\hbox{\footnotesize \it SUSY}\ea}{\Longleftrightarrow}\, \hbox{4 Majorana zinos} \,
\stackrel{\ba{c}\hbox{\footnotesize \it SUSY}\ea}{\Longleftrightarrow}\, \hbox{5 spin-0 bosons}.
\ea
$}
\ee

\vspace{1mm}\noindent
This is valid as long as supersymmetry remains unbroken in this sector.
This massive $U(1)_R$ symmetric nMSSM spectrum even presents, for the $W^\pm$ and $Z$ multiplets,  an effective \hbox{$N=2$} supersymmetry, 
with in this case the electric charge acting as a central charge  \cite{Z}, this $\,N\!=2$ supersymmetry being broken in the $W^\pm$ multiplet 
for $\tan\beta\neq 1$.

\vspace{3mm}

 The $N=2$ supersymmetry may be extended to 
gluons and gluinos, if the latter are turned into Dirac particles, and accompanied by a complex octet of \hbox{spin-0} ``sgluons'' \cite{glu}.
 If we want to pursue in this direction of $N=2$ supersymmetry we must introduce, next to the singlet $S$, adjoint 
$SU(2)$ and $SU(3)$ chiral superfields {$T$} and {$O$} with trilinear super-Yukawa couplings fixed 
by the gauge couplings, 

\vspace{-6mm}
\be
\label{lambdag2}
 \lambda_i= g_i\, \sqrt 2\,,\ \ \ \lambda'=\frac{g'}{\sqrt 2}\,.
\ee
The electroweak couplings of $H_1$ and $H_2$ to the singlet $S$ and triplet {$T$} 
are given by the superpotential (\ref{fbr0}) already encoutered for the $F$-breaking of the supersymmetry \cite{for,hyper},

\vspace{-5mm}
\be
\label{fbr}
{\cal W}\,=\, \hbox{\small$\dis \frac{1}{\sqrt 2}$}\ H_2\, (g \,\tau .T -g'S)\,H_1+\sigma S\,.
\ee

This $N=2$ superpotential  includes in particular, precisely,  the nMSSM-type one 
\be
 {\cal W}_\circ\,=\, -\ \hbox{\small$\dis \sqrt{\frac{g^2+g'^2}{2}}$}\ H_2^0H_1^0\,S_Z +\sigma S\,
\ee
for the chiral superfield $S_Z=\cos\theta \,T_3-\sin\theta \,S$ associated with the $Z$,
as written earlier in (\ref{w0}).
This provides an $N=2$ interpretation for the mass degeneracy occurring at the classical level, for unbroken supersymmetry, 
between the 5 neutral spin-0 bosons of the nMSSM, all of mass $m_Z$ for $\lambda=\sqrt{ (g^2 + g'^2)/2}\,$
\cite{invi}.

\vspace{3mm}

One must also consider 4 doublet chiral superfields 
(or $SU(5)$ quintuplets) instead of 2, by introducing $H'_1$ and $H'_2$ next to $H_1$ and $H_2$  to provide the required degrees of freedom for constructing 4 Dirac winos, so that
\be
\label{W2}
W^\pm\! \stackrel{\ba{c}\hbox{\footnotesize \it SUSY}\ea}{\Longleftrightarrow}\! \hbox{4 Dirac winos}  \!
\stackrel{\ba{c}\hbox{\footnotesize \it SUSY}\ea}{\Longleftrightarrow}\! \hbox{5 charged spin-0 bosons}.
\ee

\vspace{2mm}
\noindent
 The $W^\pm$ and $Z$ are then associated 
with 5 charged and 5 neutral spin-0 bosons, all of masses $m_W$ and $m_Z$ as long as the $N=2$ supersymmetry remains unbroken.  This ultimately  provides the spectrum for the gauge-and-BE-Higgs sector of \hbox{$\,N=2$} supersymmetric grand-unified theories \cite{2guts}, which may then be formulated in a 5 or 6-dimensional space-time \cite{gutbis}.

 \subsection{Radiative \vspace{1mm}gluino masses from {messenger quarks}}
 \vspace{-1mm} 
 
Gluinos being Majorana particles transforming chirally as in (\ref{R2}), a  continuous $R$ symmetry would forbid gluino masses, except if gluinos are turned into Dirac particles, as would be the case, precisely, within a $N=2$ theory as we discussed.
 
  \vspace{2mm}
  But let us return to $N=1$ for a moment. Gluinos are massless at lowest order, within global supersymmetry. 
  Still if one abandons the continuous $R$ symmetry one may consider generating {\it radiatively\,} gluino masses from their couplings to a new set of massive {\it messenger quarks\,} described by the chiral superfields $\cal Q$  and $\bar{\cal Q}\,$, {\it vectorially coupled\,} to standard model particles and sensitive to the source of supersymmetry-breaking,
  for messenger squarks and quarks to have different masses \cite{glu}.
  One also needs to introduce a source of breaking for the continuous $R$-symmetry, otherwise gluinos would still remain massless.
  
  \vspace{2mm}
This requires some care especially if we intend to appeal to the $F$-mechanism for breaking spontaneously supersymmetry \cite{for,for2}, as the presence of an $R$ symmetry is needed for a generic breaking of the supersymmetry as discussed in subsection \ref{subsec:role}. At the same time however,  $R$ symmetry  must be broken to get gluino masses, then leading to a massless {\it $R$-Goldstone-boson} or light {\it $R$-axion}. This may lead to prefer generating the spontaneous breaking of the supersymmetry in the messenger sector through the extra-$U(1)$ gauge interactions of the messenger (s)quarks  with a non-vanishing $<\!D\!>$, \,as done in  \cite{glu}.

 \vspace{2mm}
Let us consider a second octet of para\-gluinos $\zeta_a$, described by a chiral octet superfield $O$ with $R=0$.
 \,One may then generate radiatively a gluino mass in a $R$-symmetric way,  with the two Majorana octets transforming in opposite ways according to
  \be
  \label{rglu}
  \left\{
  \ba{ccc}
    \hbox{\it gluinos} \ \  \lambda_a & \stackrel{R}{\to} & \ e^{\gamma_5 \alpha}\ \,\lambda_a\,,
    \vspace{1mm}\\
  \hbox{\it paragluinos} \ \  \zeta_a\ &  \stackrel{R}{\to} & \ e^{-\gamma_5 \alpha}\ \,\zeta_a\,.
   \ea \right.
   \ee
   These two octets  could originate, together with their associated \hbox{spin-0} gluons,  now often  called {\it sgluons}, from an underlying $N=2$ supersymmetry \cite{hyper,Z}.
   They are described by a chiral octet superfield {$O$} with $R=0$ (rather than 2, so as to lead to (\ref{rglu})), coupled to
   {\it massive messenger quark superfields} ${\cal Q}$ and $\bar {\cal Q}$ with $R=1$, themselves vectorially coupled to standard model particles.  
    
    \vspace{2mm}
      These messenger quark superfields interact with the octet {$O$} through the  $R=2$ superpotential \cite{glu}
 \be
 {\cal W}_{\rm mess.} = m_{\cal Q}\,\bar {\cal Q}{\cal Q}+\,
 \lambda_O\,  \bar {\cal Q}\,O {\cal Q}\,,
 \ee
 color indices being omitted for simplicity.
  This generates a $R$-conserving Dirac mass term ($m_D$) for the Dirac gluino octet with $R=1$,
  \be
    \hbox{\it Dirac gluinos} \ = \lambda_{aL} + \zeta_{aR}\,,
  \ee
 through one-loop diagrams involving the massive messenger quarks and squarks, sensitive to the source of spontaneous supersymmetry
breaking, e.g.~through an extra $U(1)$.

 \vspace{2.5mm}
 
We still have to pay attention to  the two additional real octets of spin-0 gluons described by {$O$} (ÒsgluonsÓ), as one tends to acquire a negative mass$^2$ from quantum corrections.
This instability may be avoided by introducing some amount of $R$-symmetry breaking, to locally stabilize the vacuum through 
\vspace{-.1mm}
a superpotential Majorana mass term $\frac{1}{2}\,\mu_{\hbox{\tiny$O$}}\,O^2$ for the second gluino octet
$\zeta$. 
The resulting explicit breaking of $R$ symmetry may be, again, a reason to prefer breaking spontaneously supersymmetry in the messenger sector through $D$ terms, rather than through the $F$-breaking mechanism making use of $R$ symmetry.

 \vspace{2mm}
 
This leads for gluinos to both Dirac and Majorana mass terms, with a {\it see-saw\,} type  $2 \times 2$ mass matrix \cite{glu}
\be
{\cal M}_{\rm  gluinos}= \left( \ba{cc}0 &m_D\vspace{1mm}\\
m_D&\mu_{\hbox{\scriptsize$\,O$}} \ea \right)\,,
\ee
a mechanism introduced for gluinos even before it started getting widely considered for neutrinos.
To be general a direct Majorana mass term $m_3$ for ordinary gluinos could still be added, 
although the purpose of this study was to discuss how an effective $m_3$ could be generated radiatively from the above see-saw type mass matrix.

 \vspace{2mm}

This still leaves us, however, with a vacuum state that is locally-stable but only {\it metastable\,}, with a lower energy vacuum state for which color would be spontaneously broken \cite{glu}. Fortunately this metastable vacuum is in practice effectively stable. The possible interest of such metastable vacuum states, that had escaped attention at the time, was  brought back to consideration more recently \cite{iss}.
  To generate radiatively in this way a significant mass for the gluinos, which must now be $\sim$ TeV scale at least as the result of LHC experiments \cite{susyat,susycms},  it is necessary  to consider quite high values for the messenger quark masses.
  
 \vspace{2mm}
   One may also imagine gauging the $R$ symmetry, eliminating the corresponding Goldstone boson if $R$ is spontaneously broken. This would lead to {\it a new force acting only on supersymmetric particles} (still to be discovered), and therefore, presumably, on dark matter. This new force may even be long-ranged if $R$ symmetry stayed unbroken, otherwise it would have a finite range $\hbar/mc$  where $m$ is the mass of the corresponding gauge boson associated with $R$ symmetry.

    \vspace{1mm}

\section{\boldmath $N=2$ \vspace{1.5mm} supersymmetric \hbox{grand-unified} theories}

    \vspace{1mm}
    
\subsection{Moving to higher dimensions}
 
We saw that a $N\!=\!2$ supersymmetric theory with a massless matter hypermultiplet in the adjoint representation  provides the $N=4$ supersymmetric Yang-Mills theory, 
the adjoint gauge superfield interacting with 3 adjoint chiral ones 
 coupled through the trilinear superpotential (\ref{sss})
${\cal W}=g\,\sqrt 2\, f_{ijk}\, S_{1}^i\, S_{2}^j \,S_{3}^k$,
describing 1 spin-1  + 4  spin-$\frac{1}{2}$ + 6 spin-0 adjoint  fields, with a $SU(4)_R\sim O(6)$ group acting on the 4 supersymmetry generators
\cite{hyper,Z}.
These theories, also obtained from the low-energy region of the dual spinor model \cite{n4bis}, or from  the dimensional reduction of a supersymmetric Yang-Mills theory in 10 dimensions \cite{n4ter}, are remarkably elegant but very constrained, and more difficult to apply to fundamental particles and interactions. 
Still the extra dimensions of space-time may well be at the origin of the breaking of both the supersymmetry and the grand-unification symmetry, as we shall see.
 
  \vspace{2mm}
 
 Starting again from $N=1$ supersymmetry in 4 dimensions,
 with  gaugino and higgsino mass terms related by $m_1=m_2=m_3=-\mu$,
 possibly also equal to the gravitino mass $m_{3/2}$, and assuming $\tan\beta=1$ for simplicity,
 we get, directly or from (\ref{mwinos1},\ref{mzino}), remarkable mass relations like  \cite{mwinos}
\be
\label{winos}
 \left\{\
 \ba{ccc}
 m^2(\hbox{winos}) \!&=&\!m_W^2+m_{3/2}^2\,,
 \vspace{.5mm}\\
 m^2(\hbox{zinos})  \!&=&\!m_Z^2 +m_{3/2}^2\,,
  \vspace{.5mm}\\
  m(\hbox{photino})  \!&=&\!  m(\hbox{gluinos})\,=\,m_{3/2}\,,
 \ea\right.
 \ee
 up to radiative corrections.
 
 \vspace{2mm}
 
These formulas, obtained in 4 dimensions, already point to a higher-dimensional 
  \vspace{-1mm}
 origin of the $m_{3/2}^2$  
 contributions to the 4d mass$^2$, 
 with supersymmetric particles carrying momenta $\pm\, m_{3/2}$ along an extra compact dimension.
 This leads us to consider again theories with a $N=2$ extended supersymmetry applied to electroweak and strong interactions,
or with a grand-unifica\-tion  symmetry like $SU(5), \ O(10)$ or $E(6)$, before moving to higher dimensions \cite{2guts,epjc}.

\subsection{\boldmath \hspace{-1mm}Are the $N\!=2$ \,``central charges'' really central\,?}
  \vspace{-1mm}

Within $N=2$ supersymmetry the particles get again organized within massless or massive multiplets, the $W^{\pm}$ and $Z$, $X^{\pm 4/3}$ and $Y^{\pm1/3}$ gauge bosons belonging to different kinds of massive gauge multiplets. 

 \vspace{2mm}

To start with, $N=2$ theories involve a new sort of massive multiplet known as an ``hypermultiplet'', 
describing massive spin-$\frac{1}{2}$  and  spin-0 particles, with the two spin-0 fields transforming as the two components of a $SU(2)_R$ isodoublet while the Dirac spinor is an isosinglet \cite{hyper}. This massive multiplet with maximum spin-$\frac{1}{2}$
looks at first intriguing as in principle it should not exist, not being a representation of the $N=2$ supersymmetry algebra $\, \{\,Q^i\,,\,\bar Q^j\,\}=\,-\,2\,P\!\!\!\!/\ \delta^{\,ij}$.  

 \vspace{2mm}

The theory being nevertheless  invariant under the two supersymmetry generators $Q_1$ and $Q_2$, and thus under $\,\{Q^1,\,\bar Q^2\}$,
 the above $N=2$ algebra must be modified, to allow for additional bosonic symmetry generators within the expression of  
 $\,\{Q^1,\,\bar Q^2\}$. How it gets modified is quite interesting, especially in view of the spontaneous breakings of the grand-unification symmetry, in a way allowing for the electroweak breaking to occur.
 
 \vspace{2mm}

We are led to consider, in a way compatible with Lorentz symmetry (and the symmetry property of the anticommutators)  the extended algebra
 \be
 \label{2susy}
 \{\,Q^i\,,\,\bar Q^j\,\}\,=\,-\,2\,P\!\!\!\!/\  \,\delta^{ij}\,+\,2\,\epsilon^{ij}\,(Z+\gamma_5 Z')\,,
 \ee
 where one should still identify correctly the two symmetry generators $Z$ and $Z'$. They are often referred to as ``central charges'', meaning that they ought to commute with all other symmetry generators of the theory.
 
 \vspace{2mm} 
 
 Let us consider, however, a $R$-symmetric theory under which $Q^1$ and $Q^2$ transform chirally according to (\ref{R}),
 \be
\label{R1bis}
Q^i\ \stackrel{R}{\to}  \ e^{-\gamma_5 \alpha}\,Q^i\,,
\ee 
or equivalently $\,Q^i_R \to e^{i\alpha}\,Q^i_R$. The operators $Z$ and $Z'$ appearing in  (\ref{2susy}) get rotated  according to 
\be
\label{rzz}
Z-iZ'\ \stackrel{R}{\to}  \ e^{2i \alpha}\,(Z-iZ')\,,
\ee
so that \cite{2guts}
\be
\label{rzz2}
[\,R,\,Z\,]=-\,2i\,Z'\,,\ \ [\,R,\,Z'\,]=\,2i\,Z\,.
\ee
These operators $Z$ and $Z'$, although commonly referred to as ``central charges'', {\it do not belong to the center of the symmetry algebra\,!}
 
\vspace{2mm}

This may seem surprising in view of the study of all  possible supersymmetries of the $S$ matrix, according to which $Z$ and $Z'$ must commute with all symmetry generators \cite{hls}.
 This analysis, however, disregards massless particles and symmetry breaking. It is thus not directly applicable here, both massless particles and symmetry-breaking effects playing an essential role. The $Z$ and $Z'$ generated from the anticommutation relations  (\ref{2susy})  do not necessarily commute with all symmetry generators, as seen in (\ref{rzz},\ref{rzz2}) for $R$ symmetry \cite{2guts}.  
 $Z$ and $Z'$  do not in general commute between themselves nor with gauge symmetry generators, in the non-abelian case, and as such do not qualify as ``central charges''.

  \vspace{2mm}
  
Indeed in a gauge theory the anticommutators of the two supersymmetry generators may be expressed as in (\ref{2susy}) but only {\it up to} (non-abelian or abelian) {\it field-dependent gauge transformations} (and modulo field equations of motion). For a $N=2$ supersymmetric Yang-Mills theory equation (\ref{2susy}) reads \cite{Z}

\vspace{-4mm}
\be
\{\,Q^1_R,\, \overline {Q^2_L}\ \}\ =\ 2\ \hbox{\small$\dis\frac{1+i\gamma_5}{2}$}\ 
\underbrace{g\,T_i\,(a_i-ib_i)}_{\hbox{\small$Z-iZ'$}}\,,
\ee

\vspace{-2mm}
\noindent
 $a_i$ and $b_i$ being the spin-0 partners of the adjoint or singlet gauge fields $V^\mu_{\,i}$, so that
 \be
 Z= g\,T_i \,a_i\,,\ \ Z'= g\,T_i \,b_i\,.
 \ee
 The $T_i$ denote the generators of the gauge symmetry group.
$a_i$ and $b_i$, described by adjoint or singlet chiral superfields with $R=2$ (as for the nMSSM singlet $S$ in (\ref{oldrbis})), transform under $R$ according to
\be
 \label{rab}
a_i-ib_i\ \stackrel{R}{\to}  \ e^{2i \alpha}\,(a_i-ib_i)\,,
\ee
so that $Z$ and $Z'$ do actually transform according to (\ref{rzz}).
\vspace{2mm}

The spin-0 adjoints $a_i$ and $b_i$ will soon be interpreted as originating from the 5th and 6th components
$V^5\!=a$ and $V^6\!=b$  of the 6d gauge fields $V^{\hat \mu}$. The $R$ transformation  (\ref{rab}) gets then associated with a 6d rotation ${\cal R}_{56}$ in compact space, under which
\be
\label{rab2}
V_{\,i}^5-iV_{\,i}^6\ \stackrel{{\cal R}_{56}}{\rightarrow}  \ e^{2i \alpha}\,(V_{\,i}^5-iV_{\,i}^6)\,,
\ee
\vspace{-7mm}
\be
\label{rab3}
{P}^5-i{P}^6\ \stackrel{{\cal R}_{56}}{\rightarrow}  \ e^{2i \alpha}\,({P}^5-i{P}^6)\,,
\ee

\noindent
and similarly for the extra components of the covariant translation operator  ${\cal P}^{\hat \mu}$ in 6 dimensions.

 \vspace{2mm}

The translation of these adjoint gauge scalars (constrained to $\,f_{ijk}\,<\!a_j\!>\ <\!b_k\!>\ =0\,$ for the potential to be minimum) leads to a spontaneous breaking of the non-abelian gauge symmetry.
It generates in the anticommutation relations (\ref{2susy}) finite field-independent parts
\be
<Z>\ \,= \,g\,T_i <\!a_i\!> \ \ \hbox{and}\ \ <Z'>\ \,= \,g\,T_i <\!b_i\!>\,.
\ee
These  now truly deserve the name of central charges, commuting between  themselves and with all unbroken symmetry generators \cite{Z,2guts}
\be
<Z> \ \hbox{and}\  <Z'>\ \hbox{\em are the central charges}.
\ee

\vspace{1mm}

This leads to the {\it spontaneous generation of central charges} in the anticommutation relations of the $N\!=2$ super\-symmetry algebra, 
i.e.~to {\it a spontaneous modification of the graded symmetry algebra}. Note that a central charge $Z_\circ$ may already be present before \hbox{spin-0} fields get translated.
These symmetry operators act as abelian, 
commuting in particular with all gauge symme\-try generators surviving the spontaneous breaking.
In practice we shall often drop the symbols $\,<\ >\,$ and simply refer for convenience to $Z$ and $Z'$, instead of $<\!Z\!>$ and  $<\!Z'\!>$, as being the central charges.

\vspace{1mm}

\subsection {Solving the ``doublet-triplet splitting problem''}

\vspace{1mm}

Central charges  can thus be ``spontaneously generated'' in a $N=2$ supersymmetry algebra,
 one of them at least being closely connected with the spontaneous breaking of the grand-unification symmetry \cite{Z,2guts}. 
 It  could be the weak-hypercharge operator $Y$,  \,spontaneously generated in the algebra through the symmetry breaking
$\,SU(5)\to SU(3)\times SU(2)\times U(1)\,$, in a way which preserves the rank of the gauge group.

    \vspace{2mm}
    
 But $Y$ is not conserved by the electroweak breaking. 
And, conversely, this breaking cannot occur  in a $N=2$ theory with a central charge proportional to $Y$. This is easily seen as initially massless spin-0 BEH doublets with $Y=\pm \,1$ would acquire large masses $ \frac{3}{5}\,m_X$, getting unable to trigger the electroweak breaking.

    \pagebreak
    
But we can start with initially massive quintuplets of mass $m$ before the grand-unification breaking, and take advantage of the {\it flat directions for the adjoint or singlet gauge scalars} $a_i$ and $b_i$ in a $N=2$ theory.  \,Indeed the adjoint mass parameter vanishes in the superpotential, already as a consequence of $R$ symmetry, as for the singlet $S$ in the nMSSM.\,
The magnitude of the adjoint v.e.v., denoted by $V$, can then freely adjust in the weak-hypercharge direction preserving $SU(3)\times SU(2) \times U(1)$, so that the resulting doublet mass parameter $\,m_D=m-\frac{3}{2}\, gV$ vanishes. This vanishing allows  for the electroweak breaking, with  the triplet mass parameter $\,m+\, gV$ getting identical to $m_X=\frac{5}{2}\, gV\,$ \cite{2guts}:
 \be
 \label{split}
 \framebox [8.55cm]{\rule[-.6cm]{0cm}{1.38cm} $ \dis
\ m_{\rm quint.}\!\!\stackrel{\ba{c}\footnotesize \hbox{\em GUT br.}\ea}{\longrightarrow} \! \left\{ \ba{ccccl}
 m_D\!&=&\! m-\frac{3}{2}\,gV \!&\equiv&\ 0\ ,
 \vspace{1.8mm}\\
 m_T\!&=&\! m\ +\ gV \!&=&\!\frac{5}{2}\,gV\,\equiv\,m_X\,.
 \ea \right.
 $}
 \ee
 
 \vspace{3mm}

This mechanism provides an automatic and natural solution to the ``doublet-triplet splitting problem''. This one is usually considered as a serious difficulty  
for the electroweak breaking in a $N=1$ supersymmetric grand-unified theory (and even more in a non-supersymmetric one), requiring a very large and unnatural adjustment for parameters of the order of the grand-unification scale.
This severe fine-tuning problem is solved by moving to $N=2$, and from there to higher-dimensional theories. The vanishing of the doublet mass parameter  $m_D$ allows for their translation generating the electroweak breaking. Even better, this translation requires, conversely, that the doublet mass parameter $m_D$ vanishes exactly {\it by locking it to 0, for the energy to be minimum}.

 \vspace{.5mm}
  
\subsection{\boldmath The massive  \vspace{2mm}  $X^{\pm 4/3},\,Y^{\pm 1/3}, W^\pm$ and $Z$ \hbox{\ \ \ \ multiplets}, within $N=2$ supersymmetry}

 \vspace{.5mm}
All spin-1 gauge bosons must then belong to massless or massive multiplets of $N=2$ supersymmetry.
But there are {\it three different types of massive gauge multiplets}, in contrast with $N=1$. 
Type I  multiplets are appropriate to describe the $W^\pm$ and $Z$ as in (\ref{Z2},\ref{W2}). They  involve no central charge, and may be complex or real.
Type II and type III multiplets, on the other hand, have a non-vanishing value of the central charge $<Z>$ (from now on simply denoted by $Z$) and  are necessarily complex. They differ by their field content, and are appropriate to the description of grand-unification gauge bosons such as $X^{\pm 4/3}$ and $Y^{\pm 1/3}$, in a $SU(5)$ theory.
 
 \vspace{2mm}
These grand-unification bosons, which have the same weak-hypercharge $Y=\pm \,5/3$, belong to two multiplets  with the same value of the central charge
  \be
 \label{zxy}
Z(\hbox{\small $X^{\pm 4/3}$})\,=\,Z(\hbox{\small $Y^{\pm 1/3}$})\,=\,\pm \ m_X\,
 \ee
 (up to a possible convention-dependent sign).
This one includes a contribution $\pm\, \frac{3}{5}\ m_X Y$, \,spon\-ta\-neously generated \cite{Z}
 in the $N=2$ algebra when the grand-unification symmetry is spontaneously broken to $SU(3)\times SU(2)\times U(1)\,$. 
 This central charge reads
  \be
  \label{Z}
Z\,=\,Z_\circ +\,\hbox{\small$\dis\frac{3}{5}$}\,m_X\,Y\,,
 \ee
and is such that
 \be
 \label{ZD}
 Z(\hbox{\em spin-0 doublets})\,\equiv\,0\,,
 \ee
 as in (\ref{split}),
 naturally allowing for the doublet translations responsible for the electroweak breaking.
 
 \vspace{2mm}
The $X^{\pm 4/3}$  belongs to a smaller multiplet, of type II, including a single spin-0 boson $x^{\pm 4/3}$ and two Dirac xino (anti)triplets, with
$m_X=|Z|> 0$. The $Y^{\pm 1/3}$ belongs to a larger  multiplet, of type III, with 5 spin-0 bosons and 4 Dirac yino (anti)triplets,  verifying altogether \cite{Zgut}
\be
\left\{\ 
\ba{ccc}
\hbox{type II}:& \ m_X\,=\,|Z|\,> \,0\,,
\vspace{1mm}\\
\hbox{type III}: & \ m_Y> |Z| = m_X >0\,.
\ea\right.
\ee
The smaller character of  the $X^{\pm 4/3}$ multiplet  as compared to the $Y^{\pm 1/3}$ one is associated with the mass equality $m_X=|Z|$, in contrast  $m_Y>|Z|$. This may be easily understood when moving to 6 dimensions, where  the $X^{\pm 4/3}$ is massless in relation with an unbroken $SU(4)$ \,{\it electrostrong symmetry} in 6 dimensions while the $Y^{\pm 1/3}$ is already massive (with mass $m_W$) in 6 dimensions.

 \vspace{2mm}

The $Y^{\pm 1/3}$ multiplet, 
\vspace{0mm}
larger than the $X^{\pm 4/3}$ one, also accommodates the 4 triplet components from the 4 quintuplets. All of them have the same mass$^2$ $m_Y^2$, including a $m_X^2$ contribution in agreement with (\ref{split}). These 4 quintuplets describe in particular the 4 spin-0 doublets responsible for the electroweak breaking in a $N=2$ theory, also providing spin-0 partners for the $W^\pm$ and $Z$ bosons.

 \vspace{2mm}
The $Y^{\pm 1/3}$ (and associated partner) mass$^2$ originates from the two contributions generated by the adjoint and doublet v.e.v.'s $\,V$ and $v$, respectively, so that
\be
\label{myw}
m_Y^2 \,= \,\underbrace{m_X^2}_{\hbox{\small$Z^2$}} +\,m_W^2\,.
\ee

\vspace{-2mm}
\noindent
The $W^\pm$ and $Z$, on the other hand, carry no central charge $Z$ and belong to massive gauge multiplets of type I,  describing 4 inos and 5 spin-0 bosons for every spin-1 particle, as in (\ref{Z2},\ref{W2}).

\vspace{2mm}

 When supersymmetry is broken mass relations similar to (\ref{winos}), like
 \be
 \label{xinos}
 \!\!\left\{\,
 \ba{ccl}
 m^2(\hbox{xinos}) \!&=&\!m_X^2+m_{3/2}^2\,,
 \vspace{.5mm}\\
 m^2(\hbox{yinos})  \!&=&\!m_Y^2 +m_{3/2}^2= \,m_X^2 +m_W^2+ m_{3/2}^2 \,,\!\!
 \ea\right.
 \ee
 are  obtained for xinos and yinos, and interpreted in terms of momenta $\pm \,m_{3/2}$ carried along an extra compact dimension \cite{gutbis}.


 \section{Supersymmetry  \vspace{1.5mm}  and  \vspace{1.5mm}  grand-unification   \hbox{in extra dimensions}}
\label{sec:dim}

 \vspace{1mm}

\subsection{\boldmath From $N=2$ supersymmetry to 6 dimensions}

These $N=2$  theories may then be formulated in a 5 or 6 dimensional space-time \cite{epjc,gutbis}, with the ``central charges''  $Z$ and $Z'$ getting  turned into the 5th and 6th components of the (covariant) momentum  along the compact dimensions \cite{Z}.
The two spin-0 photons and spin-0 ``sgluon'' octets present in $N = 2$ supersymmetry get described by the fifth and sixth components of the photon and gluon fields $V^{\hat \mu}_{\,i}$  in 6 dimensions.~\,{\it The $W^\pm$ and $Z$ masses are already present in 6 dimensions}, where the photon and gluons are coupled with the same strength. 
 
 \vspace{2mm}
 
 Viewing  $\,Q\, \sqrt {3/8}\,$ as one of the $SU(4)$ {\it electrostrong symmetry} generators, suitably normalized
 in the same way as  for the $SU(3)$ generators,  provides in 6d   the $SU(4)$ relation 
 between the electromagnetic and strong couplings,

\vspace{-4mm}
 \be
 \ba{l}
 \!\!\!\!\hbox{ \underline{\em electrostrong} \em \,symmetry}\ \ \Longrightarrow
 \vspace{2mm}\\
 \dis\ \ \  e_{\,6d}\,=\, \sqrt{\,\frac{3}{8}}\ \ g_{3\,6d}\,, \ \ \hbox{i.e.} \ \ \alpha_{\,6d}\, = \,\frac{3}{8}\,\ \alpha_{3\,6d}\,.
  \ea
  \ee
 This relation is exact in 6d as long as we do not introduce the grand-unification breaking 
 through antiperiodic boundary conditions for GUT-odd particles, discussed in the next subsection.  
We also have, by returning to $SU(5)$ to include weak in addition to electrostrong interactions,
$\sin^2\theta=$ $ e^2/g^2=3/8 \,$ for the electroweak angle in 6d,  at the classical level.
 
  \vspace{2mm}

The central charge $Z$ of the $N=2$ algebra in 4 dimensions, essential to the discussion of the grand-unification breaking, originates from the fifth component of the (covariant) momentum along a compact dimension, according to

\vspace{-4mm}

 \be
 {\cal P}^5 
 \,=\,-\left(Z_\circ +\,\hbox{\small$\dis\frac{3}{5}$}\,m_X\,Y\right)\,
 \ee
 (up to a possible convention-dependent sign). Once we are in 5 or 6 dimensions, we only have to refer to 
 the extra components of the covariant momenta, 
 ${\cal P}^5$ and ${\cal P}^6$, rather than to the corresponding central charges $Z$ and $Z'$ of the 4d $\,N=2$ theory.

  \vspace{1.5mm}

 The particle content of the $N = 2$ multiplets are given in \cite{2guts,gutbis}.
 The massive gauge multiplet (\ref{Z2}) describing the $Z$ in 4d originates from the massive $Z$ gauge multiplet in 6d, such that
\be
\label{Z3}
\framebox [8.55cm]{\rule[-.25cm]{0cm}{1.18cm} $ \dis
Z  \stackrel{6d}{\stackrel{\ba{c}\vspace{-6mm}\\ \hbox{\footnotesize \it SUSY}\vspace{0mm}\\ \ea }{\Longleftrightarrow }}  \,\hbox{8-comp. Dirac zino}  
 \stackrel{6d}{\stackrel{\ba{c}\vspace{-6mm}\\ \hbox{\footnotesize \it SUSY}\vspace{0mm}\\ \ea }{\Longleftrightarrow }}
 \,  \hbox{3 spin-0 bosons}.
$}
\ee

  \vspace{1mm}
This multiplet reduces to (\ref{Z2}) in 4 dimensions, in which the $Z$  is associated with 4 Majorana zinos and 5 spin-0 BEH bosons. Similar expressions hold for the $W^\mp$ and $Y^{\pm 1/3}$ multiplets. This set of 5 neutral spin-0 BEH bosons associated with the $Z$ in 4 dimensions, before the breaking of the $N=2$ supersymmetry, is similar to 
the nMSSM one in (\ref{Z2}) for $\lambda = \sqrt{(g^2+g'^2)/2}\,\simeq .52$, and presumably includes the 125 GeV boson found at CERN.

  \vspace{2mm}
  
 $N=2$ theories in 4 dimensions, being vectorlike, also include mirror partners for quarks and leptons, to which they are coupled through the exchanges of spin-0 gluons and photons, in particular. But no such particles have been observed yet.
 Their presence at low energy may be avoided by considering a mirror-parity operator $M_p$ under which 
mirror particles, as well as spin-0 quarks and photons, ...\,, are $M_p$-odd. 
The $Z$ multiplet  then gets further reduced,  to include just a single spin-0 boson $z$ associated with the $Z$.

  \vspace{2mm}
Indeed among the 4 spin-0 doublets (originating from 4 quintuplets)  the usual ones $h_1$ and $h_2$ are taken as $M_p$-even so as to survive in the low-energy theory.
Their $N=2$ partners $h'_1$ and $h'_2$, being $M_p$-odd,  disappear from the low-energy theory. The definition of this $M_p$ operator involves in particular, as seen from (\ref{rab}
 -\ref{rab3}), a rotation ${\cal R}_{56}(\pi)$, equivalent to a reflexion symmetry in compact space,
$x^5\to -\,x^5,\ x^6\to -\,x^6$, \,under which 
\be
V^\mu_{\,i}\, \to\,V^\mu_{\,i}\,;\ \ \ V^5_{\,i}\, \to\, -\,V^5_{\,i}\,,\ \ V^6_{\,i}\,\to\,-\,V^6_{\,i}\, . 
\ee
 
\vspace{1mm}

 Anticipating on the supersymmetry breaking discussed in the next subsection, the 
$R$-odd zinos are present only at the compactification scale associated with the sixth dimension, 
starting with two Dirac zinos (combining a Dirac gaugino with a Dirac higgsino), at mass$^2$ \cite{gutbis}
\be
m^2(\hbox{\em zinos})\,=\,m_Z^2+\pi^2/L_6^2\ .
 \ee

 \vspace{1mm}

 We then simply remain, in the low-energy 4d theory below the compactification scales, with the field-content 
 of the standard model but for the presence of the two spin-0 doublets $h_1$ and $h_2$,
 with quartic doublet couplings fixed by
$(g^2+g'^2)/8$ and $g^2/2$ as in (\ref{susy},\ref{mhZ},\ref{lhhMSSM}). This is crucial for the gauge-BEH unification according to which \cite{R,gh,prd14}
 
 \vspace{-4mm}
\be
\label{gh3-1}
\hbox {\framebox [8cm]{\rule[-.28cm]{0cm}{.85cm} $ \dis
\hbox{\em spin-1} \ \,Z\ \  \stackrel{\ba{c}\hbox{\footnotesize \it SUSY}\ea}{\longleftrightarrow} 
\,\stackrel{\ba{c}\hbox{\footnotesize \it SUSY}\ea}{\longleftrightarrow}\!\  \ \hbox{\em spin-0 \,BEH boson}\, z\ ,
$}}
\ee
with the spin-0 $z$ having the same mass as the $Z$ before supersymmetry breaking effects get taken into account.
We also expect, in the same way, the following association

 \vspace{-4mm}
\be
\label{gh3-1w}
\hbox{\em spin-1} \ \,W^\pm\ \  \stackrel{\ba{c}\hbox{\footnotesize \it SUSY}\ea}{\longleftrightarrow} 
\,\stackrel{\ba{c}\hbox{\footnotesize \it SUSY}\ea}{\longleftrightarrow}\!\  \ \hbox{\em spin-0 \,BEH boson}\ w^\pm\ ,
\ee
for the charged spin-0 boson in  (\ref{wh}),
\be
\label{wh2}
w^\pm\, \equiv\, H^\pm=\,\sin \beta\  h_1^\pm +\cos\beta \ h_2^\pm\,.
\ee

 \vspace{2mm}

 In a grand-unified theory, with for example $SU(5)$  as the gauge group, this one gets spontaneously broken down to an $SU(4)$ 
 {\it electrostrong symmetry\,}  group. The  $X^{\pm 4/3}$ (anti)triplet remains massless in 6d, where the $Y^{\pm 1/3}$ (anti)triplet
 has the same mass $m_W$ as the $W^\mp$, with
 which they form a 
 \be
SU(4)\   \hbox{\em electrostrong antiquartet}\ \ \ \left(\ba{c}  
 Y^{+1/3}
 \vspace{0mm}\\  W^-
 \ea\! \right).
 \ee
 
 \vspace{2mm}
 
In this higher-dimensional space-time, the $SU(5)$ symmetry is broken through the BEH-quintuplet v.e.v.'s,
providing in 6 dimensions equal masses to the $ Y^{\pm 1/3}$ and $W^{\mp}$ gauge fields, according to
\be
\label{es}
\framebox [8.55cm]{\rule[-.2cm]{0cm}{1.18cm} $ \dis
SU(5) \stackrel{\hbox{\small \em EW breaking}}{\stackrel{\ba{c}\ \vspace{-6.3mm}\\ \hbox{\small \em in 6d}\ea}{\longrightarrow }} SU(4) \ \hbox {\it electrostrong gauge group,}
$}
\ee
leading to  the $SU(4)$ relation 
 $\,\alpha_{\,6d}\, = \, (3/8)\ \alpha_{s\,6d}\,$
in 6 dimensions,  and to $\sin^2 \theta=$ $3/8$, at the classical level.
 
 \vspace{2mm}
 
  This electrostrong-weak breaking in 6d {\it separating weak from electrostrong interactions} leads  in 4d to the mass relation (\ref{myw})
\be
m_Y^2 \,= m_X^2 +m_W^2 \,
\ee
found previously, with $m_X^2={P}_5^2+{P}_6^2$, for the $X$ and $Y$ gauge bosons and their susy partners. 
These relations are valid for each excitation level of the extra compact dimensions, for the $X^{\pm4/3}$ and $Y^{\pm 1/3}$ gauge fields in 6 dimensions.

 \vspace{1mm}
 
 \subsection{Grand-unification \vspace{1.5mm}  and  supersymmetry  breaking from 
 {\em \,discrete\,} boundary conditions in compact space}

\vspace{1mm}
 
The extra compact space dimensions may play an essential role in the breaking of the supersymmetry and grand-unification symmetries, together
with the mirror-parity operator $M_p$ allowing to avoid mirror quarks and leptons, spin-0 gluons and photons,
and additional spin-0 BEH bosons, 
in the low-energy spectrum. This may be done through boundary conditions involving, in an interesting way, {\it discrete} rather than continuous symmetries. They include $R$-parity, a $GUT$-parity $G_p$ and mirror parity $M_p$ acting as translation and reflexion symmetries in the compact space, thanks to its topological properties \cite{gutbis,epjc}. 
\vspace{2mm}

These three discrete symmetries naturally allow for the presence at low energy of the two spin-0 doublets $h_1$ and $h_2$, even under $R_p,\,G_p$ and $M_p$. They can thus
generate the same spontaneous electroweak breaking in 4 dimensions as already resulting in (\ref{Z3},\ref{es}) from the grand-unification breaking into the $SU(4)_{es}$ electrostrong symmetry subgroup in 6 dimensions, with the $W^\pm$ and $Z$ masses in 4 dimensions directly originating from the 6d theory.

\vspace{2mm}

The breaking of the supersymmetry may be obtained by identifying the action of travelling along a complete loop 
${\cal L}_6$
in compact space (i.e.~\,a translation 
$x^6\to x^6 +L_6$, in the simplest example of a flat torus) with a discrete $R$-parity transformation, 
$R_p=(-1)^{3(B-L)}\ (-1)^{2S}=\pm 1\,$:

\vspace{-4mm}
\be
\framebox [8.55cm]{\rule[-.53cm]{0cm}{1.27cm} $ \dis
\ba{c}
\hbox{\em travelling along a complete loop ${\cal L}_6$ in compact space}\ \ 
\vspace{1mm}\\\equiv\ \  R\hbox{\em -parity transformation}\,.
\ea $}
\ee

\vspace{1mm}
\noindent
This makes all superpartners naturally ``very heavy'', i.e. at the compactification scale:
\be
m(R\hbox{\em -odd superpartners})\  \approx\ \hbox{\em compactification scale\,.}
\ee

\vspace{2mm}
This compactification scale is unknown but has to be $\simge $ TeV scale, at least. We may thus be lucky enough to see superpartners, together with the opening of extra space dimensions, in a not-too-distant future. But  we may also have to face the eventuality that superpartner masses be considerably larger than the presently accessible 
$\approx $ TeV scale, especially if the compactification of extra dimensions also sets the scale for the grand-unification breaking. 
The latter scale, however, may be substantially reduced as compared to usual expectations, as we shall see.

\vspace{2mm}

In a similar way, the breaking of the $SU(4)$  {\it electro\-strong symmetry\,}  group in 6d 
may be obtained by identifying travelling along   a complete loop ${\cal L}_5$  (e.g.~a translation $x^5 \to x^5+L_5$ on a flat torus) 
with a discrete $Z_2$ $GUT$-parity transformation $G_p$,
\be
\framebox [8.25cm]{\rule[-.53cm]{0cm}{1.27cm} $ \dis
\ba{c}
\!\hbox{\em travelling along another loop ${\cal L}_5$  in compact space\,} 
\vspace{1mm}\\  \equiv\ \  GUT\hbox{\em -parity transformation}\,,
\ea
 $}
\ee
This one is defined from expression (\ref{Z})  of the central charge $Z$, as $e^{i\pi Z/m_X}$, or more precisely as 
\be
\label{gp}
\ba{c}
GUT\hbox{\em -parity}\ \ G_p\,=\ G'\,\times\ e^{\,i\pi\frac{3}{5}Y}\,=\,(-1)^{Z/m_X}\,=\,\pm\, 1\,.
\vspace{2mm}\\ 
\ea
\ee

\vspace{2mm}

Here $\,G'\!=e^{i\pi Z_\circ/m_X}$ is a global symmetry operator commuting with both $SU(5)$ and  supersymmetry, acting 
in particular on quark and lepton grand-unification multiplets, and \hbox{spin-0} BEH multiplets.
$G_p$ may be expressed in terms of the central charge $Z$  present in the $N\!=2\,$ supersymmetry algebra in 4 dimensions, as in (\ref{2susy},\ref{zxy}
-\ref{ZD}), part of which, proportional to the weak-hypercharge $Y$, is generated 
spontaneously during the breaking  of the grand-unification symmetry \cite{Z,2guts}.
Taking the fifth dimension as cyclic, of size $L_5$, with periodic and antiperiodic boundary conditions for 
$GUT$-even and $GUT$-odd  fields, we can identify the action of a $GUT$-parity transformation with the one
of a translation of $L_5$ on the torus, so that
\be
\ba{l}
\hbox{\em action of \ $GUT$-parity}
\vspace{2mm}\\
\ \ \ \ \,\equiv\,\hbox{\em action of}\ \ e^{\,i \,P_5L_5}\,=\,(-1)^{P_5/\frac{\pi}{L_5}}\,.
\ea
\ee

The $X^{\pm 4/3}$ and $Y^{\pm 1/3}$ gauge bosons, with $Y\!=\pm \, 5/3$, are odd under $GUT$-parity, and carry momenta
$(2n_5+1)$ $\pi/L_5$ along the fifth dimension.
This is also the case for the spin-0 triplet partners of the electroweak doublets within $SU(5)$ quintuplets. These belong to the same massive gauge multiplet of type III as the $Y^{- 1/3}$, in agreement with  (\ref{zxy},\ref{gp}) \cite{2guts,gutbis,Zgut}.

\vspace{2mm}

  This {\it triplet-doublet splitting mechanism,\,} already operating within $N=2$ supersymmetric GUTs as in (\ref{split})  \cite{2guts}, is of the same nature as the one splitting the $X^{\pm 4/3}$ and $Y^{\pm1/3}$ masses away from the gluon, photon, $W^\mp$ and $Z$ masses. It provides for their physical spin-0 triplet and doublet components the same masses as for  the $Y^{\pm 1/3}, \,W^\mp$ and $Z$, before the super\-symmetry breaking. The massless components associated with the would-be Goldstone bosons are eliminated when the 
$Y^{\pm 1/3}, \,W^\mp$ and $Z$ acquire their 6d masses $m_W,\, m_W$ and $m_Z$.

\vspace{2mm}

The $X^{\pm 4/3}, \ Y^{\pm 1/3}$ and color-(anti)triplet spin-0 bosons, being $G_p$-odd,  have no direct couplings between two ordinary (anti)quarks or (anti)leptons, even under $GUT$-parity. This is in contrast with ordinary GUTs, 
and implies that 
\be
\hbox{\it the proton is expected to be stable}\,,
\ee
at least in the simplest situations \cite{epjc,repli}. The corresponding compactification scale 
associated with the grand-unification breaking might then be lower, and possibly significantly lower,  than the $\approx 10^{16}$ GeV usually considered.

\vspace{2mm}

Altogether the spontaneous breaking of  the supersymmetry and grand-unification symmetries may both be induced through the compactification of the extra dimensions. This leads to the possibility of fixing the scales associated with these breakings in terms of the compactification scales for the extra dimensions \cite{gutbis}.  In the simplest case of two flat extra dimensions and for the lowest-lying excited states, we would get relations like
\be
\label{LL}
\left\{\ 
\ba{ccl}
m_{3/2} \!&=&\! \dis \frac{\pi}{L_6}=\, \frac{1}{2R_6}  
\vspace{1mm}\\
&& \hbox{(from \ {\em $R$-parity} $\,\equiv\,$
translation of $L_6$\,)}\,,
\vspace{2mm}\\
 m_X \!&=&\! \dis  \frac{\pi}{L_5}= \,\frac{1}{2R_5}
\vspace{1mm}\\
&&\hbox{(from \  {\em $GUT$-parity} $\,\equiv\,$
translation of $L_5$\,)}\, ,
\ea \right.
\ee

\vspace{1mm}
\noindent
up to radiative corrections. The lowest-lying superpartners, or grand-unification particles, are expected to be present at these mass scales determined by $m_{3/2}$ and $m_X$, respectively.

\vspace{2mm}

This use of {\it discrete} boundary conditions associated with a non-trivial topology, involving for supersymmetry $R$-parity rather than a continuous symmetry, allows to link rigidly these fundamental supersymmetry and grand-unification breaking parameters $m_{3/2}$ and $m_X$ to the compactification scales.   This approach contrasts with the initial one in \cite {ss} disregarding fields corresponding to excited states that become infinitely massive
when the size of the compact space is shrunk to zero (in particular states of masses proportional to $\pi/L$, essential here). We obtain instead {\it quantized mass parameters\,} fixed in terms of the compactification scales, with the {\it geometry\,}  now determining the masses of the new particles in which we are interested.

\subsection{Implications for the compactification scales}

The resulting 4d theory has, in its simplest version, {\it the same content as the standard model\,} at low-energy {\it but for 
the two spin-0 doublets $h_1$ and $h_2$}, \,while still allowing 
for the gauge/BE-Higgs unification that is one of the most interesting features of supersymmetric theories.  
The new (sixth) dimension opens up at the compactification scale $m_{3/2}$, 
 i.e.~$\pi /L_6$ in the simplest case.  Superpartners, mirror particles, spin-0 gluons etc., as for a $N=2$ theory in 4 dimensions \cite{2guts}, would  then appear at or above this threshold.
They now originate from a $N=1$ theory in 5 dimensions, its mass spectrum resulting from the discrete boundary conditions involving the $R_p$ and $M_p$ symmetries.

\vspace{2mm}
Let us assume  $m_{3/2}$ smaller than $m_X$.  Below $m_{3/2}$ the theory has in its simplest version
the same field content as the standard model, but for the second spin-0 doublet.
The evolution of the gauge couplings 
(or simply of the differences $g_i^{-2}-g_j^{-2}$) in the 4d theory between $m_W$ and $m_{3/2}$ is only slightly modified as compared to the standard model. It does not lead a grand-unification of these three couplings below the compactification scale.

\vspace{2mm}

Above $m_{3/2}$ the theory gets 5-dimensional, non-renor\-malisable, with gauge couplings having the dimension of mass$^{-1/2}$.
We can no longer discuss as usual the running of the gauge couplings.  
One may still feel tempted to continue evaluating an evolution of effective couplings with energy,
taking into account only a finite number of states up to a cut-off mass $\Lambda$,
but one should be cautious before drawing conclusions.
In addition the asymptotic freedom of QCD is expected to be ruined 
owing to the extra mirror families of quarks and leptons, unless one considers that quarks, leptons and mirror partners do not have excited states for the compact dimensions.

\vspace{2mm}

We recall that the dimension ($x^6$)  responsible for the evolution of (effective) gauge couplings 
\vspace{-.4mm}
between $m_{3/2}$ and $m_X$ 
is distinct from the one  ($x^5$) responsible for the breaking of the GUT symmetry at the higher grand-unification scale. The latter, $m_X=\pi/L_5$, however, may be only slightly larger than  $m_{3/2}=\pi/L_6$.

\vspace{2mm}

$\sin^2\theta$, evaluated in a 4d theory, is in particular sensitive to the number of \hbox{spin-0} doublets and associated higgsinos, usually $\,h_1,h_2$ and $\tilde h_1,\tilde h_2$ in a $N=1$  theory (counting very much as for 6 spin-0 doublets) with the field content of the (N/n)MSSM. \,Only the 2 doublets $h_1$ and $h_2 $, without their accompanying higgsinos, are here present in the 4d theory below $m_{3/2}$.
The evolution of the three gauge couplings, if extrapolated up to a unified value within the 4d theory,  would necessitate a too small $\sin^2\theta$, much as for  the standard model.

\vspace{2mm}
This may be more than compensated, however,  by the 4 spin-0 doublets $\,h_1,h_2,\,h'_1,h'_2 \ +$  associated higgsinos  
(twice as much as in the MSSM) present in the 5d theory between $m_{3/2}$ and $m_X$. These extra doublet degrees of freedom (taking also into account the extra adjoint gaugino and spin-0 fields) tend to lead to a too large value of $\sin^2 \theta$.  This indicates that the correct value of $\sin^2\theta$ may be obtained from a balance between these two effects,
presumably with $m_{3/2}$ not far below $m_X$. 

\vspace{2mm}

If the proton is indeed stable $m_X$ may be much lower than the usual $\approx 10^{16}$ GeV scale, especially with a faster evolution of effective gauge couplings in the 5d theory between $m_{3/2}$ and $m_X$. Their unification may then occur for a rather low value of the grand-unification scale $m_X$. One may even imagine that these unification and compactification scales 
 \be
 m_{3/2}=\,\frac{\pi}{L_6}\ <\ m_X=\,\frac{\pi}{L_5}
 \ee 
 may be not so far above the $\approx$ few TeV scale soon accessible at LHC. This would provide new perspectives for a possible discovery of superpartners, new space dimensions and maybe grand-unification particles, in a not-too-distant future.

\vspace{2mm}

This set of jointly-operating mechanisms, based on supersymmetry, extra dimensions and discrete symmetries, allows for the electroweak
breaking 
 to already occur in 6d dimensions, where it leaves unbroken an electrostrong symmetry group.  It provides in 4d the electroweak breaking  induced by $h_1$ and $h_2$ at low energies, even in the presence of  significantly larger scales associated with grand-unification and possibly (in a more remarkable way) supersymmetry breaking. 
In particular
\be
\hbox {\framebox [7.6cm]{\rule[-.53cm]{0cm}{1.25cm} $ \dis
\ba{c}
\hbox {\em no fine-tuning between GUT-scale parameters}
\vspace{1mm}\\
\hbox{\em is  required,}
\ea
$}}
\ee
and the electroweak breaking in the low-energy theory appears largely insensitive to the behavior of the higher-dimensional theory.

\vspace{2mm}
To each of the three conserved symmetries $R_p,\ G_p$ and $M_p$ acting in compact space is associated a stable particle, possible candidate for the non-baryonic dark matter of the Universe. The LSP and LGP, lightest  supersymmetric and lightest grand-unification particles, are directly associated with the excitation of the compact sixth and fifth dimensions.   The mirror-parity operator $M_p$, associated with the reflexion of the compact coordinates  (or rotation of $\pi$ in compact space) leads to the 
lightest $M$-odd particle or LMP. This one, to be found among mirror quarks and leptons, and spin-0 gluons and photons or other neutral spin-0 gauge bosons, \,...~, is also associated with the excitation of the compact dimensions.

\vspace{-1mm}


\section{Conclusion}
 
\vspace{1mm}

In addition to superpartners, supersymmetric theories lead to an extended set of spin-0 bosons $H^\pm, H,h,A, ...\,$. \ Some of them appear as extra states for massive \hbox{spin-1} gauge bosons,
providing a relation between spin-1 mediators of gauge
 interactions and spin-0 particles associated with symmetry breaking and mass generation.

\vspace{2mm}

Searches for supersymmetric particles started in the late seventies, first looking for light gluinos and associated $R$-hadrons, light charged sleptons, etc., often relying on the missing-energy momentum carried away by unobserved neutralino or gravitino LSP's, at the modest energies accessible at the time  \cite{ff,ff2}. 
Considerable work has been done since throughout the world, most notably at PETRA (DESY) and PEP  (SLAC),  LEP (CERN) and at the Tevatron (Fermilab). These searches are now at the forefront of particle physics with the restart of LHC experiments at CERN.

\vspace{2mm}

All this could not be discussed here, nor the status of the lightest supersymmetric particle, presumably a neutralino, as a possible dark matter candidate in a $R$-parity conserving theory. We know now that strongly-interacting squarks and gluinos should be heavier than about 1 TeV at least. We refer the reader to the original results from the ATLAS and CMS collaborations at LHC
 \cite{hat,hcms,susyat,susycms}, and to the other articles in this book to complete this theoretical description with the presentation of experimental results and constraints on supersymmetric particles and additional spin-0 BEH bosons.

\vspace{2mm}

The next run of LHC experiments,  with an energy increased from 8 to 13 TeV, may well allow for the direct production of supersymmetric particles, and of an extended system of spin-0 bosons including a charged $H^\pm$. Will this energy be sufficient, and at which energy scale should the new superpartners be found~? Is it indeed not too far from the electroweak scale, and accessible at LHC~? 
\,Or still significantly larger, as it could happen for superpartner masses determined by the very small size of an extra dimension
($L \simle 10^{-17}$ cm corresponding to $\pi \hbar /Lc \,\simge\, 6$ TeV$/c^2$)\,?

\vspace{2mm}

In any case the 125 GeV boson observed at CERN may well be interpreted, up to a mixing angle induced by supersymmetry breaking, 
as 
 the spin-0 partner of the $Z$ under
  two supersymmetry transformations,
  \vspace*{-2mm}
\be
\label{gh3}
\hbox{\em spin-1} \ \,Z\ \  \stackrel{\ba{c}\hbox{\footnotesize \it SUSY}\ea}{\longleftrightarrow} 
\,\stackrel{\ba{c}\hbox{\footnotesize \it SUSY}\ea}{\longleftrightarrow}\!\  \ \hbox{\em spin-0 \,BEH boson}\,,
\ee
\noindent
 i.e.~as a $Z$ that would be deprived of its spin.
 This provides within a theory of electroweak and strong interactions 
 the first example of two known fundamental particles of different spins that may be related by supersymmetry.

\vspace{0mm}

Even if $R$-odd superpartners were still to remain out of reach for some time, possibly due to large momenta along very small space dimensions,
supersymmetry could  still be tested in the gauge-and-BEH sector at present and future colliders,
in particular through the properties 
of the new spin-0 boson and the search for additional \ ones.

  \end{document}